\documentclass[12pt]{iopart}
\usepackage{iopams}
\usepackage{graphicx}

\begin{document}
 
\title[Spherical spaces]{How well-proportioned are lens and prism spaces?}
\author{R.\ Aurich and S.\ Lustig}
 
\address{Institut f\"ur Theoretische Physik, Universit\"at Ulm,\\
Albert-Einstein-Allee 11, D-89069 Ulm, Germany}

\begin{abstract}
The CMB anisotropies in spherical 3-spaces
with a non-trivial topology are analysed with a focus on lens and
prism shaped fundamental cells.
The conjecture is tested that well proportioned spaces lead to a suppression
of large-scale anisotropies according to the observed
cosmic microwave background (CMB).
The focus is put on lens spaces $L(p,q)$
which are supposed to be oddly proportioned.
However, there are inhomogeneous lens spaces whose shape of the
Voronoi domain depends on the position of the observer within the manifold.
Such manifolds possess no fixed measure of well-proportioned
and allow a predestined test of the well-proportioned conjecture.
Topologies having the same Voronoi domain are shown to possess
distinct CMB statistics which thus provide a counter-example to the
well-proportioned conjecture.
The CMB properties are analysed in terms of cyclic subgroups $Z_p$,
and new point of view for the superior behaviour of the Poincar\'e
dodecahedron is found.
\end{abstract}


\submitto{\CQG}

\section{Introduction.}
\label{sec:intro}

This paper studies the cosmic microwave background (CMB) anisotropies
of spherical 3-spaces which tile the 3-sphere ${\cal S}^3$.
A deck group $\Gamma$ of order $|\Gamma|$ tessellates
the 3-sphere ${\cal S}^3$ in $|\Gamma|$ domains that are identified.
In this way multi-connected manifolds are constructed
which provide models for the spatial structure of our Universe.
The CMB anisotropies on large angular scales differ from
that of the simply-connected ${\cal S}^3$ and
can serve as a signature for topology.
The main motivation for cosmic topology is the
low power of the CMB anisotropies
that is observed \cite{Hinshaw_et_al_1996,Spergel_et_al_2003}
at large angular scales on the microwave sky.
This feature can naturally be explained by suitably chosen
multi-connected manifolds.

An introduction to the cosmic topology can be found
in \cite{Lachieze-Rey_Luminet_1995,Luminet_Roukema_1999,Levin_2002,%
Reboucas_Gomero_2004,Luminet_2008}.
An extended discussion of spherical multi-connected manifolds is provided by
\cite{Gausmann_Lehoucq_Luminet_Uzan_Weeks_2001}.
Spherical spaces attracted attention by the paper
\cite{Luminet_Weeks_Riazuelo_Lehoucq_Uzan_2003}
which claims that the low power in CMB anisotropies on large scales
can be described by the Poincar\'e dodecahedral topology.
In the sequel this result was investigated and
extended to other spherical topologies such as the
binary octahedral space and the binary tetrahedral space
with respect to their statistical CMB properties, see e.\,g.\
\cite{Roukema_et_al_2004,Aurich_Lustig_Steiner_2004c,Gundermann_2005,%
Aurich_Lustig_Steiner_2005a,Aurich_Lustig_Steiner_2005b,Lustig_2007,%
Lew_Roukema_2008,Roukema_et_al_2008a,Roukema_Kazimierczak_2011}.
The latter are also termed truncated cube and octahedron.
Another class of spherical spaces is provided by the lens spaces 
$L(p,q)={\cal S}^3/Z_p$ specified by a cyclic group $Z_p$
that are studied in \cite{Uzan_Riazuelo_Lehoucq_Weeks_2003}.
The fundamental domains of the lens spaces can be visualised by
a lens-shaped solid
where the two lens surfaces are identified by a $2\pi q/p$ rotation
for integers $p$ and $q$
that do not possess a common divisor greater 1 and obey $0<q<p$.
For more restriction on $p$ and $q$,
see  below and \cite{Gausmann_Lehoucq_Luminet_Uzan_Weeks_2001}.

In this paper we analyse the so-called
``well-proportioned'' conjecture \cite{Weeks_Luminet_Riazuelo_Lehoucq_2005}
which states that manifolds that stretch in all directions by
roughly the same amount yield a stronger suppression of
the CMB anisotropies on large angular scales
than oddly shaped manifolds.
This is a purely geometric criterion that only uses the shape of
the fundamental domain,
but it ignores how the faces of the fundamental domain are identified.
A genuine test of this hypothesis would be provided by two deck groups
$\Gamma_1$ and $\Gamma_2$
that possess fundamental domains with the same geometric shape
but with distinct identifications of their faces.
Fortunately, such examples exist among spherical multi-connected manifolds.

We put the focus on the lens spaces $L(p,q)$ with 
$p=4n$ and $q=p/2-1=2n-1$ for the integers $n=2,3,4,\dots$.
These lens spaces are inhomogeneous in the sense
that the geometric shape of their fundamental domain,
defined as a Voronoi domain,
and their statistical properties of the CMB anisotropies
vary with the position of the CMB observer.
This allows a test of the well-proportioned conjecture
only in dependence of the position of the CMB observer
since all cosmological parameters are held fixed.
More important is, however, the fact
that the lens spaces $L(p,p/2-1)$ possess for two special observer positions
a Voronoi domain identical to those of two homogeneous spherical manifolds.
The first case is the homogeneous lens space $L(p,1)$ and
second one is the prism space ${\cal D}_p= {\cal S}^3/D^\star_p$
generated by the binary dihedral group $D^\star_p$.
The prism space ${\cal D}_p$ is also termed binary dihedral space.
For these two observer positions in $L(p,p/2-1)$
one has thus a genuine test for the well-proportioned conjecture.
One can test how far the geometry of 
the Voronoi domain is reflected in the CMB anisotropies.

The paper \cite{Weeks_Luminet_Riazuelo_Lehoucq_2005}
that put forward well-proportioned conjecture
does not provide a quantitative measure for the property of well-proportioned.
Instead, their verbal definition states that ``a well-proportioned space
being one whose three dimensions are of similar magnitudes.''
It is the aim of this paper to put this conjecture on a firmer footing
by proposing a quantitative measure for the well-proportioned property.
For a number of lens spaces $L(p,q)$ and prism spaces ${\cal D}_p$,
the CMB anisotropies are analysed with respect to this shape measure.
For comparison, we also study the Poincar\'e dodecahedral topology.

In the first step,
one has to give a definition of the fundamental domain ${\cal F}$
with respect to the deck group $\Gamma$ of the multi-connected space.
A fundamental domain ${\cal F}$ is defined as a domain
which contains every physical point only once.
Thus, there is no pair of distinct points $x,x'\in {\cal F}$
that can be mapped by elements $g\in\Gamma$ onto each other.
The natural definition of a fundamental domain is that of the Voronoi domain
which is used in this paper.
For cosmological applications it is usual as well as natural
to place the CMB observer in the centre of the coordinate system.
This facilitates the computation of the CMB anisotropies by
exploiting the spherical symmetry of the problem.

Therefore,
given a realisation of a deck group $\Gamma$ in the chosen coordinates,
it is natural to define the fundamental cell ${\cal F}$ in such a way
that the fundamental cell ${\cal F}$ does not contain points
which can be transformed by a element $g\in\Gamma$ any closer to the
observer sitting in the centre $x_o$,
i.\,e.\
\begin{equation}
x\in {\cal F} \hspace{5pt} \hbox{ if } \hspace{5pt}
d(x_o, x) \; \leq \; d(x_o, g(x))
\hspace{10pt} \hbox{ for all } g \in \Gamma
\hspace{10pt} ,
\end{equation}
where $d(x_1,x_2)$ 
measures the spherical distances between the points
$x_1,x_2\in{\cal S}^3$.
A fundamental domain ${\cal F}$ constructed in this natural way
is called Voronoi domain.

Let us now turn to the question
whether a manifold is homogeneous or inhomogeneous.
Consider two observers such that the first observer position
can be mapped by a transformation $M \notin \Gamma$ onto the second one.
Assume that the first observer determines his Voronoi domain by the
group elements $g\in \Gamma$,
then the second observer gets his Voronoi domain by
a similarity transformation of the group $M^{-1} g M$.
If $M$ and $g$ commute for all $g\in \Gamma$,
then both observers construct the same Voronoi domain,
i.\,e.\ the manifold is homogeneous.
On the other hand, if $M$ and $g$ do not commute,
the constructed Voronoi domain depends on the position of the observer.
In this case the manifold is called inhomogeneous.

In \cite{Uzan_Riazuelo_Lehoucq_Weeks_2003}
the inhomogeneous lens spaces $L(p,q)$ with $q>1$
are investigated only for one special observer
which is located at the centre of the spherical lens.
For this observer position,
none of these lens spaces show small correlations of the CMB on large scales
as seen in the data of COBE \cite{Hinshaw_et_al_1996}
and WMAP \cite{Copi_Huterer_Schwarz_Starkman_2008}.
This leads to the question
whether the exploitation of the inhomogeneity of lens spaces
leads to viable models.
This paper extends our previous work \cite{Aurich_Kramer_Lustig_2011}
which analyses the CMB anisotropy of the lens space $L(8,3)$ in comparison
to that of the lens space $L(8,1)$ 
and the prism space ${\cal D}_8$.
In this paper the lens spaces $L(p,1)$ and $L(p,p/2-1)$
as well as the prism spaces ${\cal D}_p$
are compared with respect to the well-proportioned conjecture.
This comparison is very interesting
because it shows how strong the influence of the geometry of 
the Voronoi domain on the CMB anisotropies is.
The CMB anisotropies are also analysed in terms of the cyclic subgroups
of Clifford translations of the deck groups, 
of the transformation behaviour of the deck group
on the sphere $S_{\hbox{\scriptsize sls}}$ of last scattering,
and of the general observer position.

\section{Specification of the spherical manifolds $L(p,q)$ and ${\cal D}_p$}
\label{sec:spherical_lens}

In order to specify spherical manifolds,
one has to define the representation of the 3-sphere ${\cal S}^3$.
The 3-space  ${\cal S}^3$ is embedded in the four-dimensional
Euclidean space described by the coordinates
$$
\vec{x} \; = \; (x_0,x_1,x_2,x_3)^T\in {\cal S}^3
\hspace{10pt} \hbox{ with the constraint } \hspace{10pt}
|\vec x\,| = 1
\hspace{10pt} ,
$$
i.\,e.\ the 3-space  ${\cal S}^3$ is considered as the
manifold with $x_0^2+x_1^2+x_2^2+x_3^2 = 1$.
The literature offers several possibilities to choose intrinsic coordinates
which in turn determine the representation of the deck group.
The real representation operates with $\hbox{SO}(4,\mathbb{R})$ matrices
which describe transformations as four-dimensional rotations
in the four-dimensional Euclidean space.
In this paper, we choose complex coordinates
$z_1 := x_0 + \hbox{i} x_3$ and $z_2 := x_1 + \hbox{i} x_2$
which are used to define the coordinate matrix
\begin{equation}
\label{Eq:coordinates_u}
u \; = \;
\left(\begin{array}{cc}
z_1 & \hbox{i} z_2 \\ \hbox{i} \overline{z}_2 & \overline{z}_1
\end{array}\right)
\; = \;
\left(\begin{array}{cc}
x_0+\hbox{i}x_3& -x_2+\hbox{i}x_1\\
x_2+\hbox{i}x_1&x_0-\hbox{i}x_3
\end{array}\right)
\in \hbox{SU}(2,\mathbb{C}) \equiv {\cal S}^3
\hspace{10pt} .
\end{equation}
The intrinsic coordinates $(\rho,\alpha,\epsilon)$ are related to $\vec x$  by
\begin{equation}
\label{Eq:coordinates_rho}
\left(\begin{array}{c} x_0 \\ x_1 \\ x_2 \\ x_3 \end{array}\right)
\; = \;
\left(\begin{array}{c}
\cos\rho \; \cos\alpha \\
\sin\rho \; \cos\epsilon \\
\sin\rho \; \sin\epsilon \\
\cos\rho \; \sin\alpha
\end{array}\right)
\hspace{10pt} , \hspace{10pt}
0\leq \rho \leq \frac\pi 2 \; \; , \; \;
0\leq \alpha, \epsilon \leq 2\pi
\hspace{10pt} ,
\end{equation}
that is $z_1 = \cos\rho \, e^{\hbox{\scriptsize i}\alpha}$
and $z_2 = \sin\rho \, e^{\hbox{\scriptsize i}\epsilon}$.
Form these coordinates, one obtains the spherical distance of
the point $(\rho,\alpha,\epsilon)$ from the origin
\begin{equation}
\label{Eq:coordinates_radial}
\tau(\hat n) \; = \; \arcsin r(\hat n) 
\hspace{10pt} \hbox{ with } \hspace{10pt}
r(\hat n) \; = \; \sqrt{x_1^2+x_2^2+x_3^2}
\hspace{10pt} .
\end{equation}

In the complex representation,
the transformations are determined by two $\hbox{SU}(2,\mathbb {C})$ matrices
denoted as the pair $(g_a, g_b)$
that acts on the points $u \in \hbox{SU}(2,\mathbb{C})$ of the 3-sphere
${\cal S}^3 \equiv \hbox{SU}(2,\mathbb{C})$ by left and right multiplication
\begin{equation}
\label{Ref:act}
(g_a, g_b): \, u \rightarrow g_a^{-1}\,u\, g_b
\hspace{10pt} .
\end{equation}
For the group of transformations $g=(g_a, g_b)$ there is the isomorphism
$\hbox{SO}(4,\mathbb{R}) \equiv (\hbox{SU}_a(2,\mathbb{C})\otimes
\hbox{SU}_b(2,\mathbb{C}))/\{\pm ({\bf 1},{\bf 1})\}$.
The identity ${\bf 1}$ is the $2\times 2$ unit matrix in this representation.
Compared to the real representation,
the complex representation has the advantage of revealing the
inhomogeneity of the manifold immediately.

After having defined the transformations $g=(g_a, g_b)$ on
${\cal S}^3 \equiv \hbox{SU}(2,\mathbb{C})$,
we can turn to the specification of the spherical lens spaces $L(p,q)$,
where $p$ and $q$ have to be relatively prime with $1\leq q<p$.
The deck group $\Gamma$ is generated by $g_{L(p,q)} = (g_a, g_b)$ with
\begin{equation}
\label{Def:L_p_q}
g_a \; = \;
\hbox{diag}(e^{-\hbox{\scriptsize i}\Psi_a},e^{\hbox{\scriptsize i}\Psi_a})
\hspace{10pt} \hbox{ and } \hspace{10pt}
g_b \; = \;
\hbox{diag}(e^{-\hbox{\scriptsize i}\Psi_b},e^{\hbox{\scriptsize i}\Psi_b})
\end{equation}
with $\Psi_a=\left(\frac{q+1}p\right)\pi$ and
$\Psi_b=\left(\frac{q-1}p\right)\pi$.
The complete set of group elements $g_n\in\Gamma$ is then obtained by
\begin{equation}
\label{Def:L_group}
g_n \; = \; \Big( (g_a)^n, (g_b)^n \Big)
\hspace{10pt} \hbox{ for } \hspace{10pt}
n = 1, \dots, p
\hspace{10pt} .
\end{equation}

Two spaces $L(p,q)$ and $L(p',q')$ are homeomorphic
if and only if $p=p'$ and either $q=\pm q' (\hbox{mod } p)$ or
$q\,q' = \pm 1 (\hbox{mod } p)$ \cite{Gausmann_Lehoucq_Luminet_Uzan_Weeks_2001}.
For example, the lens spaces $L(p,q)$ and $L(p,p-q)$ are mirror images.
Then the above restrictions on $p$ and $q$ leave as homogeneous lens spaces
only $L(p,1)$ where the generator of the deck group (\ref{Def:L_p_q}) is given by 
$g_a \; = \; \hbox{diag}(e^{-\hbox{\scriptsize i}\frac {2 \pi} p},e^{\hbox{\scriptsize i}\frac {2 \pi} p})$
and $g_b \; = \; {\bf 1}$ .

Finally we specify the deck group $D^\star_p$ of the prism space ${\cal D}_p$. 
The two generators $g_1=(g_{a1 },{\bf 1})$ and $g_2=(g_{a2},{\bf 1})$
of the deck group $D^\star_p$ 
can be represented by 
\begin{equation}
\label{Def:D_p}
g_{a1} \; = \;
\hbox{diag}(e^{-\hbox{\scriptsize i}\Psi_{az}},e^{\hbox{\scriptsize i}\Psi_{az}})
\hspace{10pt} \hbox{ and } \hspace{10pt}
g_{a2} \; = \;
\left(\begin{array}{cc}
\cos(\Psi_{ay}) & -\sin(\Psi_{ay})\\ \sin(\Psi_{ay})& \cos(\Psi_{ay}) 
\end{array}\right)
\end{equation}
with $\Psi_{az}=2\pi\left(\frac{2}p\right)$ and
$\Psi_{ay}=2\pi\left(\frac{1}4\right)$.
The deck group $D^\star_p$ contains as cyclic subgroups $Z_{p/2}$ and $Z_4$
with multiplicity 1 and $p/4$, respectively,
for more details see \cite{Gausmann_Lehoucq_Luminet_Uzan_Weeks_2001}.


\section{Transformation of the CMB observer}
\label{sec:trafo_obs}


Let us now address the question how the group elements $g\in \Gamma$
transform under a change of the observer position,
whereby each observer naturally puts his position at the origin of his
coordinate system.
The behaviour under such transformations will lead to the distinction between
homogeneous and inhomogeneous manifolds.
By applying an arbitrary isometry $t$ to the coordinates
\begin{equation} 
\label{Eq:trafo_coordinate}
u \rightarrow u'= u\, t \; \; , \; \; t \in \hbox{SU}(2,\mathbb{C})
\hspace{10pt} ,
\end{equation}
the origin of the coordinate system can be transformed to every point
on ${\cal S}^3$.
Now consider a given point which coordinate matrices with respect to
two observers $o$ and $o'$ are related by $u' = u t$.
Assume that the group elements of the observer $o$ 
are given by $g_n=(g_{an},g_{bn})$, $n=1, 2, \dots$.
Every point $u$ on the 3-sphere is mapped by $g_n$ to points $\tilde{u}_n$
which are to be identified $\tilde{u}_n\equiv (g_{an})^{-1}\,u\,g_{bn}$.
The transformation of the points $\tilde{u}_n$
into the coordinate system of observer $o'$ results in
\begin{eqnarray} \nonumber
\tilde{u}_n \rightarrow \tilde{u}'_n = \tilde{u}_n\,t 
& = &(g_{an})^{-1}\,u\,g_{bn}\,t \\
\label{Eq:trafo_coordinate_general_group}
& = & (g_{an})^{-1}\,u\,t\,(t^{-1}\,g_{bn}\,t)
= (g_{an})^{-1}\,u'\,(t^{-1}\,g_{bn}\,t)
\hspace{10pt}.
\end{eqnarray}
The observer $o'$ uses the equation $\tilde{u}'_n= (g'_{an})^{-1}\,u'\,g'_{bn}$
in order to identify points on the 3-sphere. 
Comparing this equation with eq.\,(\ref{Eq:trafo_coordinate_general_group}),
one gets the deck transformations
\begin{equation}
\label{Eq:trafo_group}
g'_n \; = \; (g'_{an},g'_{bn}) \; = \;(g_{an},t^{-1}\,g_{bn}\,t)
\hspace{10pt} , \hspace{10pt}
n=1, 2,\dots
\hspace{10pt} ,
\end{equation}
with respect to the observer $o'$.
Since the coordinate transformation $t$ is given as right action,
the left action $g_{an}$ of a deck transformation $g_n$ does not chance,
but the right action $g_{bn}$ of a deck transformation $g_n$ 
chances under a coordinate transformation $t$, in general.

In the case of the manifolds $L(p,1)$ and ${\cal D}_p$
the action of the group can be given as a pure left action.
This is the reason why their group elements and
their Voronoi domains are unchanged under an isometry $t$,
and these manifolds are called homogeneous.
In contrast the deck groups of the lens spaces $L(p,q)$ with $q>1$
are always given by right and left action.
Therefore, the action of the isometry $t$ changes
the group elements as well as the Voronoi domains
of these inhomogeneous manifolds.


\begin{figure}
\begin{center}
\vspace*{0pt}
\hspace*{0pt}
\hspace{-1.8cm}\includegraphics[width=9cm]{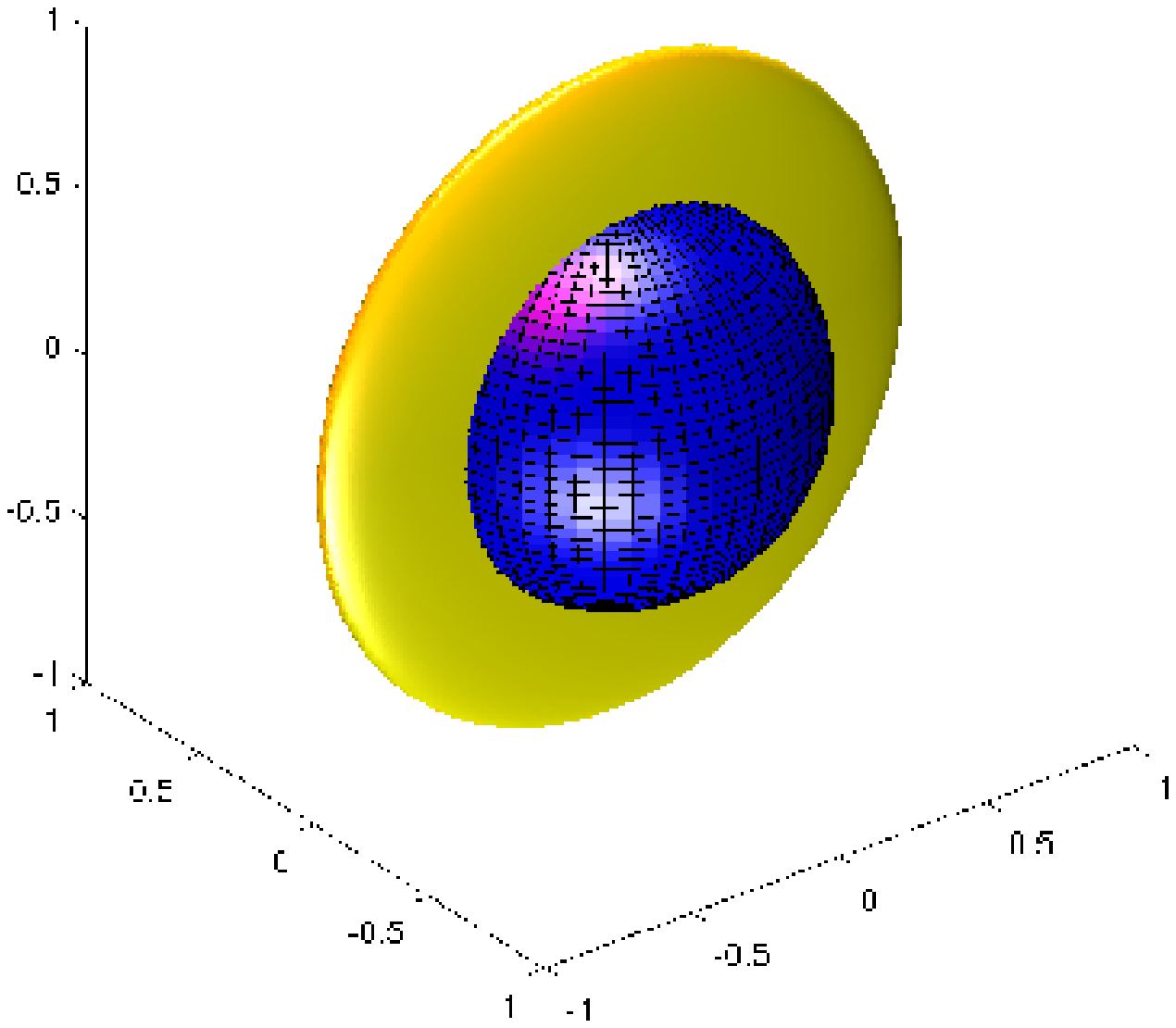}
\hspace{-1.0cm}\includegraphics[width=9cm]{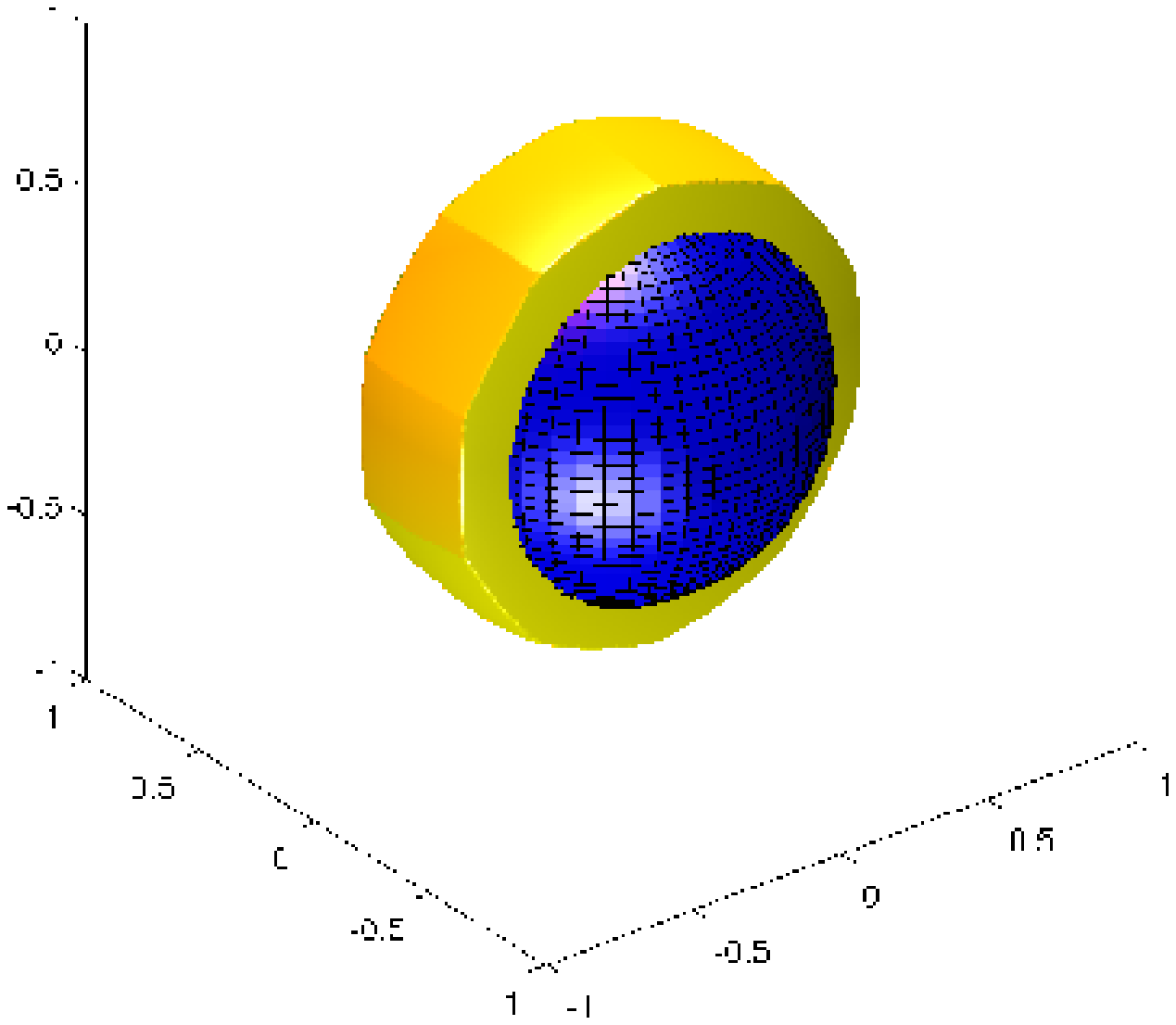}
\put(-405,175){(a)}
\put(-302,80){$\rho= 0$}
\put(-190,175){(b)}
\put(-72,80){$\rho= \pi/4$}
\vspace*{-22pt}
\end{center}
\caption{\label{fig:Voronoi_L20_9}
The Voronoi domain for the inhomogeneous $L(20,9)$ space is shown in yellow.
In panel (a) the observer position is $\rho=0$ and
the Voronoi domain is identical to that of the
homogeneous $L(20,1)$ lens space.
Choosing for the observer position $\rho=\pi/4$
the prism shaped Voronoi domain shown in panel (b) is obtained
which is identical to that of the prism space ${\cal D}_{20}$.
In both figures the sphere $S_{\hbox{\scriptsize sls}}$ of last scattering
is depicted as a blue sphere having a radius $\tau_{\hbox{\scriptsize sls}}=0.6$.
This value corresponds to $\Omega_{\hbox{\scriptsize tot}}=1.04$.
Smaller values of $\Omega_{\hbox{\scriptsize tot}}$ lead to
even smaller spheres, see figure \ref{fig:S60_P3}(a).

}
\end{figure}


In the following the position of the observer is shifted using 
the parameterisation
\begin{equation}
\label{Eq:coordinate_t_rho_alpha_epsilon}
t(\rho,\alpha,\epsilon) \; = \;
\left( \begin{array}{cc}
\cos(\rho)\, e^{+\hbox{\scriptsize i}\alpha} &
\hbox{i}\sin(\rho)\, e^{+\hbox{\scriptsize i}\epsilon} \\
\hbox{i}\sin(\rho)\, e^{-\hbox{\scriptsize i}\epsilon} &
\cos(\rho)\, e^{-\hbox{\scriptsize i}\alpha}
\end{array} \right)
\end{equation}
with $\rho \in [0,\frac{\pi}{2}]$, $\alpha, \epsilon \in [0,2\pi]$.
The transformation (\ref{Eq:coordinate_t_rho_alpha_epsilon})
maps the group elements (\ref{Def:L_group}) generated by (\ref{Def:L_p_q})
according to (\ref{Eq:trafo_group}) onto $g'_n \; = \; (g'_{an},g'_{bn})$ with
\begin{eqnarray}
\label{Def:trafo_group_Lpq}
g'_{an} = 
\hbox{diag}(e^{-\hbox{\scriptsize i}\,n\Psi_a},e^{\hbox{\scriptsize i}\,n\Psi_a})\\
\nonumber
\hspace{10pt} \hbox{ and } \hspace{10pt}\\
\nonumber
g'_{bn} =
\left(\begin{array}{ll}
\cos(n\Psi_{b})-\hbox{i}\sin(n\Psi_{b})\cos(2\rho) & 
\sin(2\rho)e^{\hbox{i}\left(\epsilon-\alpha\right)}\sin(n\Psi_{b})\\
-\sin(2\rho)e^{-\hbox{i}\left(\epsilon-\alpha\right)}\sin(n\Psi_{b})& 
\cos(n\Psi_{b})+\hbox{i}\sin(n\Psi_{b})\cos(2\rho)
\end{array}\right)
\hspace{4pt}.
\end{eqnarray}
$\Psi_a$ and $\Psi_b$ are defined below eq.\,(\ref{Def:L_p_q}).

The above description of the spherical space also allows a
visualisation of the Voronoi domains by projecting $\vec x$,
see eq.\,(\ref{Eq:coordinates_rho}),
down to the three-dimensional space $(x_1,x_2,x_3)^T$,
i.\,e.\ by simply omitting the $x_0$ coordinate.
The Voronoi cell is computed with respect to an observer at the
origin $\vec x_0=(1,0,0,0)$ by using the group elements 
(\ref{Def:trafo_group_Lpq}) of $L(p,q)$ for $\alpha, \epsilon=0$.
For the space $L(20,9)$ two Voronoi domains are visualised in this way
in figure \ref{fig:Voronoi_L20_9},
where two different observer positions are chosen.
To give an impression of the size of the sphere $S_{\hbox{\scriptsize sls}}$
of last scattering,
it is depicted for a radius $\tau_{\hbox{\scriptsize sls}}=0.6$
which corresponds to the largest sphere which is compatible
with the present cosmological parameters.


\section{The eigenmodes of spherical manifolds}


\subsection{The eigenmodes of the 3-sphere ${\cal S}^3$}

The eigenmodes of the simply-connected spherical manifold ${\cal S}^3$
can be generated by the Lie algebra of 
$\hbox{SO}(4, \mathbb{R}) \equiv
(\hbox{SU}_a(2,\mathbb{C})\otimes \hbox{SU}_b(2,\mathbb{C}))
/ \{\pm({\bf 1},{\bf 1})\}$.
The abstract generators satisfy the relations
\begin{equation}
\label{Eq:Lie_algebra}
[J_{ax},J_{ay}] \; = \; \hbox{i} \, J_{az}
\hspace{10pt} , \hspace{10pt}
[J_{bx},J_{by}] \; = \; \hbox{i} \, J_{bz}
\hspace{10pt} \hbox{and cyclic permutations.}
\end{equation}
The components of
$\vec J_a = (J_{ax},J_{ay},J_{az}) \in \hbox{SU}_a(2,\mathbb{C})$
commute with all components of
$\vec J_b = (J_{bx},J_{by},J_{bz}) \in \hbox{SU}_b(2,\mathbb{C})$. 
The eigenmodes of $\vec J_a$ obey
\begin{eqnarray}
\label{Eq:eigenmodes_J2}
{\vec J_a}^{\;2} |j_a,m_a\rangle & = & j_a(j_a+1) |j_a,m_a\rangle
\hspace{10pt} , \hspace{10pt}
j_a \in\frac{\mathbb{N}_0}2 \\
J_{az} |j_a,m_a\rangle & = & m_a |j_a,m_a\rangle
\hspace{10pt} , \hspace{10pt}
|m_a| \leq j_a
\hspace{10pt} ,
\end{eqnarray}
and similarly for $\vec J_b$.
The complete basis in respect of $\hbox{SO}(4,\mathbb{R})$ factorises as
\begin{equation}
\label{Eq:SO4_Basis}
| j; m_a, m_b \rangle \; := \; |j,m_a\rangle \, |j,m_b\rangle
\hspace{10pt} .
\end{equation}
The Laplace-Beltrami operator $\Delta$ on the 3-sphere ${\cal S}^3$
can be written as
$$
-\Delta \; = \; 2 \Big( {\vec J_a}^{\;2} + {\vec J_b}^{\;2} \Big)
\hspace{10pt} .
$$
Thus, the eigenmodes with the eigenvalue $E_j := 4j(j+1) = \beta^2-1$ of 
the operator $-\Delta$ are given by eq.\,(\ref{Eq:SO4_Basis})
where we defined $\beta:=2j+1$.
An equivalent representation of the eigenmodes $| j; m_a, m_b \rangle$
is given by $| j; l, m \rangle$ where $l$
is the eigenvalue of $\vec L := \vec J_a + \vec J_b$.
These two representations of the eigenmodes are connected by
\begin{eqnarray}
\label{Eq:Trafo_Clebsch_Gordan}
| j; m_a, m_b \rangle & = & \sum_l \langle j m_a j m_b | lm \rangle
\; | j; l, m \rangle \\
\nonumber
| j; l, m \rangle & = & \sum_{m_a} \langle j m_a j m_b | lm \rangle
\; | j; m_a, m_b \rangle
\hspace{10pt} 
\end{eqnarray}
where the $\langle j m_a j m_b | lm \rangle$ are the
Clebsch-Gordan coefficients \cite{Edmonds_1964}.
In general, $\langle j m_a j m_b | lm \rangle\neq 0$ only for $0\leq l\leq 2j$
and $m_a+m_b=m$.


\subsection{The eigenmodes of the lens spaces $L(p,q)$}


The action of the generator (\ref{Def:L_p_q}) of the lens space $L(p,q)$
on the eigenmodes (\ref{Eq:SO4_Basis}) can be written as 
$U_g=e^{\hbox{\scriptsize i} 2\psi_aJ_{az}}e^{\hbox{\scriptsize i} 2\psi_bJ_{bz}}$
with $\psi_a$ and $\psi_b$ defined below eq.\,(\ref{Def:L_p_q}).
The eigenmodes on the lens space $L(p,q)$ invariant under the action of $U_g$
are
\begin{equation}
\label{Eq:eigenmodes_L_p_q}
|j; m_a, m_b \rangle \hspace{10pt} \hbox{with}\hspace{10pt}
(q+1)\,m_a+(q-1)\,m_b \equiv 0\;  {\rm mod}\; p
\hspace{10pt},
\end{equation}
where $\left|m_a \right|,\left|m_b \right| \le j$.
In general, similar eigenmodes on the lens spaces are reported 
in \cite{Lachieze_Rey_2004b}
and equivalent sets of eigenmodes on the lens spaces in 
\cite{Lehoucq_Uzan_Weeks_2002,Lehoucq_Weeks_Uzan_Gausmann_Luminet_2002}. 


\subsection{The eigenmodes of the prism spaces ${\cal D}_p$}


In the chosen notation the action of the two generators (\ref{Def:D_p}) 
of the prism space ${\cal D}_p$ are given by
$U_{g_1}=e^{\hbox{\scriptsize i} 2\psi_{az}J_{az}}$ and 
$U_{g_2}=e^{\hbox{\scriptsize i} 2\psi_{ay}J_{ay}}$ 
with $\psi_{az}$ and $\psi_{ay}$ defined below eq.\,(\ref{Def:D_p}).
The eigenmodes on the prism space ${\cal D}_p$
invariant under the action of $U_{g_1}$ and $U_{g_2}$ 
are
\begin{eqnarray}
\label{Eq:eigenmodes_D_p}
\begin{array}{ll}
\frac{1}{\sqrt{2}}\left(|j; m_a, m_b \rangle +(-1)^{m_a}\,|j; -m_a, m_b \rangle\right) &
: \; j \;\hbox{even}, m_a>0  \\
|j; m_a, m_b \rangle &
: \; j \;\hbox{even}, m_a=0 \\
\frac{1}{\sqrt{2}}\left(|j; m_a, m_b \rangle-(-1)^{m_a}\,|j; -m_a, m_b \rangle\right) &
: \; j \;\hbox{odd}, m_a>0
\end{array}
\hspace{10pt},
\end{eqnarray}
where $j\in  \mathbb{N}_0 \setminus \{1,3,\dots,2 [\frac{p}{8}]-1\}$, $m_b \in \mathbb{Z}$,
$m_a \in \mathbb{N}_0$, $m_a\equiv 0\;\hbox{mod}\;p/4$, 
and $m_a,\left|m_b\right|\le j$.
In general, similar eigenmodes on these manifolds are given 
in \cite{Lehoucq_Uzan_Weeks_2002,Lachieze_Rey_2004b,Weeks_2005}.


\subsection{The observer dependence of the eigenmodes 
on lens spaces $L(p,q)$}


The operator, which corresponds to
the transformation to a new observer as discussed in sec.\,\ref{sec:trafo_obs},
can be given by 
\begin{equation}
\label{Eq:operator_new_obs}
D(t) \; = \;
D(\alpha+\epsilon,2\rho,\alpha-\epsilon) \; = \;
e^{\hbox{\scriptsize i}(\alpha+\epsilon) J_{bz}} \,
e^{\hbox{\scriptsize i}(2\rho) J_{by}} \,
e^{\hbox{\scriptsize i}(\alpha-\epsilon)  J_{bz}}
\hspace{10pt},
 \end{equation}
where the coordinates (\ref{Eq:coordinate_t_rho_alpha_epsilon}) are used for 
the observer.
The action of this operator on the eigenmodes results in
\begin{eqnarray}
\label{Eq:eigenmodes_L_p_q_new_obs}
D(t^{-1})|j; m_a, m_b \rangle
\;& = &\;
\sum_{\tilde{m}_b=-j}^{j} D^{\,j}_{\tilde{m}_b,m_b}(t^{-1}) |j; m_a,\tilde{m}_b \rangle
\\
\nonumber& &\hbox{with}\hspace{10pt}
(q+1)\,m_a+(q-1)\,m_b \equiv 0\;  {\rm mod}\; p
\end{eqnarray}
where the completeness relation 
$\sum_{\tilde{m}_b=-j}^{j}| j,  \tilde{m}_b \rangle \langle j, \tilde{m}_b | = 1$
and the definition of the $D$-function, also called Wigner polynomial,
\begin{eqnarray}
\label{Eq:D_function_rho_alpha_epsilon}
\nonumber
D^{\,j}_{\tilde{m}_b,m_b}(t)
\; &:=& \; \langle j, \tilde{m}_b |D(t)| j,  m_b \rangle \\
\;& =&\; 
e^{\hbox{i}\,(\alpha + \epsilon)\,\tilde{m}_b}
d^{\,j}_{\tilde{m}_b, m_b}(2 \rho)
e^{\hbox{i}\,(\alpha - \epsilon)\, m_b}
\hspace{10pt}
\end{eqnarray}
are used.
The numerical values of the $d$ function
are computed by the algorithm described in \cite{Risbo_1996}.
With eq.\,(\ref{Eq:Trafo_Clebsch_Gordan}) the expansion of the 
eigenmodes\,(\ref{Eq:eigenmodes_L_p_q_new_obs}) on the lens spaces $L(p,q)$
with respect to the spherical basis
$|j; l, m \rangle$ yields
\begin{eqnarray}
\label{Eq:eigenfunction_L_p_q_exp_sph}
D(t^{-1})|j; m_a, m_b \rangle
\nonumber &=&
\sum_{l=0}^{2j}\sum_{m=-l}^{l} \xi^{j,i(m_a,m_b)}_{lm}(L(p,q);t)\,
|j; l, m \rangle
\;\;\\
\xi^{j,i(m_a,m_b)}_{lm}(L(p,q);t)
&=& \langle jm_aj\tilde{m}_b|lm\rangle \,D^{\,j}_{\tilde{m}_b,m_b}(t^{-1})\\
\nonumber& &\hbox{with}\hspace{10pt}
(q+1)\,m_a+(q-1)\,m_b \equiv 0\;  {\rm mod}\; p
\hspace{10pt}. 
\end{eqnarray}
Here, the abbreviation $\tilde{m}_b:=m-m_a$ is introduced, and
$1 \le i(m_a,m_b)\le r^{L(p,q)}(\beta)$ counts the 
multiplicity $r^{L(p,q)}(\beta)$ of the eigenvalue $E_j$ of the
Laplace-Beltrami operator on $L(p,q)$ for $j \in \frac{\mathbb{N}_{0}}{2}$.
The transformation $t$ characterises the position of the observer.

For homogeneous spherical manifolds
it is shown in \cite{Aurich_Kramer_Lustig_2011} that
the same set of eigenmodes can be chosen for all observer position.
For example on $L(p,1)$ one can choose the above coefficients 
to $\xi^{j,i(m_a,m_b)}_{lm}(L(p,1)) = \langle jm_ajm_b|lm\rangle$.
In contrast for an inhomogeneous lens space $L(p,q)$, $q>1$,
such a choice is not possible due to the restriction on $m_b$,
and one has to use eq.\,(\ref{Eq:eigenfunction_L_p_q_exp_sph})
which depends on the observer position.


\begin{figure}
\begin{center}
\vspace*{-30pt}
\hspace*{0pt}
\hspace{-2.0cm}\includegraphics[width=9cm]{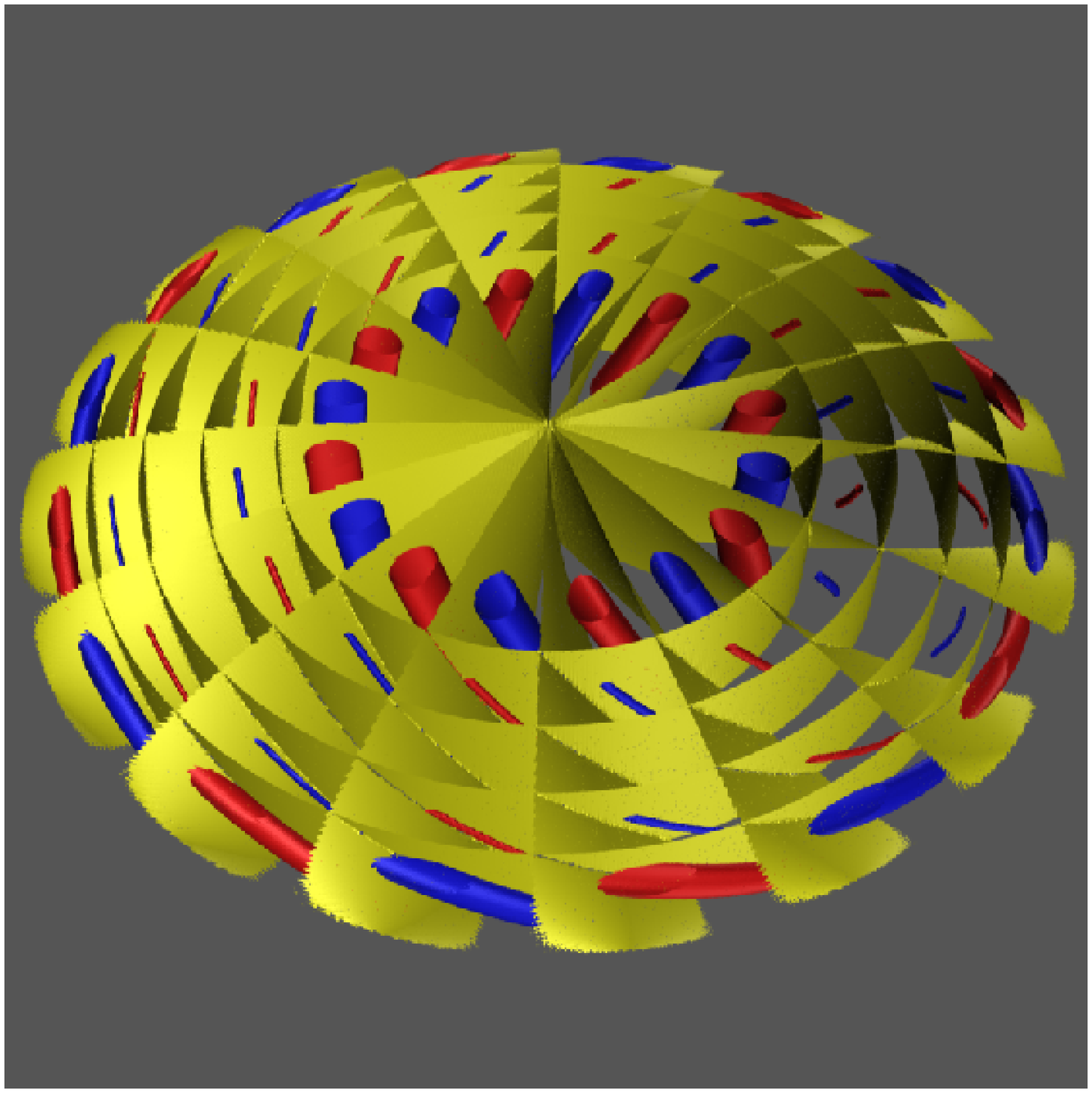}
\hspace{-0.7cm}\includegraphics[width=9cm]{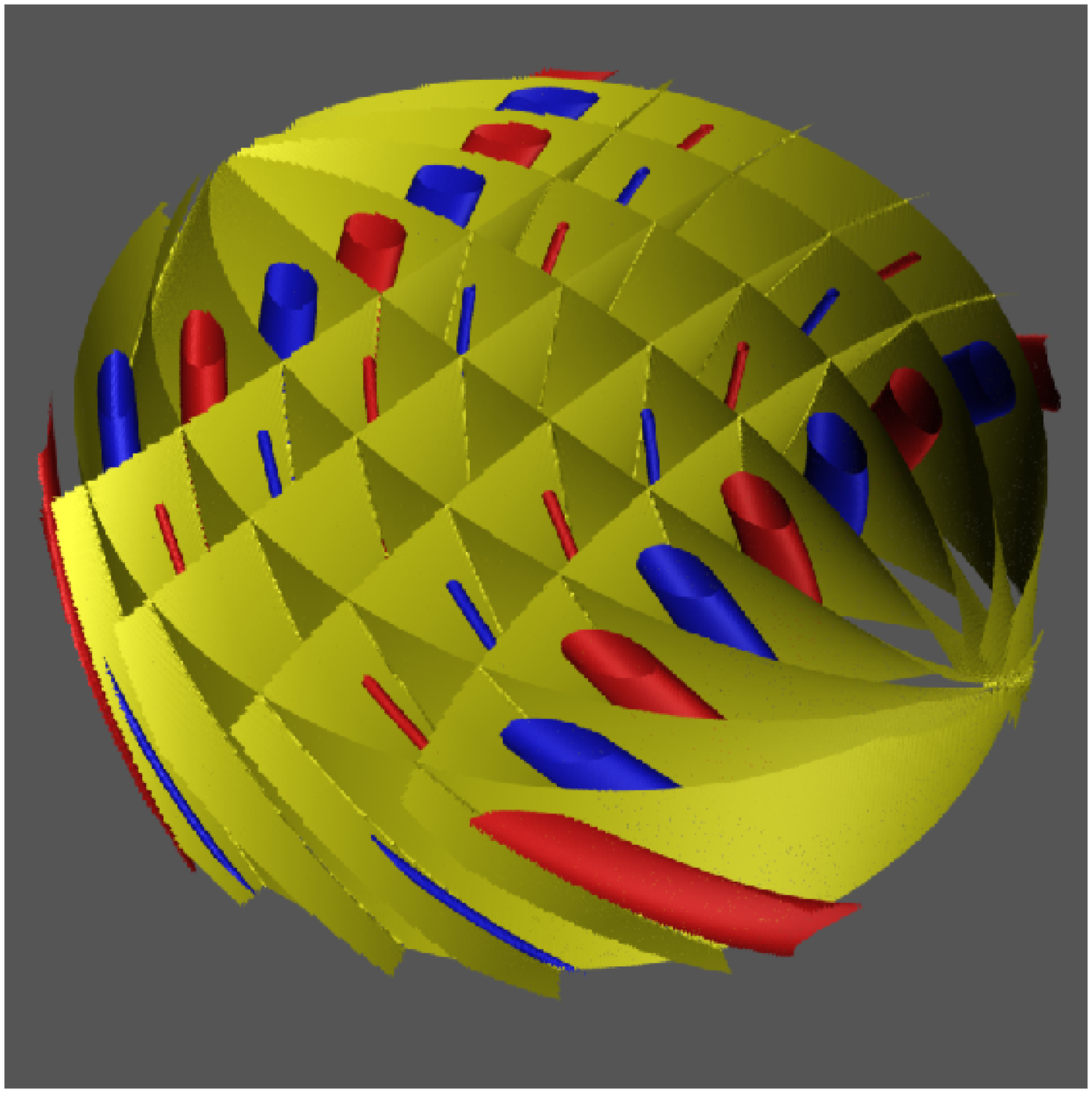}
\vspace*{-70pt}
\end{center}
\caption{\label{fig:eigenmode}
An eigenmode for the inhomogeneous lens space $L(20,9)$ for
$\beta=25$ is shown for the lens shaped Voronoi domain $\rho=0$ (left)
and for the prism shaped Voronoi domain $\rho=\pi/4$ (right).
For intermediate values of $\rho$ the structure of the eigenmode
changes continuously between the two cases shown above.
The yellow surfaces represents the nodal surface,
blue and red corresponds to a negative and a positive equipotential surface.
}
\end{figure}


The eigenmodes $|j; m_a, m_b \rangle$ are only constructed
in an abstract way since that is all what is needed.
However, it is instructive to compute the eigenmodes
as a function of the coordinates $u$ of the spherical space.
To that aim, the Wigner basis \cite{Kramer_2010,Aurich_Kramer_Lustig_2011}
\begin{equation}
\label{Eq:wigner_basis}
\psi(j,m_a,m_b)(u) \, = \, \frac{2j+1}{\sqrt{2\,\pi^2}} \;
\langle j-m_ajm_a|0 0\rangle \; D^j_{-m_am_b}(u)
\hspace{5pt} , \; \beta=2j+1 \; ,
\end{equation}
is introduced which is normalised on the 3-sphere ${\cal S}^3$.
For $\alpha=\epsilon=0$ the eigenmode $\psi^{L(p,q)}_{\rho,\beta,i}(u)$ is 
obtained by projecting
eq.\,(\ref{Eq:eigenmodes_L_p_q_new_obs}) onto ${\cal S}^3$
\begin{eqnarray}
\label{Eq:eigenfunction}
\psi^{L(p,q)}_{\rho,\beta,i}(u) & = & \langle u | D(t^{-1})|j; m_a, m_b \rangle
\\ \nonumber & = &
\sum_{\tilde{m}_b=-j}^{j} D^{\,j}_{\tilde{m}_b,m_b}(t^{-1}) \,
\langle u |j; m_a,\tilde{m}_b \rangle
\\ \nonumber & = &
\sum_{\tilde{m}_b=-j}^{j} D^{\,j}_{\tilde{m}_b,m_b}(t^{-1}) \,
\psi(j,m_a,\tilde{m}_b)(u)
\end{eqnarray}
with the restriction $(q+1)\,m_a+(q-1)\,m_b \equiv 0\; {\rm mod}\; p$.
The dependence on the observer position $\rho$ is determined by
the Wigner polynomial $D^{\,j}_{\tilde{m}_b,m_b}(t^{-1})$,
eq.\,(\ref{Eq:D_function_rho_alpha_epsilon}).
Thus, one can take an eigenmode $\psi^{L(p,q)}_{\rho,\beta,i}(u)$
and visualise it for various values of $\rho$.
For the two interesting positions $\rho=0$ and $\rho=\pi/4$,
an eigenmode of $L(20,9)$ for $\beta=25$ is depicted
in figure \ref{fig:eigenmode}.


\subsection{The quadratic sum of eigenmodes on spherical spaces}
\label{sec:quadratic_sum}


The calculation of the ensemble average of
the temperature 2-point correlation function (\ref{Eq:C_theta})
or the multipole spectrum (\ref{Eq:Cl_ensemble_L_p_q})
on inhomogeneous lens spaces ($q>1$) requires 
the evaluation of the following quadratic sum:
\begin{eqnarray}
\label{Eq:quadratic_sum_L_p_q}
\nonumber
\frac{1}{2l+1}\sum_{m=-l}^{l}\sum_{m_a,m_b}
\left|\xi^{j,i(m_a,m_b)}_{lm}(L(p,q);t)\right|^2
\hspace{-150pt} & & \\
\nonumber
& &=\frac{1}{2l+1}\sum_{m=-l}^{l}\sum_{m_a,m_b}
\left|\langle jm_aj\tilde{m}_b|lm\rangle\,D^{\,j}_{\tilde{m}_b,m_b}(t^{-1})\right|^2 \\
& &=\frac{1}{2l+1}\sum_{m=-l}^{l}\sum_{m_a,m_b}
\left|\langle jm_aj\tilde{m}_b|lm\rangle\,d^{\,j}_{\tilde{m}_b,m_b}(-2 \rho)\right|^2
\\
 \nonumber& &\hbox{with}\hspace{10pt}
(q+1)\,m_a+(q-1)\,m_b \equiv 0\;  {\rm mod}\; p
\hspace{10pt}.
\end{eqnarray}
In the derivation of eq.\,(\ref{Eq:quadratic_sum_L_p_q}) we have used
eq.~(\ref{Eq:D_function_rho_alpha_epsilon})
where the coordinates of the observer position on the lens space $L(p,q)$ are 
parameterised by eq.\,(\ref{Eq:coordinate_t_rho_alpha_epsilon}).
The above quadratic sum depends only on the coordinate $\rho$,
and the analysis of the CMB statistics can be restricted to observer positions
with $\alpha=\epsilon=0$ and $\rho \in [0,\frac{\pi}{2}]$.
A further reduction of the observer positions to the smaller interval
$\rho \in [0,\frac{\pi}{4}]$ is possible on lens spaces $L(p,q)$
which satisfy the condition $p=4n$ and $q=p/2-1=2n-1$ for $n=2,3,4,\dots$.
This reduction is shown in \cite{Aurich_Kramer_Lustig_2011}
for $L(8,3)$ (see eq.\,(53) in \cite{Aurich_Kramer_Lustig_2011}).

\begin{table}[!htbp]
\centering
 \begin{tabular}{|c|c|c|}
 \hline
 manifold ${\cal M}$ & wave number spectrum $\{\beta\}$  &
 multiplicity $r^{\cal M}(\beta)$ \\
 \hline
      &               &           \\
 ${\cal S}^3$ & $\mathbb{N}$  & $\beta^2$ \\
       &               &           \\
 \hline
     &               &           \\
 $L(p,1)$,  & $\{1,3,5,\dots,p\}$ &
 $\beta \big( 2 \,\big[\frac{\beta-1}{2p}\big]+1 \big)$ for $\beta$ odd  \\
 $p$ odd $\ge 1$    &  $\cup \{n|n\in \mathbb{N}, n\ge p+1\}$             &  
$2 \beta \big( \big[\frac{\beta-1}{p}\big]- \big[\frac{\beta-1}{2p}\big] \big)$ for $\beta$ even         \\
    &               &           \\
 \hline
     &               &           \\
 $L(p,1)$,& $2\mathbb{N}-1$ &
 $\beta \big( 2 \, \big[\frac{\beta-1}p\big]+1 \big)$ \\
  $p$ even $\ge 2$   &               &           \\
 \hline
     &               &           \\
 ${\cal D}_p$, &
 $\{1,5,9,\dots,4\left[\frac{p}8\right]+1\}$ &
 $\beta   \big( \big[\frac{2(\beta-1)}p \big]+2 \left[\frac{\beta-1}4\right] - \frac{\beta-3}2 \big)$ \\
  $p/4\ge 2$   & $\cup
 \{2n+1|n\in \mathbb{N},n\ge 2\left[\frac{p}8\right]+1\}$               &           \\
 \hline
     &               &           \\
 ${\cal T}$ & $\{1,7,9\}$ &
 $\beta\left( 2 \left[\frac{\beta-1}6\right] + \left[\frac{\beta-1}4\right] -
 \frac{\beta-3}2\right)$ \\
     &   $\cup \{2n+1|n\in \mathbb{N},n\ge 6\}$             &           \\
 \hline
     &               &           \\
 ${\cal O}$ & $\{1,9,13,17,19,21\}$ &
 $\beta\left( \left[\frac{\beta-1}8\right] + \left[\frac{\beta-1}6\right] +
 \left[\frac{\beta-1}4\right] - \frac{\beta-3}2\right)$ \\
     &    $\cup \{2n+1|n\in \mathbb{N}, n\ge 12\}$           &           \\
 \hline
     &   $\{1,13,21,25,31,33,37\}$            &           \\
 ${\cal I}$ & $\cup \{41,43,45,49,51,53,55,57\}$ &
 $\beta\left( \left[\frac{\beta-1}{10}\right] + \left[\frac{\beta-1}6\right] +
 \left[\frac{\beta-1}4\right] - \frac{\beta-3}2 \right)$ \\
 & $\cup \, \{2n+1| n\in \mathbb{N}, n \ge 30\}$ & \\
 \hline
 \end{tabular}
 \caption{\label{Tab:Spectrum}
The spectrum of the eigenvalues $E_{\beta}=\beta^2-1$ 
of the Laplace-Beltrami operator on homogeneous spherical manifolds ${\cal M}$
and their multiplicities $r^{\cal M}(\beta)$ are given. 
In this table an error in
\cite{Ikeda_1995,Aurich_Lustig_Steiner_2005a,Lustig_2007} is corrected,
and an alternative description of the multiplicity is given
for the lens spaces $L(p,1)$ and for the prism spaces ${\cal D}_p$.
This description leads to a faster numerical handling of the multiplicity.
The bracket $[x]$ denotes the integer part of $x$.
}
\end{table}

In the case of a homogeneous spherical manifold ${\cal M}$
this quadratic sum is independent of the observer position and is given by 
\cite{Aurich_Lustig_Steiner_2004c,Gundermann_2005,Aurich_Lustig_Steiner_2005a,%
Bellon_2006,Lustig_2007}
\begin{equation}
\label{Eq:quadratic_sum_hom_mani}
\frac{1}{2l+1}\sum_{m=-l}^{l}\sum_{m_a,m_b}
\left|\xi^{j,i(m_a,m_b)}_{lm}({\cal M})\right|^2
\; = \; \frac{r^{{\cal M}}(\beta)}{\beta^2}
\hspace{10pt},
\end{equation}
where $r^{\cal M}(\beta)$ is the multiplicity
of the eigenvalue of the Laplace-Beltrami operator on the spherical 
manifold ${\cal M}$.
Since 1995 the multiplicities of the eigenvalues of the
Laplace-Beltrami operator on all homogeneous spherical manifolds
${\cal M}$ are known \cite{Ikeda_1995}, see also table\,\ref{Tab:Spectrum}.
Thus it is not necessary to calculate the 
expansion coefficients $\xi^{j,i(m_a,m_b)}_{lm}({\cal M})$
for homogeneous spherical manifolds.
This advantage is exploited in the case of the Poincare dodecahedron 
${\cal I}={\cal S}^3/I^\star$,
the binary octahedral space ${\cal O}={\cal S}^3/O^\star$, 
and the binary tetrahedral space ${\cal T}={\cal S}^3/T^\star$.
Here $I^\star$, $O^\star$, and $T^\star$ are the binary icosahedral,
the binary octahedral, and the binary tetrahedral groups 
\cite{Gausmann_Lehoucq_Luminet_Uzan_Weeks_2001,Aurich_Lustig_Steiner_2005a}.
The binary octahedral space ${\cal O}$ is also called truncated cube.
The binary tetrahedral space ${\cal T}$ is also named octahedron
because of the geometry of its Voronoi domain
\cite{Gausmann_Lehoucq_Luminet_Uzan_Weeks_2001},
however, this notation can be misleading.
Eq.\,(\ref{Eq:quadratic_sum_hom_mani}) is also used for the
prism space ${\cal D}_p={\cal S}^3/D_p^\star$ and
the homogeneous lens space $L(p,1) ={\cal S}^3/Z_p$.


\section{Measures of the shape}
\label{sec:shape_measure}


In order to verify the well-proportioned conjecture quantitatively,
one has to define a measure of the shape.
Since it is conjectured that shapes with equal dimensions are
preferred with respect to a large scale suppression of CMB anisotropies,
the shape measure should single out the sphere.
For that reason we require that the measure is zero for a sphere,
and the larger, the more oddly shaped the Voronoi domain is.
With respect to the observer sitting in the centre of the coordinate system,
the mean radius $\langle \tau \rangle$ of the Voronoi domain is
\begin{equation}
\label{Eq:mean_radius}
\langle \tau \rangle \; := \;
\frac{\int d\Omega \, \sin^2(\tau(\hat n)) \; \tau(\hat n)}
{\int d\Omega \, \sin^2(\tau(\hat n))}
\hspace{10pt} , \hspace{10pt}
d\Omega = \sin\theta\, d\theta\, d\phi
\hspace{10pt} ,
\end{equation}
where $\tau(\hat n)$ is the spherical distance to the surface
of the Voronoi domain in the direction $\hat n$
defined by the angles $\theta$ and $\phi$.
The distance $\tau(\hat n)$ is defined in eq.\,(\ref{Eq:coordinates_radial}).
The variance $\sigma_\tau^2$ can be taken as a shape measure
\begin{equation}
\label{Eq:variance_radius}
\sigma_\tau^2 \; := \;
\frac{\int d\Omega \, \sin^2(\tau(\hat n)) \,
\Big(\langle \tau \rangle - \tau(\hat n)\Big)^2}
{\int d\Omega \, \sin^2(\tau(\hat n))}
\hspace{10pt} .
\end{equation}
For a hypothetical spherical domain, the variance vanishes.
Since the value of the variance $\sigma_\tau^2$ increases
with increasing asymmetry of the Voronoi domain,
this variance can serve as a measure of the shape.

The variance (\ref{Eq:variance_radius}) can only measure
the geometric shape of the Voronoi domain.
It ignores, however, the connection of the points on its surface completely.
The connection is defined by the deck group which specifies
how the faces of the Voronoi domain are mapped onto each other.
To every point $\tau(\hat n)$ of the surface belongs a point $g\tau(\hat n)$
that also lies on the surface but is obtained from the former by
applying one special chosen group element $g\in\Gamma$.
The spherical distance between this pair of identified points
encodes also the topology and its symmetry.
To this end, one defines
\begin{equation}
\label{Eq:variance_topo_trafo}
D \; := \;
\frac{\int d\Omega \; d( \tau(\hat n),g\tau(\hat n) )}
{\int d\Omega}
\hspace{10pt} .
\end{equation}
We use this quantity with respect to the well-proportioned conjecture
but find that models with large CMB anisotropy suppression on
large angular scales are not singled out by this measure.


\begin{figure}
\begin{center}
\vspace*{0pt}
\hspace*{0pt}
\hspace{-1.8cm}\includegraphics[width=9cm]{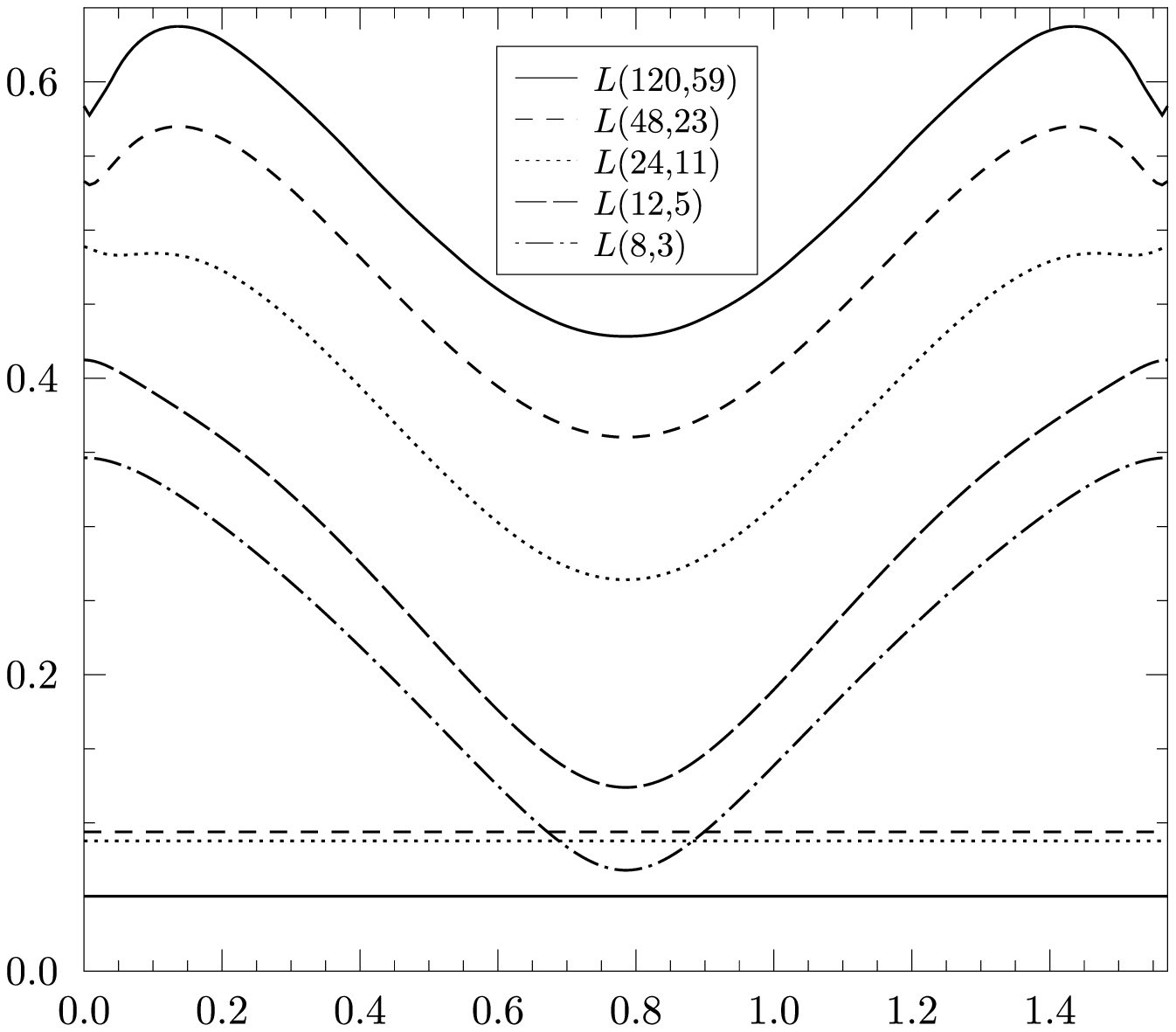}
\hspace{-1.0cm}\includegraphics[width=9cm]{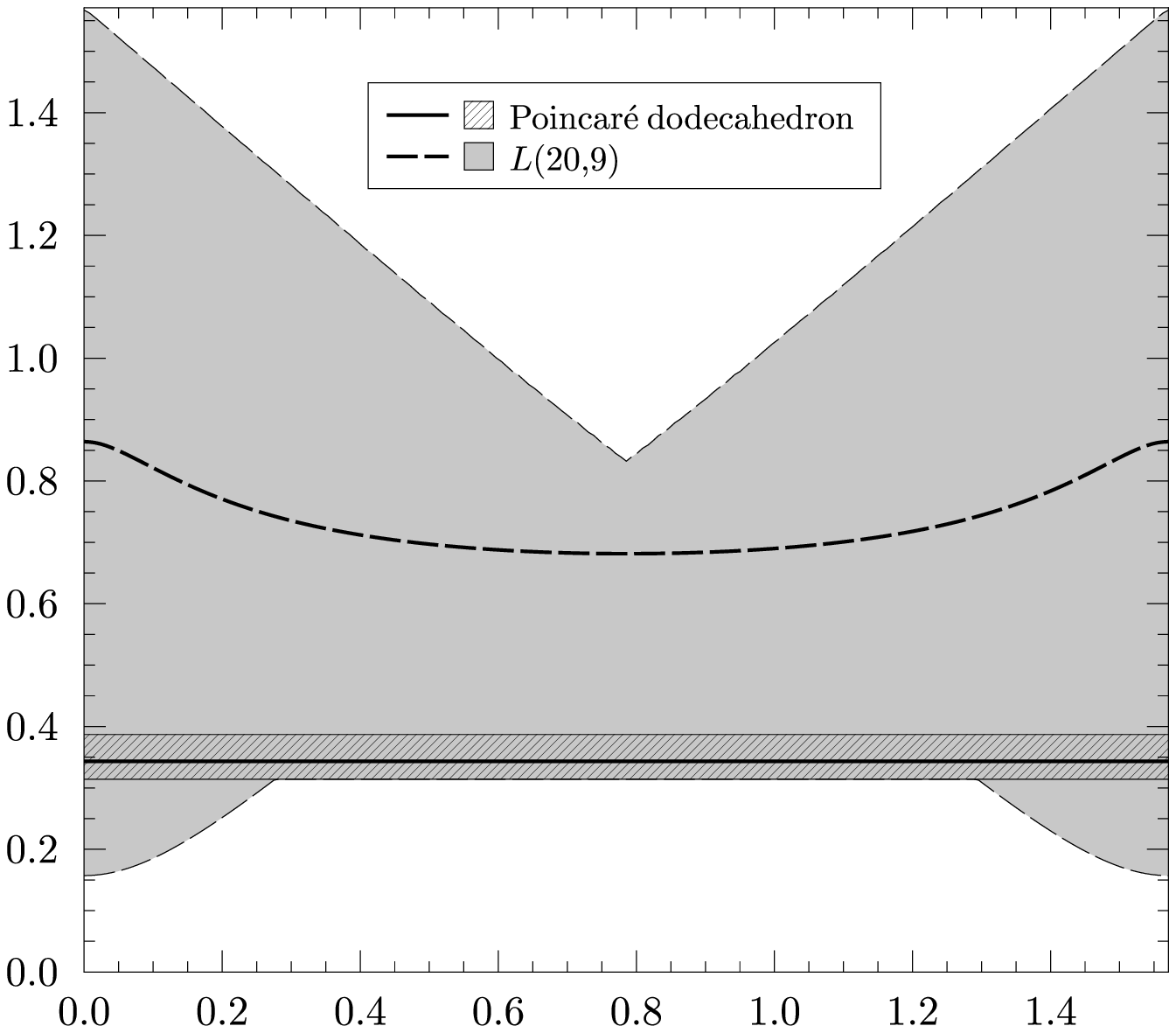}
\put(-405,175){(a)}
\put(-465,195){$\frac{\sigma_{\tau}}{\langle \tau \rangle }$}
\put(-282,22){$\rho$}
\put(-190,175){(b)}
\put(-230,185){$\tau$}
\put(-52,22){$\rho$}
\vspace*{-25pt}
\end{center}
\caption{\label{fig:sigma_dist_sph}
In panel (a) the relative variation $\sigma_{\tau}/\langle \tau \rangle $ of 
the geometric shape of the Voronoi domain 
is shown as a shape measure.
It is shown for five lens spaces $L(p,p/2-1)$ 
as a function of the observer position which is parameterised by $\rho$. 
For $\rho=0$ these lens spaces possess a Voronoi domain of the same geometry
as the homogeneous lens spaces $L(p,1)$ and for $\rho=\pi/4$ as 
the homogeneous prism spaces ${\cal D}_p={\cal S}^3/D^\star_{p}$.
For this reason the ratios $\sigma_{\tau}/\langle \tau \rangle$ have the same 
value for $L(p,p/2-1)$ at $\rho=0$ and $L(p,1)$, and also for $L(p,p/2-1)$ 
at $\rho=\pi/4$ and ${\cal D}_p$.
The horizontal lines give the corresponding values of 
the Poincar\'e dodecahedron ${\cal I}$ (full line), 
of the binary octahedral space ${\cal O}$ (dashed line), 
and of the binary tetrahedral space ${\cal T}$ (dotted line).
In panel (b) the mean values $\langle \tau \rangle $ for the lens space 
$L(20,9)$ and for the Poincar\'e dodecahedron are diagrammed.
In addition the minimal distance $\tau_{\hbox{\scriptsize min}}$ and the maximal 
distance $\tau_{\hbox{\scriptsize max}}$ from the observer to the surface 
of the Voronoi domain for these manifolds are represented. 
The interval between the minimal and the maximal distances is shown 
as a shaded band for the Poincar\'e dodecahedron and as a grey area
for the lens space $L(20,9)$.
Between $\rho$ about 0.275 and 1.296 the lens space $L(20,9)$ has the same
minimal distance $\tau_{\hbox{\scriptsize min}}$ as the Poincar\'e dodecahedron. 
}
\end{figure}


In the following we use $\sigma_\tau$, 
i.\,e.\ the square root of the variance (\ref{Eq:variance_radius}),
as a shape measure.
Since the manifolds of interest possess Voronoi domains
with different volumina,
we consider the normalised quantity $\sigma_\tau/\langle \tau \rangle$
where $\langle \tau \rangle$ is the mean radius of the Voronoi domain.
We investigate this ratio for the lens spaces $L(p,p/2-1)$
in dependence on the observer position which is parameterised by $\rho$.
The relative variation $\sigma_\tau/\langle \tau \rangle$
is diagrammed in figure \ref{fig:sigma_dist_sph}(a) for the lens spaces
belonging to $p=8$, 12, 24, 48, and 120.
One observes that the values of $\sigma_\tau/\langle \tau \rangle$
increase with the order $p$ of the deck group.
For the considered lens spaces $L(p,p/2-1)$ the minimum of
$\sigma_\tau/\langle \tau \rangle$ is always found at $\rho=\pi/4$. 
This is a special position of the observer since the lens space $L(p,p/2-1)$
possesses for $\rho=\pi/4$ a Voronoi domain of the same geometrical shape
as the prism space ${\cal D}_p$.
Thus, the lens space $L(p,p/2-1)$ at $\rho=\pi/4$ and the prism space 
${\cal D}_p$ have the same value for the introduced measure of shape.
In addition all lens spaces $L(p,q)$, $q>1$, have the same shape of the 
Voronoi domain at $\rho=0$ as the homogeneous lens space $L(p,1)$, 
namely a spherical lens.
Therefore, they have also the same value for the ratio 
$\sigma_{\tau}/\langle \tau \rangle$ which does not contain information
about the identification of points on the surface of the lens. 

The Poincar\'e dodecahedron ${\cal I}$,
the binary octahedral space ${\cal O}$,
and the binary tetrahedral space ${\cal T}$
can describe very good the CMB anisotropies at large scales
\cite{Gundermann_2005,Aurich_Lustig_Steiner_2005a}. 
For this reason we use the shape measure of these manifolds
as reference values. 
Since these three manifolds are homogeneous,
their values of $\sigma_{\tau}/\langle \tau \rangle$ have no $\rho$ dependence
and are represented by the horizontal lines in
figure \ref{fig:sigma_dist_sph}(a).
The figure reveals that only $L(8,3)$ around $\rho=\pi/4$
has values for the shape measure of the same magnitude as
the binary polyhedral spaces ${\cal I}$, ${\cal O}$, and ${\cal T}$.
If a small value of the ratio $\sigma_{\tau}/\langle \tau \rangle $ 
implies a good characterisation of the CMB anisotropies, 
then the results of figure \ref{fig:sigma_dist_sph}(a) would suggest
that a good description of the CMB is given only 
in the case of the lens space $L(8,3)$ and 
the prism space ${\cal D}_8$. 
In sec. \ref{subsec:cmb_correlation_and_well_prop}
and \ref{subsec:CMB_for_general_rho}
we investigate how far this measure of shape can really reflect the properties
of the CMB anisotropies on these manifolds. 

To give an impression how the value of $\langle \tau \rangle$ depends on
the observer position in the case of the lens spaces $L(p,p/2-1)$, 
the mean value $\langle \tau \rangle$ for the lens space $L(20,9)$
is shown in figure \ref{fig:sigma_dist_sph}(b)
together with the minimal value $\tau_{\hbox{\scriptsize min}}$ and
the maximal value $\tau_{\hbox{\scriptsize max}}$.
It turns out that the maximal value of $\langle \tau \rangle$
for $L(p,q)$ is obtained for the usual lens shaped domain
which corresponds in our description to $\rho=0$.
The maximal value of $\langle \tau \rangle$ is then the average of the
largest and the smallest values of $\tau(\hat n)$.
With $\tau_{\hbox{\scriptsize max}}=\pi/2$ and
$\tau_{\hbox{\scriptsize min}}=\pi/p$ for $\rho=0$, this value is
\begin{equation}
\label{Eq:variance_radius_max}
\langle \tau \rangle_{\hbox{\scriptsize max}} \; = \;
\frac 12 \,\left( \tau_{\hbox{\scriptsize max}}+\tau_{\hbox{\scriptsize min}}\right)
\; = \;
\frac \pi 2 \,\left(\frac 12+\frac 1p\right)
\hspace{10pt} .
\end{equation}
In addition,
the values for $\langle \tau \rangle$, $\tau_{\hbox{\scriptsize min}}$,
and $\tau_{\hbox{\scriptsize max}}$ of the Poincar\'e dodecahedron
are diagrammed in figure \ref{fig:sigma_dist_sph}(b).
An interesting point is that between $\rho$ about 0.275 and 1.296 
the lens space $L(20,9)$ has the same value for $\tau_{\hbox{\scriptsize min}}$ 
as the Poincar\'e dodecahedron.
Why this is important will become clear in
sec. \ref{subsec:subgroups}.

Before we discuss the introduced shape measure with respect to the
suppression of the CMB anisotropy on large scales,
some remarks on the temperature 2-point correlation 
$C(\vartheta)$ are in order.


\section{The CMB anisotropy on spherical spaces and
the well-proportioned conjecture}
\label{sec:cmb_correlation_observ}



\subsection{Calculation of the CMB temperature correlations
on spherical spaces}


The temperature correlations of the CMB sky with respect to their
separation angle $\vartheta$ are an important diagnostic tool.
The correlations at large angles $\vartheta$,
where the topological signature is expected,
are most clearly revealed by the temperature 2-point correlation function
$C(\vartheta)$.
It is defined as
\begin{equation}
\label{Eq:C_theta}
C(\vartheta) \; := \; \left< \delta T(\hat n) \delta T(\hat n')\right>
\hspace{10pt} \hbox{with} \hspace{10pt}
\hat n \cdot \hat n' = \cos\vartheta
\hspace{10pt} ,
\end{equation}
where $\delta T(\hat n)$ is the temperature fluctuation in
the direction of the unit vector $\hat n$.
The 2-point correlation function $C(\vartheta)$ is related
to the multipole moments $C_l$ by
\begin{equation}
\label{Eq:C_theta_Cl}
C(\vartheta) \; = \; 
\sum_l\,\frac{2l+1}{4\pi}\,C_l\,P_l\left(\cos\vartheta\right)
\hspace{10pt}.
\end{equation}
The ensemble average of $C_l$ can be expressed for a lens space $L(p,q)$ 
by the quadratic sum of the expansion coefficients 
$\xi^{\beta,i}_{lm}(L(p,q);t)$ as discussed in sec.\,\ref{sec:quadratic_sum}
\begin{eqnarray}
\label{Eq:Cl_ensemble_L_p_q}
C_l & := &
\frac{1}{2l+1}\sum_{m=-l}^l\left\langle\left|a_{lm}\right|^2\right\rangle
\\ & = &\nonumber
\sum_{\beta}\frac{ T_l^2(\beta) \; P(\beta)}{2l+1}\sum_{m=-l}^{l}\sum_{i}\left|\xi^{\beta,i}_{lm}(L(p,q);t)\right|^2
\hspace{10pt} ,
\end{eqnarray}
with the initial power spectrum $P(\beta)\sim 1/(E_{\beta}\,\beta^{2-n_{\hbox{\scriptsize s}}})$
where $E_{\beta}=\beta^2-1$ are the eigenvalues of the Laplace-Beltrami
operator on the considered spherical manifold ${\cal M}$, 
$\beta=2j+1$. 
Within the framework of this paper the spectral index $n_{\hbox{\scriptsize s}}$ 
is chosen to $n_{\hbox{\scriptsize s}}=0.961$.
$T_l(k)$ is the transfer function containing the full Boltzmann physics,
e.\,g.\ the ordinary and the integrated Sachs-Wolfe effect, 
the Doppler contribution, the Silk damping and the reionisation 
are taken into account.
The reionisation model of \cite{Aurich_Janzer_Lustig_Steiner_2007}
is applied with the reionisation parameters $\alpha=0.4$ and $\beta=9.85$.
Using the expression (\ref{Eq:quadratic_sum_L_p_q}) for the 
expansion coefficients $\xi^{\beta,i}_{lm}(L(p,q);t)$ 
we get for the ensemble average of $C_l$ on $L(p,q)$
\begin{equation}
\label{Eq:Cl_ensemble_L_p_q_final}
C_l \; = \;
\sum_{\beta}\frac{ T_l^2(\beta) \; P(\beta)}{2l+1}\sum_{m=-l}^{l}\sum_{m_a,m_b}
\left|\langle jm_aj\tilde{m}_b|lm\rangle\,d^j_{\tilde{m}_b,m_b}(-2\rho)\right|^2
\end{equation}
which depends only on the distance $\rho$.
Therefore, also the ensemble average of the 2-point correlation function 
on $L(p,q)$ with $q>1$ depends only on $\rho$.

The analogous formula for the ensemble average of $C_l$ on
homogeneous spherical manifolds ${\cal M}$ is given by
\begin{equation}
\label{Eq:Cl_ensemble_homogeneous}
C_l \; = \;
\sum_{\beta} T_l^2(\beta) \; P(\beta)\;\frac{r^{{\cal M}}(\beta)}{\beta^2}
\hspace{10pt} ,
\end{equation}
where we have used the sum relation (\ref{Eq:quadratic_sum_hom_mani})
for the quadratic sum of the expansion coefficients 
$\xi^{\beta,i}_{lm}({\cal M})$.

In this paper we restrict our analysis to lens spaces $L(p,p/2-1)$
and homogeneous manifolds ${\cal M}$ which have even order $p$ of the 
deck group.
In these cases it is possible to speed up the calculations of the 
ensemble average of $C_l$ 
using for $\beta>50$ the spectrum of the projective space 
${\cal P}^3$ divided by $V_{L(p,q)}/V_{{\cal P}^3}=p/2$.
We have numerically checked that this is a good approximation 
to the complete sum (\ref{Eq:Cl_ensemble_L_p_q_final}).
This approximation can be used in a similar way for all 
manifolds which tessellate the 3-sphere under a group of 
deck transformations of even order.

In order to quantify the power at large angular scales by a scalar measure,
the $S(60^{\circ})$ statistic
\begin{equation}
\label{Eq:S_statistic_60}
S(60^\circ)\; := \; \int^{\cos(60^\circ)}_{-1}
d\cos\vartheta \; |C(\vartheta)|^2
\hspace{10pt}
\end{equation}
has been introduced \cite{Spergel_et_al_2003},
which measures the power in the correlation function $C(\vartheta)$
on scales larger than $60^{\circ}$.
The arbitrary angle of $60^{\circ}$ is adapted to the observed missing power
of $C(\vartheta)$ for angles larger than this one.
The measure $d\cos\vartheta$ implies
that the $S(60^{\circ})$ statistic is insensitive to the behaviour
of the correlation function $C(\vartheta)$ at $\vartheta=180^\circ$.
It is more sensitive for variations of $C(\vartheta)$ in the range
$60^{\circ} \lesssim \vartheta \lesssim 120^{\circ}$.

\begin{figure}
\begin{center}
\vspace*{0pt}
\hspace*{0pt}
\hspace{-1.8cm}\includegraphics[width=9cm]{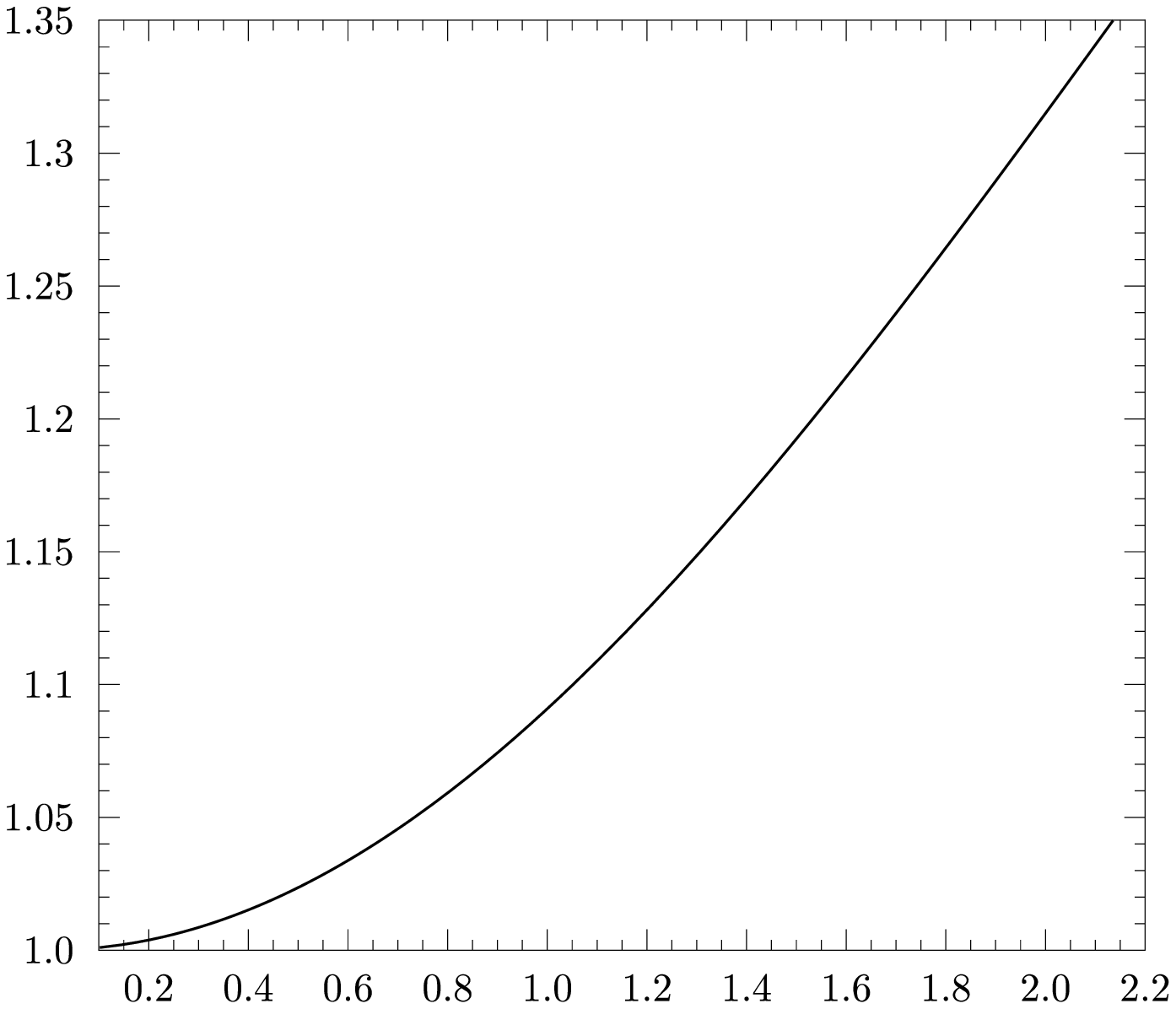}
\hspace{-1.0cm}\includegraphics[width=9cm]{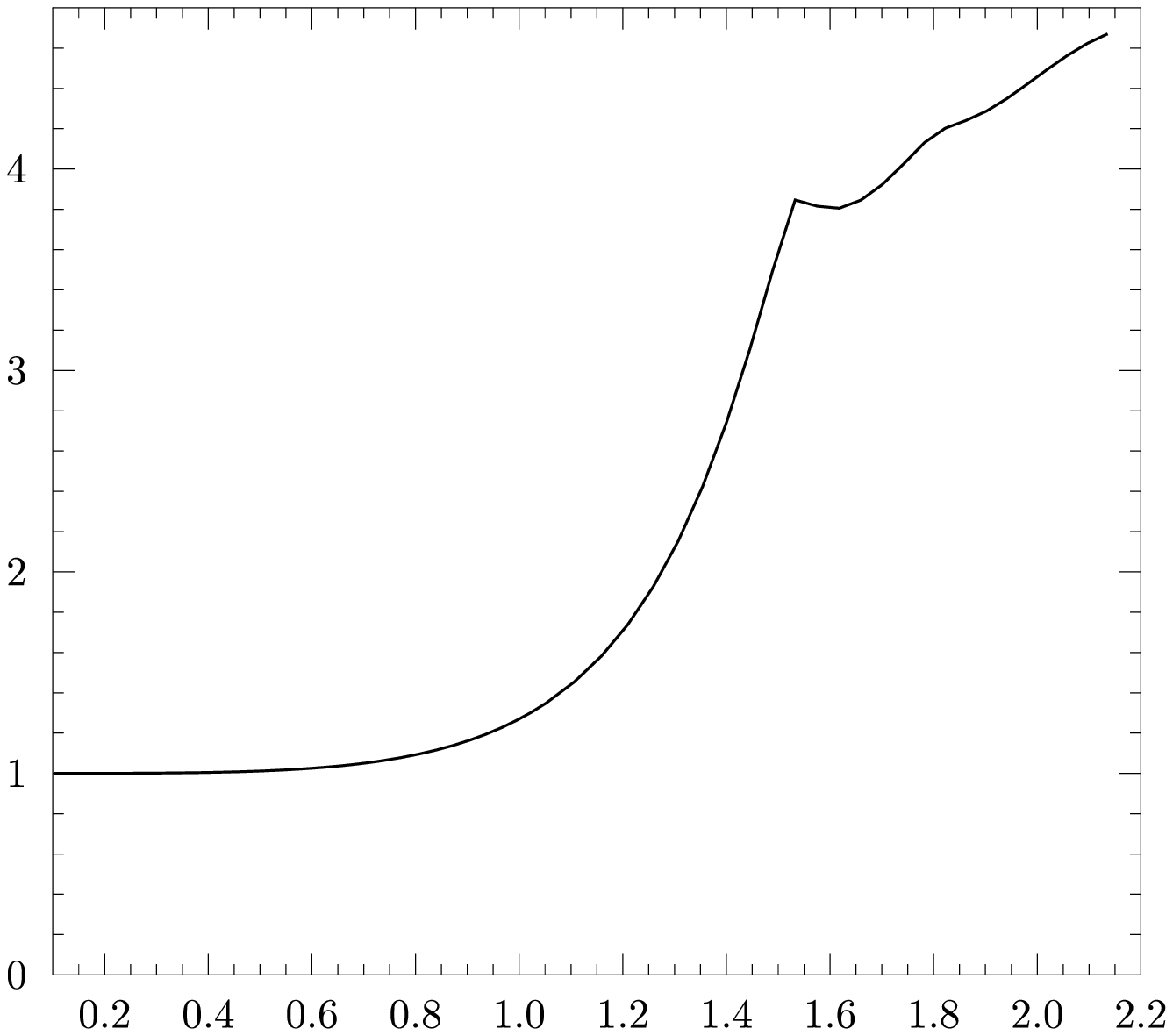}
\put(-435,165){(a)}
\put(-480,178){$\Omega_{\hbox{\scriptsize tot}}$}
\put(-285,18){$\tau_{\hbox{\scriptsize sls}}$}
\put(-205,165){(b)}
\put(-255,178){$\frac{S_{{\cal P}^3}(60^{\circ})}{S_{{\cal S}^3}(60^{\circ})}$}
\put(-55,18){$\tau_{\hbox{\scriptsize sls}}$}
\vspace*{-25pt}
\end{center}
\caption{\label{fig:S60_P3}
In panel (a) the connection of $\Omega_{\hbox{\scriptsize tot}}$ with the
distance to the surface of last scattering $\tau_{\hbox{\scriptsize sls}}$
is plotted for the density parameter of 
the cold dark matter $\Omega_{\hbox{\scriptsize cdm}} = 0.238$,
the density parameter of 
the baryonic matter $\Omega_{\hbox{\scriptsize bar}} = 0.0485$, and 
the Hubble constant $h=0.681$.
The density parameter of the cosmological constant 
$\Omega_{\scriptsize \Lambda }$ is changed according to get the desired total 
density parameter $\Omega_{\hbox{\scriptsize tot}}$.
Panel (b) shows the $S_{{\cal P}^3}(60^{\circ})$ statistics of the projective
space ${\cal P}^3\equiv L(2,1)$ normalised to the $S_{{\cal S}^3}(60^{\circ})$ 
statistics of the 3-sphere ${\cal S}^3$ depending on the distance 
$\tau_{\hbox{\scriptsize sls}}$. 
The deviation of the $S_{{\cal P}^3}(60^{\circ})$ statistics of the projective
space from the $S_{{\cal S}^3}(60^{\circ})$ statistics of the 3-sphere 
is smaller than 1 percent for $\tau_{\hbox{\scriptsize sls}}\lesssim 0.5$ and
smaller than 10 percent for $\tau_{\hbox{\scriptsize sls}}\lesssim 0.8$.
}
\end{figure}

We prefer to analyse the $S(60^{\circ})$ statistics as a function of 
the distance to the surface of last scattering $\tau_{\hbox{\scriptsize sls}}$
instead of a cosmological parameter such as the total density parameter 
$\Omega_{\hbox{\scriptsize tot}}$.
This emphasises the geometric aspect and allows the comparison
of the dimensions of the fundamental cell with respect to the distance
$\tau_{\hbox{\scriptsize sls}}$.
For the convenience of the reader,
the relation between the distance to the surface of last scattering
$\tau_{\hbox{\scriptsize sls}}$ and  the total density parameter 
$\Omega_{\hbox{\scriptsize tot}}$ is plotted in figure \ref{fig:S60_P3}(a)
for the cosmological parameters specified in the caption.
The current cosmologically viable range is only the part 
$\Omega_{\hbox{\scriptsize tot}} \lesssim 1.04$
which corresponds to $\tau_{\hbox{\scriptsize sls}} \lesssim 0.6$
according to figure \ref{fig:S60_P3}(a).
In the following we also investigate multi-connected universes
for values of $\tau_{\hbox{\scriptsize sls}} > 0.6$
in order to reveal the influence of the geometry of the Voronoi domain
and, more generally, the influence of the topology of the universe
on the CMB anisotropy.

We normalise the large scale power $S(60^{\circ})$
of a given manifold ${\cal M}$ to the corresponding power
of the projective space ${\cal P}^3$,
i.\,e.\ to $S_{{\cal P}^3}(60^{\circ})$.
As we will see this is a good choice to emphasise our point of interest.
The behaviour of the large scale power of the projective space ${\cal P}^3$
relative to that of the 3-sphere ${\cal S}^3$
is shown in figure \ref{fig:S60_P3}(b).
For $\tau_{\hbox{\scriptsize sls}} \lesssim 0.5$
the ratio $S_{{\cal P}^3}(60^{\circ})/S_{{\cal S}^3}(60^{\circ})$ is almost one,
and the projective space behaves roughly as the 3-sphere.
For values of $\tau_{\hbox{\scriptsize sls}} \gg 0.5$ the correlation function of 
the projective space reveals much more power at large scales
than the correlation function  of the 3-sphere,
but this is not the range which is interesting 
in the context of the present-day cosmology.

\begin{figure}
\vspace*{-15pt}
\begin{minipage}{18.0cm}
\hspace*{-15pt}
\begin{minipage}{9.0cm}
{
\hspace*{-15pt}
\includegraphics[width=9.0cm]{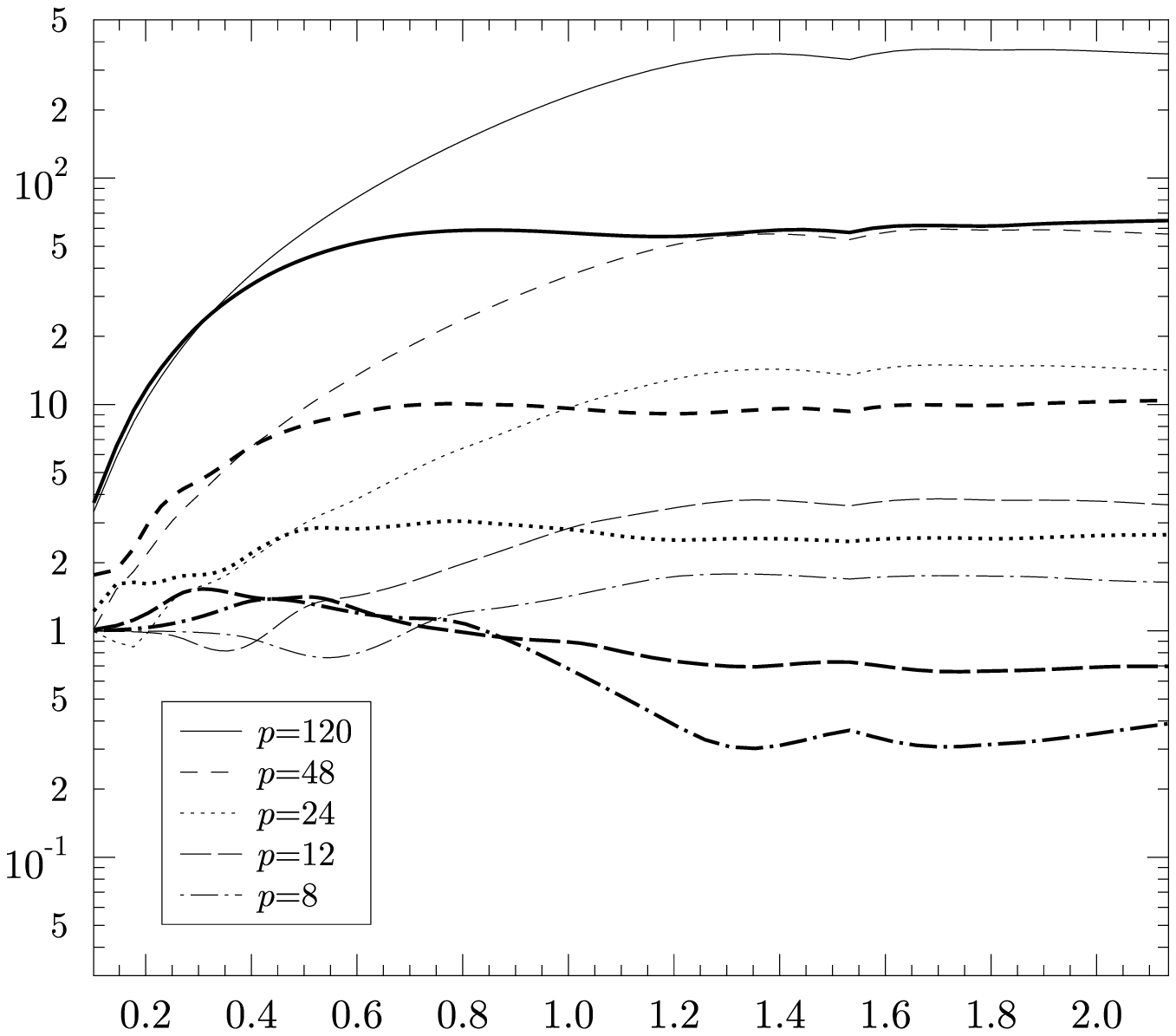}
\put(-262,190){$\frac{S_{\cal M}(60^{\circ})}{S_{{\cal P}^3}(60^{\circ})}$}
\put(-45,20){$\tau_{\hbox{\scriptsize sls}}$}
\put(-170,58){(a)\,$L(p,1)$ vs.}
\put(-160,42){$L(p,p/2-1)$ at $\rho=0$}
\vspace*{-45pt}
}
{
\includegraphics[width=9.0cm]{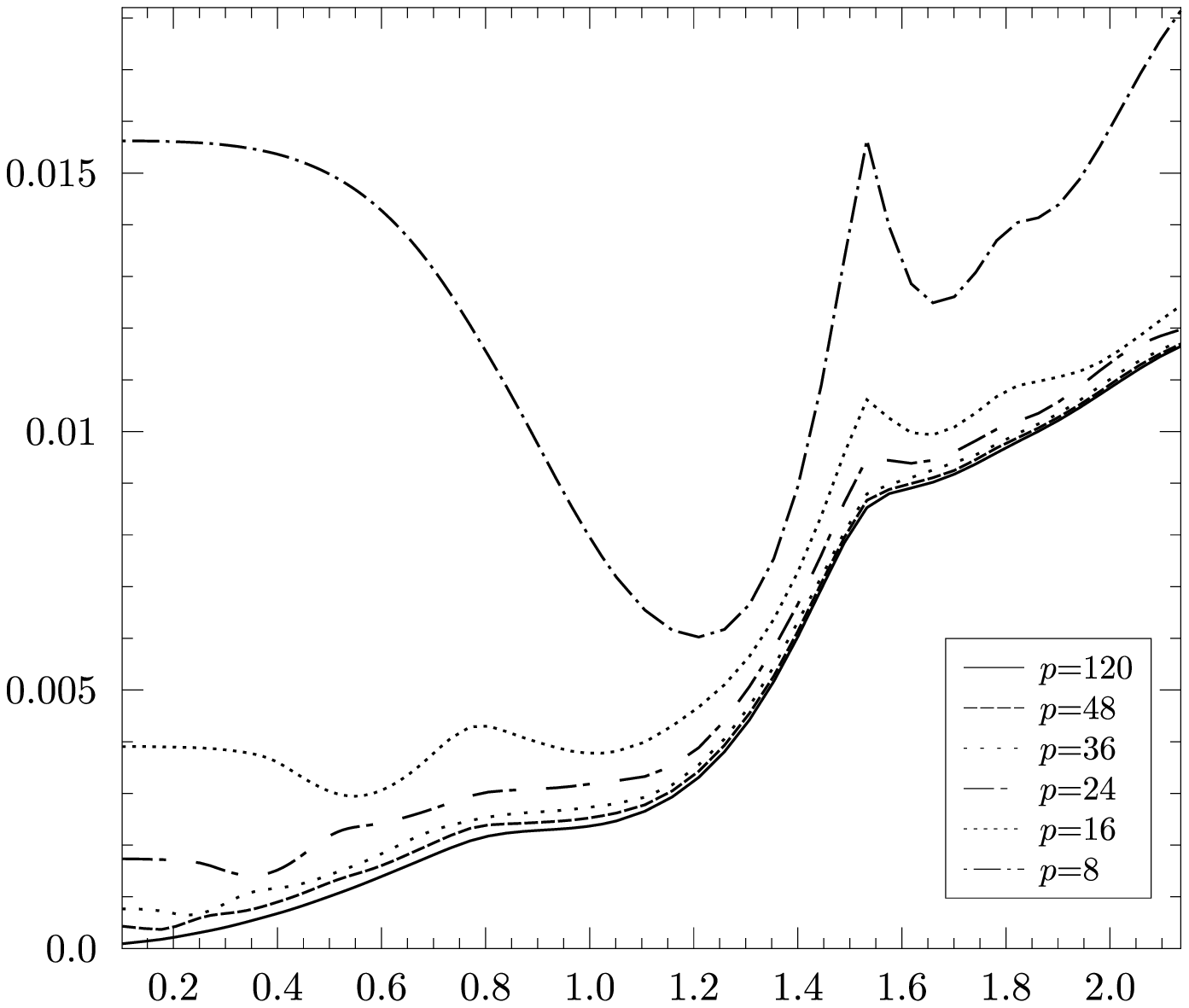}
\put(-262,172){$\frac{S_{\cal M}(60^{\circ})}{p^2}$}
\put(-45,20){$\tau_{\hbox{\scriptsize sls}}$}
\put(-205,177){(c)\,{$L(p,p/2-1)$ at $\rho=\pi/4$}}
}
\end{minipage}
\hspace*{-35pt}
\begin{minipage}{9.0cm}
{
\includegraphics[width=9.0cm]{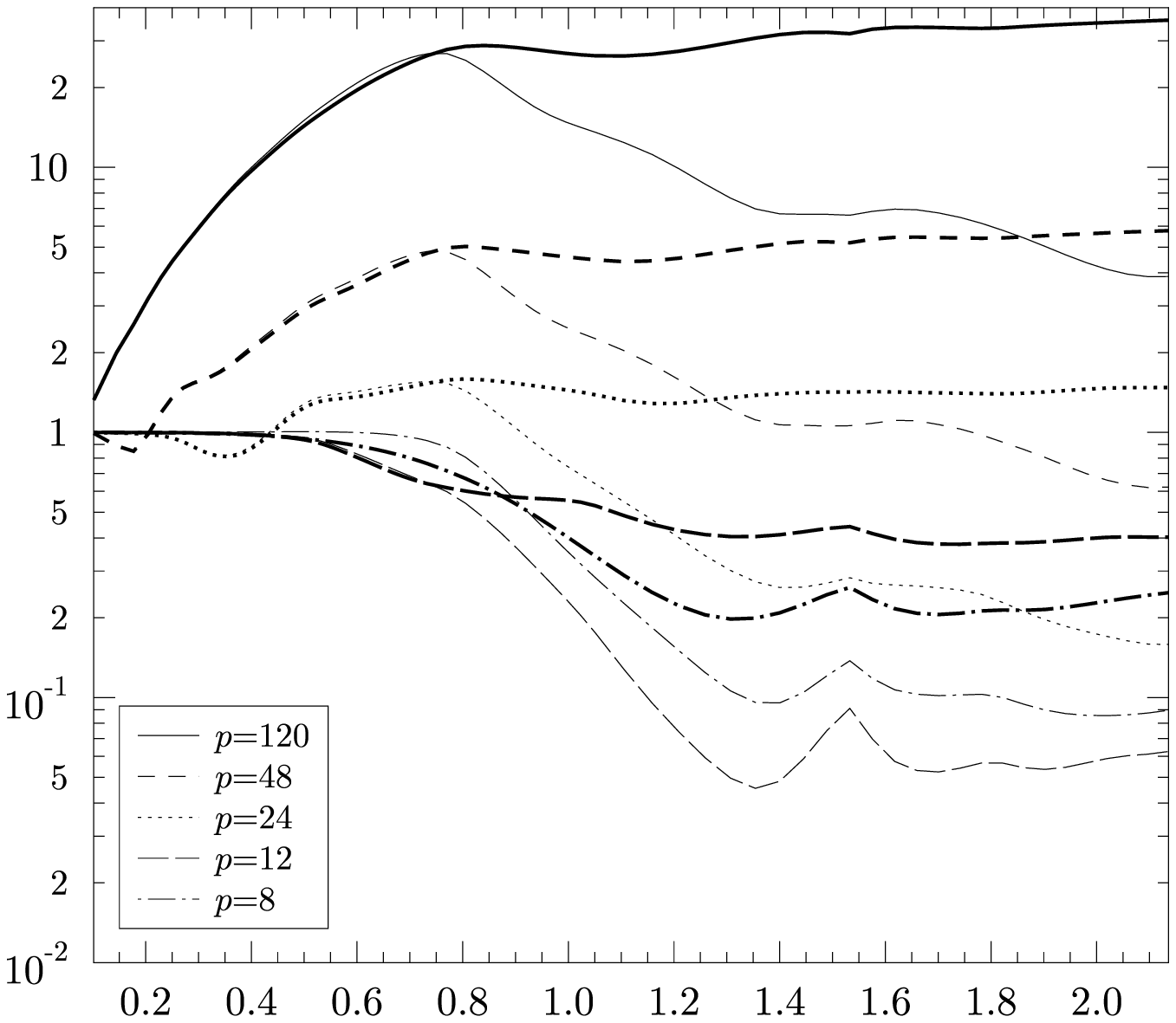}
\put(-258,190){$\frac{S_{\cal M}(60^{\circ})}{S_{{\cal P}^3}(60^{\circ})}$}
\put(-45,20){$\tau_{\hbox{\scriptsize sls}}$}
\put(-170,58){(b)\,${\cal D}_p$ vs.}
\put(-170,42){$L(p,p/2-1)$ at $\rho=\pi/4$}
\vspace*{-45pt}
}
{
\includegraphics[width=9.0cm]{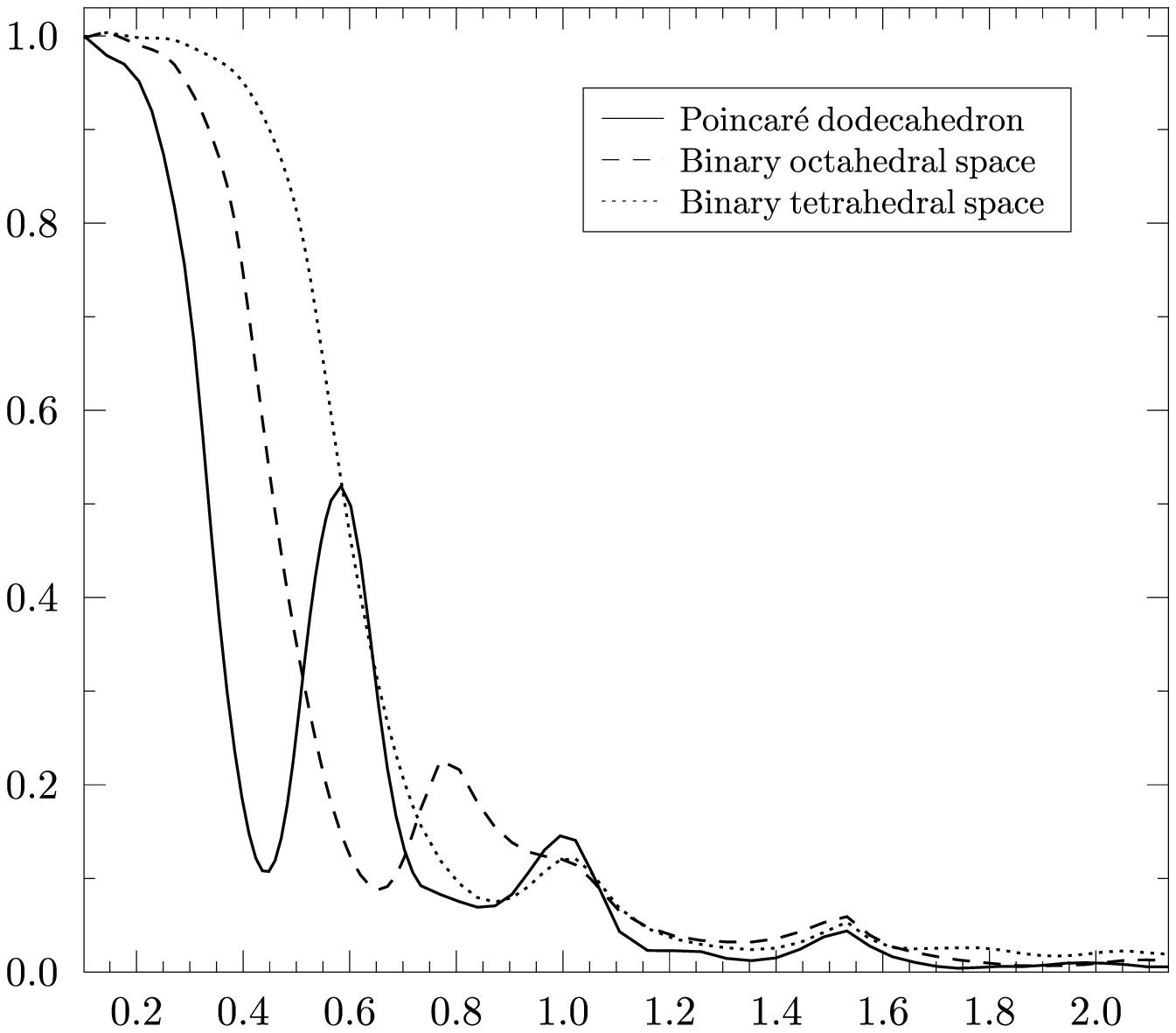}
\put(-262,172){$\frac{S_{\cal M}(60^{\circ})}{S_{{\cal P}^3}(60^{\circ})}$}
\put(-45,20){$\tau_{\hbox{\scriptsize sls}}$}
\put(-185,177){(d)}
}
\end{minipage}
\end{minipage}
\caption{\label{fig:comp_S60_geom}
In panel (a) the $S(60^{\circ})$ statistics of the lens spaces $L(p,p/2-1)$ 
with the observer at $\rho=0$ is compared with the
$S(60^{\circ})$ statistics of the homogeneous lens spaces $L(p,1)$.
The corresponding results for the lens spaces $L(p,p/2-1)$ 
with the observer at $\rho=\pi/4$ and for the prism spaces 
${\cal D}_p$ are shown in panel (b).
In both panels the thick lines belong to the lens spaces $L(p,p/2-1)$. 
In panels (a) and (b) the curves are normalised to the $S(60^{\circ})$ 
statistics of the projective space ${\cal P}^3$.
In panel (c), however, the quantity $S(60^{\circ})/p^2$ is shown for 
the lens spaces $L(p,p/2-1)$ at $\rho=\pi/4$
to emphasise the universal behaviour for a large order of the deck group $p$.
The $S(60^{\circ})$ statistics of the Poincare dodecahedron ${\cal I}$,
the binary octahedral space ${\cal O}$, and the binary tetrahedral space ${\cal T}$
are diagrammed in (d).
These three curves are again normalised to the $S(60^{\circ})$ statistics of
the projective space ${\cal P}^3$.
Notice that the scale is different for the vertical axis in all four graphics.
}
\end{figure}


\subsection{The large scale CMB anisotropy of lens and prism shaped
Voronoi domains} 
\label{subsec:cmb_correlation_and_well_prop}


Let us now address the question how the geometry of the Voronoi domain
affects the CMB properties.
As discussed in the Introduction and in sect.\,\ref{sec:shape_measure},
it is most advantageous to consider topologies
which possess the same Voronoi domain but with different connections
of their faces.
The Voronoi domain of $L(p,p/2-1)$ for $\rho=0$ is identical
to that of the space $L(p,1)$, 
and for $\rho=\pi/4$ to that of the prism space ${\cal D}_p$.
The $S(60^{\circ})$ statistics of the lens shaped Voronoi domains $(\rho=0)$
is shown in figure \ref{fig:comp_S60_geom}(a)
for the group orders $p=8$, 12, 24, 48, and 120.
One recognises that the CMB behaviour differs although each pair
for a given $p$ of $L(p,p/2-1)$ and $L(p,1)$ possesses the same Voronoi domain.
This demonstrates that, in addition to the shape,
the connection rules of the Voronoi domains affect the CMB properties.
In both cases the $S(60^{\circ})$ statistics growths with the group order $p$.
This behaviour could have been anticipated from the increasing asymmetry
$\sigma_\tau/\langle \tau \rangle$
of the Voronoi domains as shown in figure \ref{fig:sigma_dist_sph}(a)
by assuming the validity of the well-proportioned conjecture.
In addition, for $\tau_{\hbox{\scriptsize sls}}>0.8$,
the $S(60^{\circ})$ statistics takes on larger values for
the homogeneous manifold $L(p,1)$ than for
the corresponding lens shaped space $L(p,p/2-1)$.
Interestingly, this behaviour is reversed for the prism shaped Voronoi domains.
In figure \ref{fig:comp_S60_geom}(b) a comparison of
the prism shaped space $L(p,p/2-1)$ at $\rho=\pi/4$ with the
prism space ${\cal D}_p$ reveals this reversed behaviour for
$\tau_{\hbox{\scriptsize sls}}>0.8$.
In addition, their behaviour for sufficiently small values of
$\tau_{\hbox{\scriptsize sls}}$ is very similar.
However, the fact that the curves diverges for 
$\tau_{\hbox{\scriptsize sls}}\gtrsim 0.8$ demonstrates again
that the geometry of the Voronoi domain cannot suffice
in order to solely explain the CMB anisotropy suppression.
Thus, there are counter-examples to the well-proportioned conjecture.

\begin{figure}
\vspace*{-15pt}
\begin{minipage}{18.0cm}
\hspace*{-15pt}
\begin{minipage}{9.0cm}
{
\includegraphics[width=9.0cm]{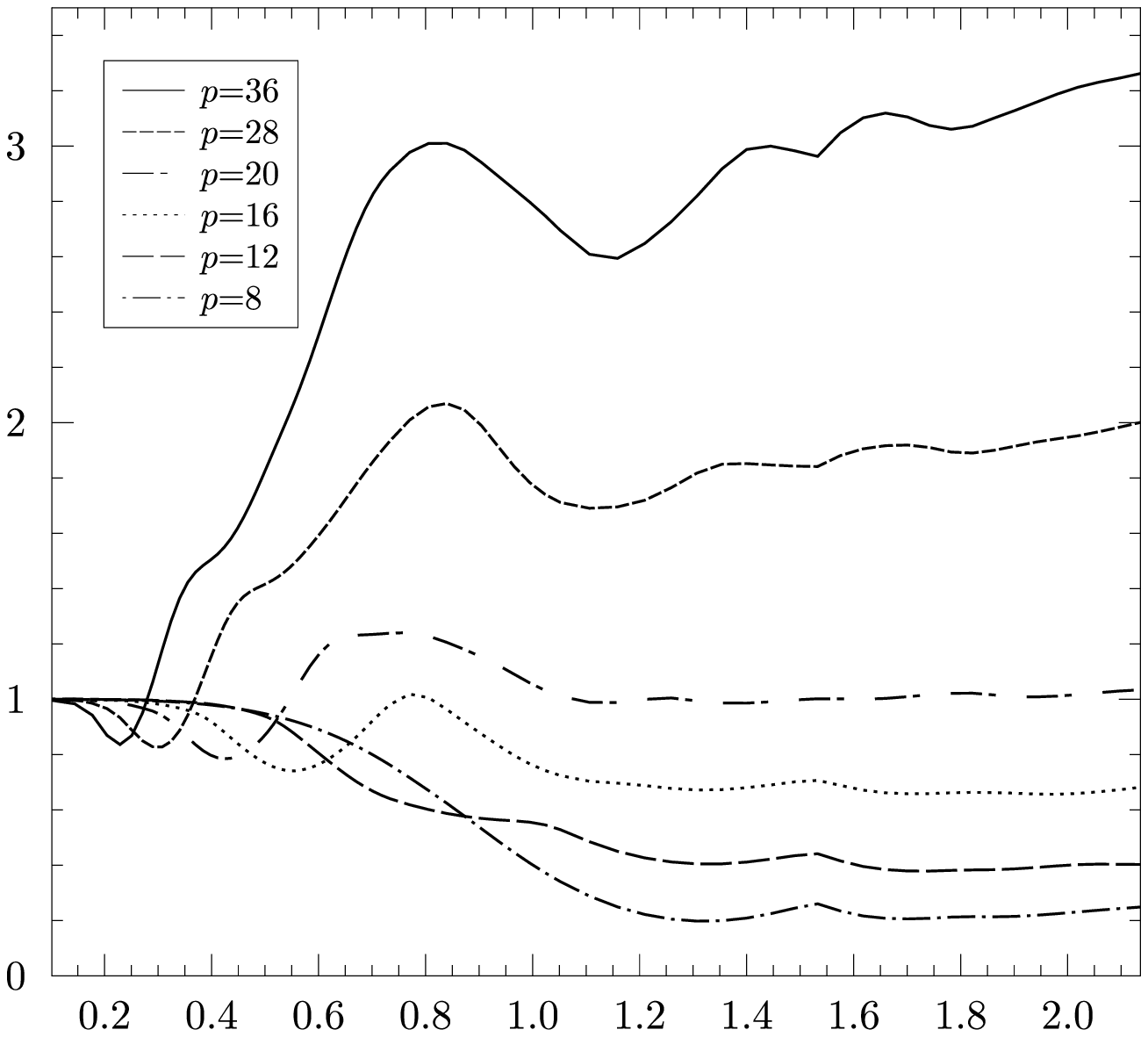}
\put(-256,185){$\frac{S_{\cal M}(60^{\circ})}{S_{{\cal P}^3}(60^{\circ})}$}
\put(-45,23){$\tau_{\hbox{\scriptsize sls}}$}
\put(-128,142){(a) $L(p,p/2-1)$}
\put(-108,127){at $\rho=\pi/4$}
\vspace*{-45pt}
}
{
\includegraphics[width=9.0cm]{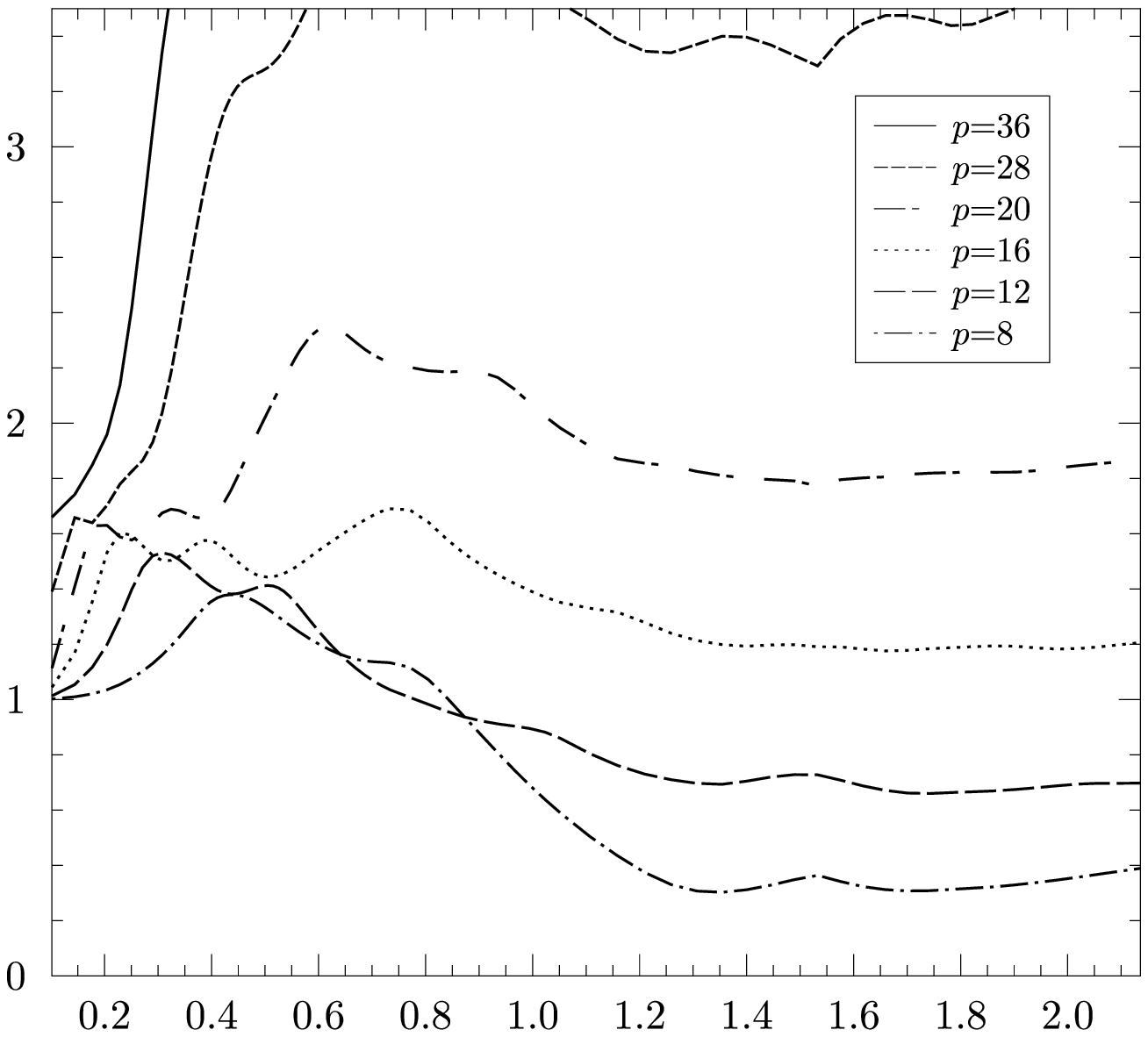}
\put(-256,185){$\frac{S_{\cal M}(60^{\circ})}{S_{{\cal P}^3}(60^{\circ})}$}
\put(-45,23){$\tau_{\hbox{\scriptsize sls}}$}
\put(-175,173){(c) $L(p,p/2-1)$}
\put(-157,158){at $\rho=0$}
}
\end{minipage}
\hspace*{-35pt}
\begin{minipage}{9.0cm}
{
\includegraphics[width=9.0cm]{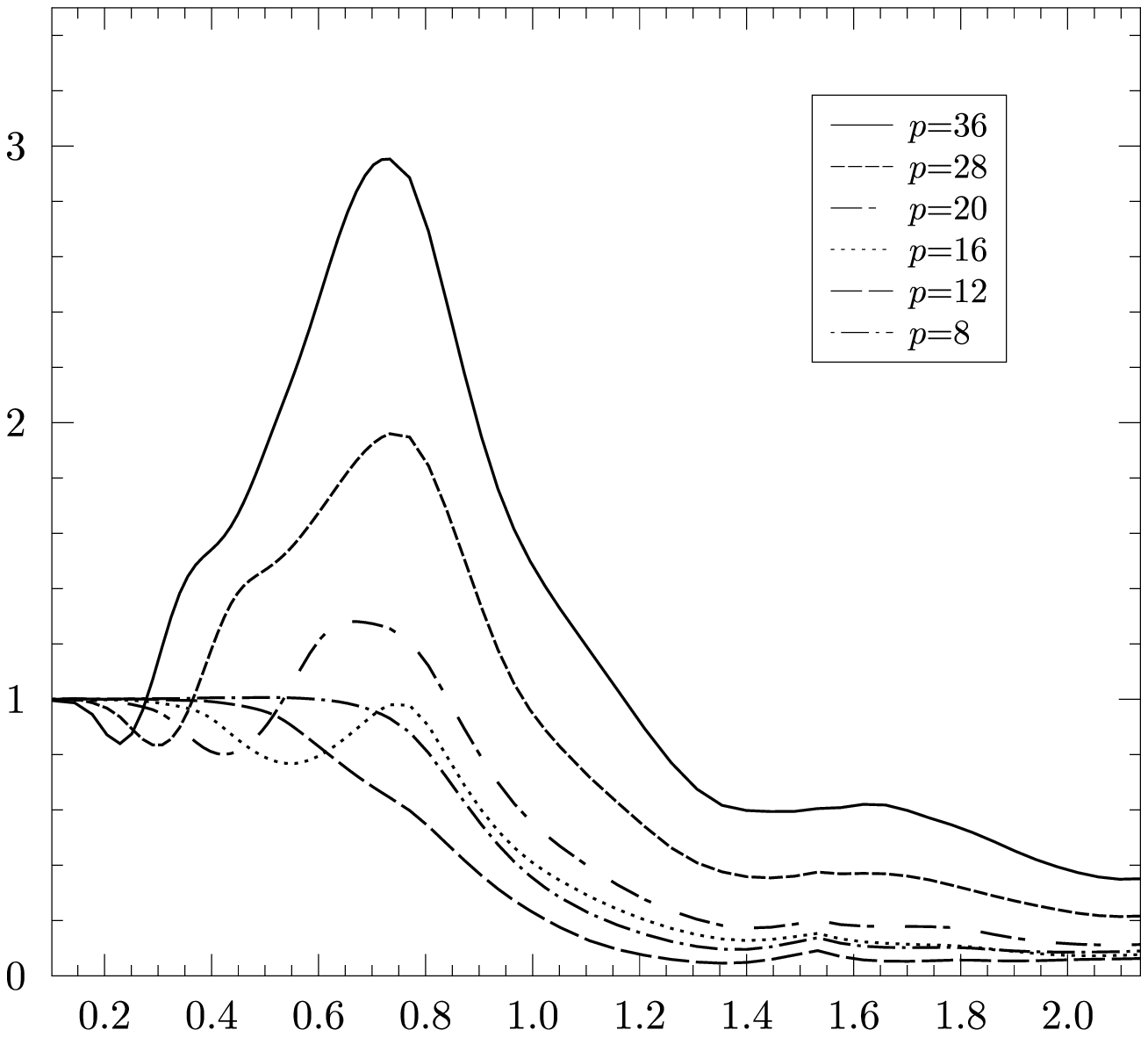}
\put(-256,185){$\frac{S_{\cal M}(60^{\circ})}{S_{{\cal P}^3}(60^{\circ})}$}
\put(-45,23){$\tau_{\hbox{\scriptsize sls}}$}
\put(-100,120){(b) ${\cal D}_p$}
\vspace*{-45pt}
}
{
\includegraphics[width=9.0cm]{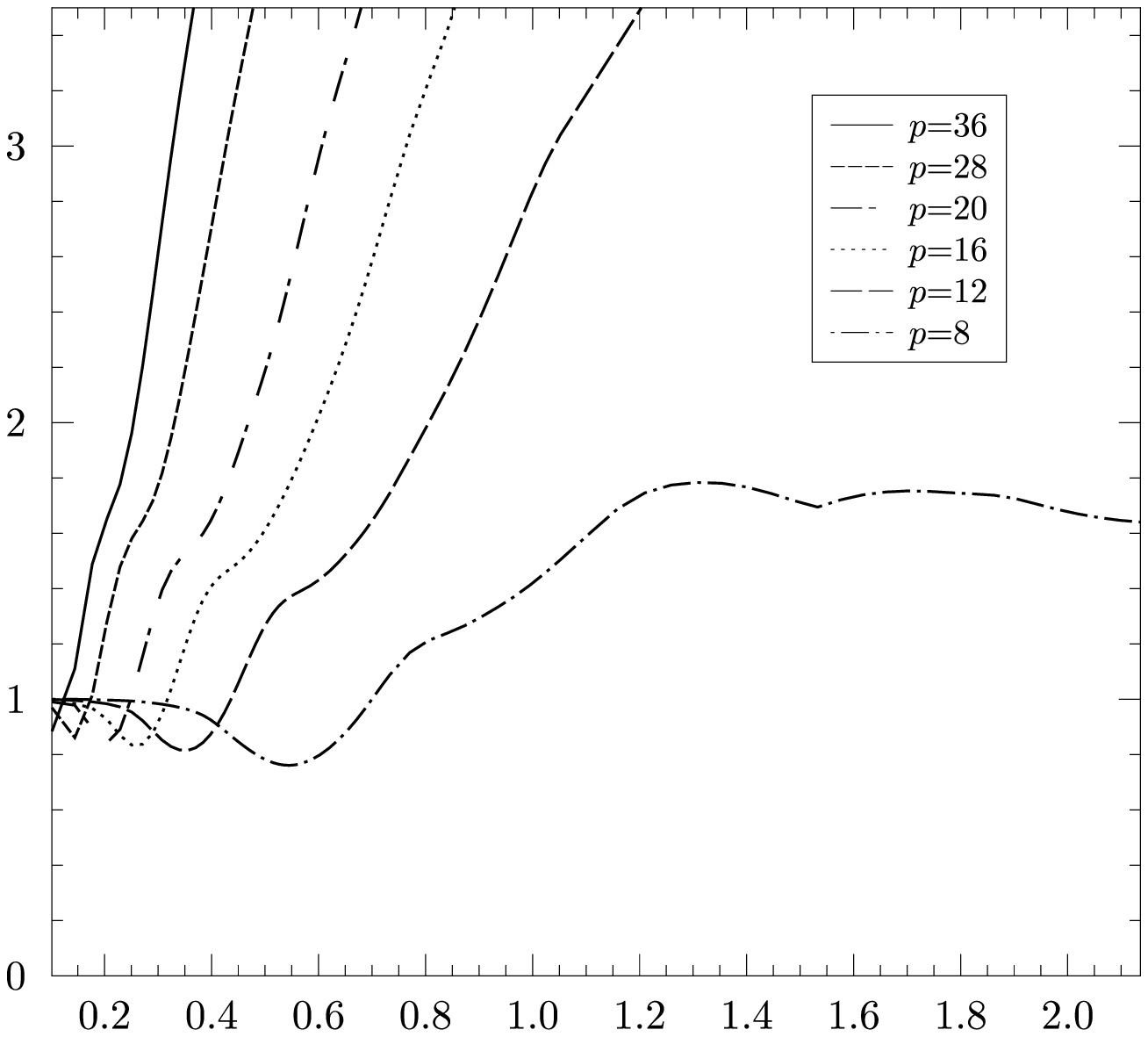}
\put(-256,185){$\frac{S_{\cal M}(60^{\circ})}{S_{{\cal P}^3}(60^{\circ})}$}
\put(-45,23){$\tau_{\hbox{\scriptsize sls}}$}
\put(-100,120){(d) $L(p,1)$}
}
\end{minipage}
\end{minipage}
\caption{\label{fig:comp_S60_geom_part_2}
The $S(60^{\circ})$ statistics is compared between models having
the same Voronoi domain.
The selected group orders are $p=8$, 12, 16, 20, 28, and 36.
The panels (a) and (b) refer to the prism shaped domains
for $L(p,p/2-1)$ with the observer at $\rho=\pi/4$ and
the prism spaces ${\cal D}_p$.
The CMB correlations of the lens shaped domains of $L(p,p/2-1)$
with the observer at $\rho=0$
and of the lens spaces $L(p,1)$ are displayed in panels (c) and (d).
}
\end{figure}

As figure \ref{fig:sigma_dist_sph}(a) reveals,
the shape measure $\sigma_{\tau}/\langle \tau \rangle$ is lower for
the prism space ${\cal D}_8$ than for ${\cal D}_{12}$.
Thus, one would expect that the power in the CMB anisotropy is also
lower for ${\cal D}_8$.
But the reverse behaviour is observed in figures \ref{fig:comp_S60_geom}(b)
and \ref{fig:comp_S60_geom_part_2}(b). 
Hence, this is a further counter-example to the well-proportioned conjecture. 

As just mentioned, the values of the $S(60^{\circ})$ statistics increase
with the group order $p$.
It turns out that the proportionality is $1/p^2$
as shown in figure \ref{fig:comp_S60_geom}(c)
where the quantity $\frac{S_{\cal M}(60^{\circ})}{p^2}$
is plotted for the prism shape $L(p,p/2-1)$ at $\rho=\pi/4$.
With the exception of the space $L(8,3)$,
a convergence of the curves with increasing group order $p$ is revealed.
The space $L(8,3)$ at $\rho=\pi/4$ is a special case
since it is a spherical Platonic space,
see \cite{Aurich_Kramer_Lustig_2011}.
A similar scaling behaviour occurs at $\rho=0$ where the Voronoi domains
possess the shape of a lens.
Thus, a universal behaviour emerges at large order $p$ of the deck group.

The $S(60^{\circ})$ statistics of the binary polyhedral spaces ${\cal I}$,
${\cal O}$, and ${\cal T}$
is presented in figure \ref{fig:comp_S60_geom}(d).
This allows the comparison of these well studied models with the
multi-connected spaces shown in the other figures.
The minimum at $\tau_{\hbox{\scriptsize sls}}\simeq 0.45$ for
the Poincar\'e dodecahedron ${\cal I}$ is the due to the
famous CMB correlation suppression at
$\Omega_{\hbox{\scriptsize tot}}\simeq 1.02$.

A further interesting detail of the $S(60^{\circ})$ statistics 
is revealed in figure \ref{fig:comp_S60_geom_part_2}.
The figures \ref{fig:comp_S60_geom_part_2}(a) and
\ref{fig:comp_S60_geom_part_2}(b) display
the CMB behaviour for the prism shaped models
$L(p,p/2-1)$ at $\rho=\pi/4$ and ${\cal D}_p$, respectively.
A comparison of the $S(60^{\circ})$ statistics for the models with
$p=36$ (full curve) shows that $L(36,17)$ and ${\cal D}_{36}$
possess for $\tau_{\hbox{\scriptsize sls}}\lesssim 0.7$ a similar statistics
but completely different values for larger $\tau_{\hbox{\scriptsize sls}}$.
The cosmological interesting minima are below
$\tau_{\hbox{\scriptsize sls}}\simeq 0.6$ and, thus,
both models give the same statistical description for our Universe.
The same trend is seen for all the other group orders $p$.
This might be a point in favour of the well-proportioned hypothesis.
However, a counter-example is provided by the lens shaped models
$L(p,p/2-1)$ at $\rho=0$ and $L(p,1)$ shown in figures
\ref{fig:comp_S60_geom_part_2}(c) and \ref{fig:comp_S60_geom_part_2}(d).
Despite identical Voronoi domains,
their statistics are dissimilar even for small distances
$0.1 \lesssim \tau_{\hbox{\scriptsize sls}}\lesssim 0.6$.
While the homogeneous models $L(p,1)$ have minima
comparable to those of the prism spaces,
the lens shaped models $L(p,p/2-1)$ at $\rho=0$ have none at all.
The latter possess even more CMB anisotropy power than the projective
space ${\cal P}^3$ for cosmological interesting values of
$\tau_{\hbox{\scriptsize sls}}$.
For small values of $\tau_{\hbox{\scriptsize sls}}$
the homogeneous lens spaces $L(p,1)$
display lower power of the correlation function $C(\vartheta)$ than
the projective space ${\cal P}^3$ and the 3-sphere ${\cal S}^3$.
The minima of the $S(60^{\circ})$ statistics occur between
$\tau_{\hbox{\scriptsize sls}}=\pi/p$ and $2\,\pi/p$.

\begin{figure}
\vspace*{-15pt}
\begin{minipage}{18.0cm}
\hspace*{-15pt}
\begin{minipage}{9.0cm}
{
\includegraphics[width=9.0cm]{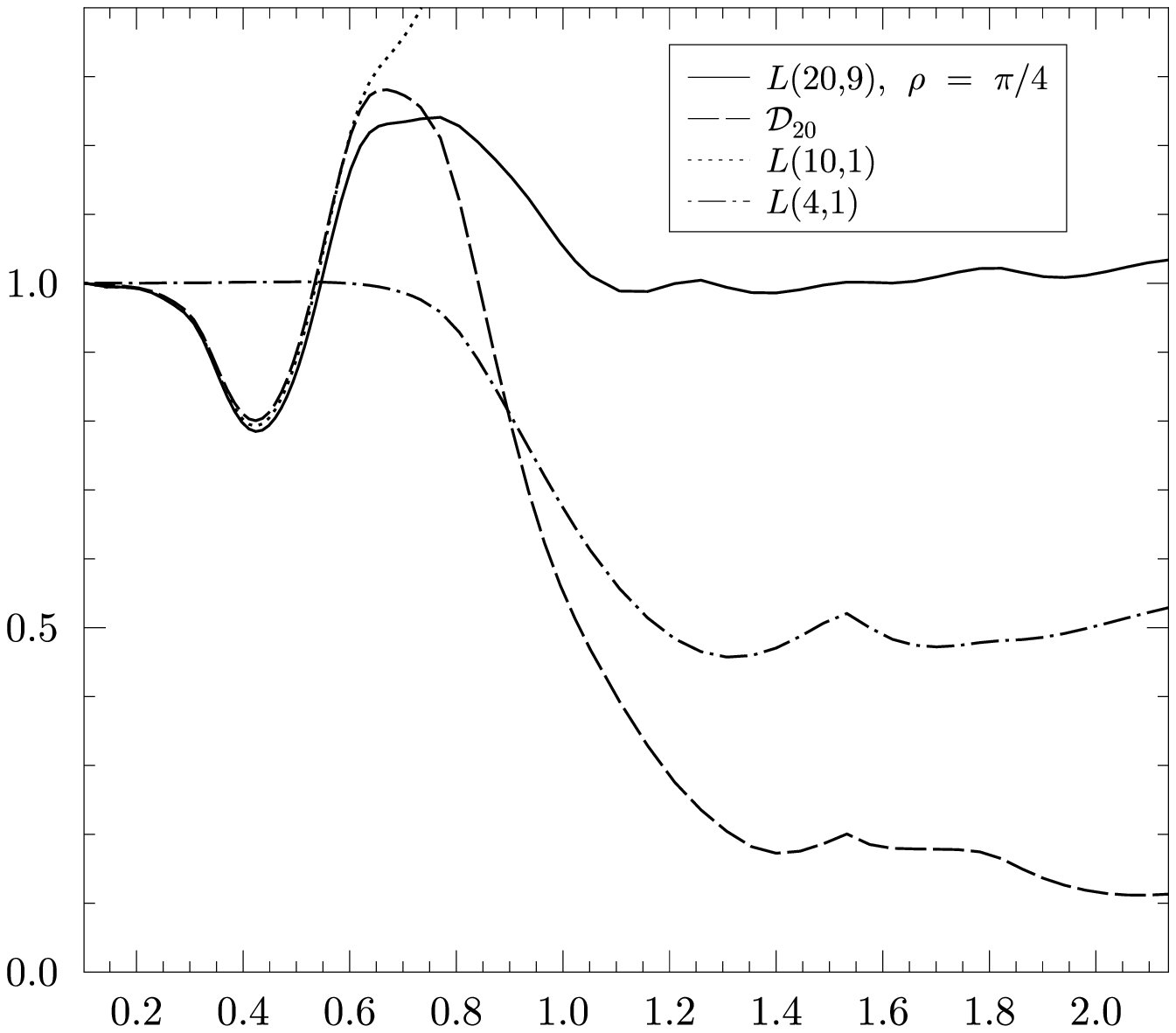}
\put(-255,176){$\frac{S_{\cal M}(60^{\circ})}{S_{{\cal P}^3}(60^{\circ})}$}
\put(-45,23){$\tau_{\hbox{\scriptsize sls}}$}
\put(-205,177){(a)}
}
\end{minipage}
\hspace*{-35pt}
\begin{minipage}{9.0cm}
{
\includegraphics[width=9.0cm]{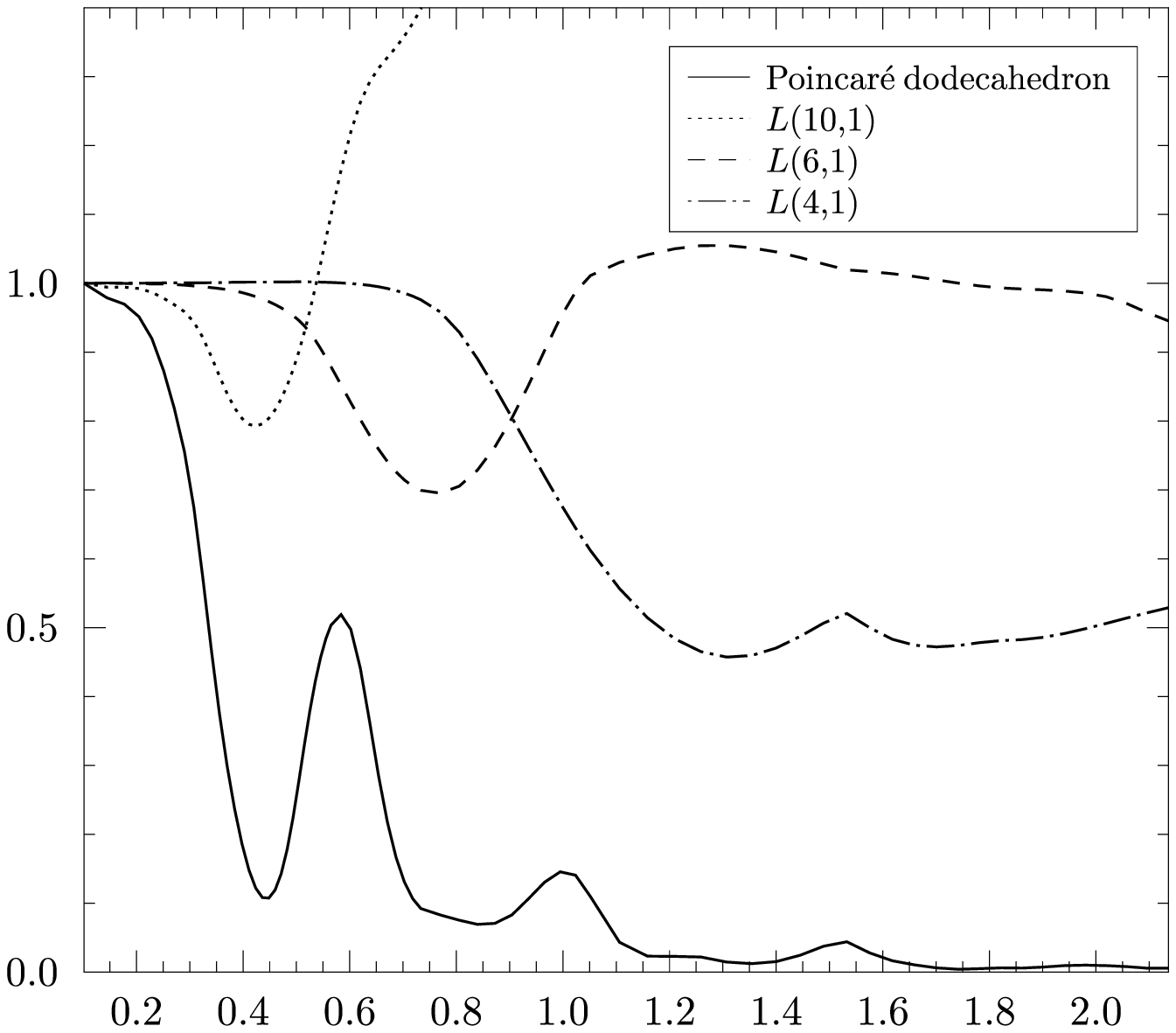}
\put(-255,176){$\frac{S_{\cal M}(60^{\circ})}{S_{{\cal P}^3}(60^{\circ})}$}
\put(-45,23){$\tau_{\hbox{\scriptsize sls}}$}
\put(-205,177){(b)}
}
\end{minipage}
\end{minipage}
\caption{\label{fig:comp_S60_group}
In panel (a) the influence of the cyclic subgroups 
on the $S(60^{\circ})$ statistics of the prism space ${\cal D}_{20}$ 
and of the lens space $L(20,9)$ with the observer at $\rho=\pi/4$ is studied.  
For this purpose also the $S(60^{\circ})$ statistics for the lens spaces 
$L(10,1)$ and $L(4,1)$ are shown.
The analogous influence is studied in panel (b) for
the Poincar\'e dodecahedron.
The decomposition of its deck group given in table \ref{Tab:subgroups}
requires the comparison with the $S(60^{\circ})$ statistics of the
lens spaces $L(10,1)$, $L(6,1)$, and $L(4,1)$.
}
\end{figure}


\subsection{The role of cyclic subgroups $Z_p$ of Clifford translations}
\label{subsec:subgroups}


The geometrical interpretation of the role of cyclic subgroups 
of Clifford translations is that such a subgroup determines 
the gluing rule for two opposing faces of the Voronoi domain.
Because Clifford translations shift all points on the 3-sphere ${\cal S}^3$ 
by the same distance,
the largest of those subgroups is related to the closest 
constant distance between such opposing faces
which in turn determines the order of the wavelength of a perturbation
in that  direction and thus the CMB anisotropy suppression.
Therefore, the subgroup $Z_p$ belonging to the largest value of $p$
is the most interesting from our point of view.
Subgroups of the largest subgroup $Z_p$,
which have a common transfomation direction, are ignored in the following.
The origin of a similar or a diverging behaviour
for models with an identical Voronoi domain can be understood in
terms of these subgroups.

The deck groups $\Gamma$ can possess common subgroups of Clifford translations.
The trivial subgroup is the cyclic subgroup $Z_1$
which is common to all spherical models.
The cyclic subgroup $Z_2$ generates the projective space ${\cal P}^3$.
Since we consider the normalised
$\frac{S_{\cal M}(60^{\circ})}{S_{{\cal P}^3}(60^{\circ})}$ statistics,
the contribution of the subgroup $Z_2$ drops out of the curves
shown in our figures.
Thus, both subgroups are ignored in the following.
Since only even group orders $p$ are considered in this paper,
the first possible interesting common subgroup is $Z_4$.
This  subgroup of homogeneous translations occurs if the order of 
the deck group $p$ is a multiple of four.
Let us put the focus on the case $p=20$.
The CMB anisotropy of the prism shaped models $L(20,9)$ at $\rho=\pi/4$
and ${\cal D}_{20}$ is shown in figure \ref{fig:comp_S60_group}(a)
together with those of $L(4,1)$ and $L(10,1)$
belonging to the subgroups $Z_4$ and $Z_{10}$.
The inspection of figure \ref{fig:comp_S60_group}(a) reveals
that for $\tau_{\hbox{\scriptsize sls}}\lesssim 0.6$ the
curves for $L(20,9)$, ${\cal D}_{20}$, and $L(10,1)$ are
nearly indistinguishable.
This shows that the behaviour for $L(20,9)$ and ${\cal D}_{20}$
is dominated by the common subgroup $Z_{10}$.
Omitting $Z_5\subset Z_{10}$, since it acts in the same direction as
the subgroup $Z_{10}$, the next common subgroup is $Z_4$.
The anisotropy of $L(4,1)$ differs only for
$\tau_{\hbox{\scriptsize sls}}\gtrsim 0.8$ from the projective space ${\cal P}^3$.
Since the subgroup $Z_4$ occurs in the deck groups of $L(20,9)$ and
${\cal D}_{20}$ with different multiplicities,
their behaviour splits for those values of $\tau_{\hbox{\scriptsize sls}}$.
As listed in table \ref{Tab:subgroups},
the subgroup $Z_4$ occurs in $L(20,9)$ only once
whereas the multiplicity of $Z_4$ in ${\cal D}_{20}$ is five.
Note that the subgroup $Z_{10}$ has in $L(20,9)$ as well as in ${\cal D}_{20}$
the multiplicity one and, therefore, leads to a common behaviour.
A diverging behaviour can thus take place when a common cyclic subgroup
occurs with different multiplicities with respect to the deck groups.

The homogeneous space $L(20,1)$ possesses as the largest subgroup
of Clifford translations the cyclic group $Z_{20}$.
This is in contrast to $L(20,9)$ where the subgroup $Z_{20}$ is realised
by inhomogeneous translations and largest subgroup of Clifford translations
is realised by $Z_{10}$, see table \ref{Tab:subgroups}.
The figures \ref{fig:comp_S60_geom_part_2}(c) and
\ref{fig:comp_S60_geom_part_2}(d)
display this difference for small values of $\tau_{\hbox{\scriptsize sls}}$.

A subgroup influences the CMB anisotropy when the smallest dimension
of its Voronoi domain is at most of the order of the surface of last
scattering.
Otherwise the surface of last scattering is contained completely inside
the Voronoi domain.
This requires $\tau_{\hbox{\scriptsize sls}} > \pi/p$
such that a subgroup can modify the CMB.
The subgroup with the largest group order determines the CMB anisotropy
for the smallest values of $\tau_{\hbox{\scriptsize sls}}$.
Since the largest subgroup occurs usually in both spaces with the
same multiplicity, i.\,e.\ multiplicity one,
the common anisotropy is explained in this way.
With increasing values of $\tau_{\hbox{\scriptsize sls}}$
the subgroups with decreasing group order influence the CMB.
Thus, as long as the common subgroups occur with the same multiplicity,
the $S_{\cal M}(60^{\circ})$ statistics of those spaces is very similar.
The different behaviour is enforced by the largest subgroup
that occurs in both spaces with a different multiplicity.
Another possibility is that a subgroup occurs in only one of the two spaces.


\begin{table}
\begin{center}
\begin{tabular}{|c|p{1cm}|p{1cm}|p{1cm}|p{1cm}|p{1cm}|}
\hline

manifold ${\cal M}$ & \multicolumn{5}{|c|}{\# of the cyclic subgroups} \\
\hline
                  & $Z_{20}$  & $Z_{10}$ & $Z_{8}$ & $Z_{6}$  & $Z_{4}$ \\
\hline
$L(20,1)$ &1&-&-&-&-\\
\hline
$L(20,9)$&- &1&-&-&1\\
\hline
${\cal D}_{20}$ &- &1&-&-&5\\
\hline
${\cal T}$ &-&-&-&4&3\\
\hline
${\cal O}$ &-&-&3&4&6\\
\hline
${\cal I}$ &-&6&-&10&15 \\
\hline
 \end{tabular}
\end{center}
 \caption{\label{Tab:subgroups}
The multiplicities of the cyclic subgroups $Z_{p}$, $p=4,6,8,10,20$,
of the deck groups $\Gamma$ are listed for
the lens space $L(20,1)$, the lens space $L(20,9)$, 
the prism space ${\cal D}_{20}$,
the binary tetrahedral space ${\cal T}$,
the binary octahedral space ${\cal O}$,
and the Poincar\'e dodecahedron ${\cal I}$.
From the subgroups of Clifford translations acting in the same direction,
only the largest independent subgroups are listed in the table.
The cyclic subgroups $Z_{p} \subset \Gamma$ of Clifford translations
are related to the deck groups of the homogeneous lens spaces $L(p,1)$.
}
\end{table}


In figure \ref{fig:comp_S60_group}(b)
the CMB anisotropy of the Poincar\'e dodecahedron ${\cal I}$
is decomposed in terms of its subgroups.
This space provides the best description of the large scale
CMB anisotropy with respect to spherical models.
Among the subgroups of Clifford translations 
of the Poincar\'e dodecahedron ${\cal I}$ are
the cyclic groups $Z_4$, $Z_6$, and $Z_{10}$,
see table \ref{Tab:subgroups}.
Consequently, figure \ref{fig:comp_S60_group}(b) displays the
CMB anisotropies of the homogeneous lens spaces $L(4,1)$, $L(6,1)$,
and $L(10,1)$.
The famous minimum in the CMB anisotropy of the Poincar\'e dodecahedron
${\cal I}$ in the range $\tau_{\hbox{\scriptsize sls}}=0.4\dots 0.5$ is due to
the subgroup $Z_{10}$ as the comparison with $L(10,1)$ reveals.
That the Poincar\'e dodecahedron possesses a much stronger anisotropy
suppression than the $L(10,1)$ space is enforced by the fact
that the $Z_{10}$ subgroup in the dodecahedral space has a multiplicity of 6.
The second minimum is caused by the subgroup $Z_6$ at higher values
of $\tau_{\hbox{\scriptsize sls}}$.
This subgroup also belongs to the binary octahedral group $O^\star$
and the binary tetrahedral group $T^\star$.
Since it is the largest subgroup of the binary tetrahedral group $T^\star$, 
the binary tetrahedral space ${\cal T}$ has its first minimum at the position
of the minimum of $L(6,1)$ as figure \ref{fig:comp_S60_geom}(d) confirms.
The first minimum in the case of the binary octahedral space ${\cal O}$
is dictated by the subgroups $Z_8$ and $Z_6$,
and it is a superposition of those of $L(8,1)$ and $L(6,1)$.
In the same way one can also explain why the prism space ${\cal D}_{12}$ 
displays lower power in the CMB anisotropy than the prism space ${\cal D}_8$
for $\tau_{\hbox{\scriptsize sls}}\gtrsim \pi/6$, 
see figure \ref{fig:comp_S60_geom_part_2}(b).
The deck group of the prism space ${\cal D}_8$ is composed of three cyclic 
subgroups $Z_4$ which determine the minimum of the $S(60^\circ)$ statistics.
In addition, the deck group $D^\star_{12}$ contains a further subgroup $Z_6$.
Thus, the influence of the additional subgroup $Z_6$ within the deck group 
of the prism space ${\cal D}_{12}$ is the explanation for the lower power
of the CMB anisotropy compared to the manifold ${\cal D}_8$.


\subsection{The transformation behaviour of $\Gamma$ on the sphere
$S_{\hbox{\scriptsize sls}}$ of last scattering}
\label{subsec:transfomation_on_sls}


The above discussion of the role of the subgroups of Clifford translations
explains the similarities of the $S_{\cal M}(60^{\circ})$ statistics
for small values of $\tau_{\hbox{\scriptsize sls}}$
as revealed by figures \ref{fig:comp_S60_geom_part_2}(a) and
\ref{fig:comp_S60_geom_part_2}(b)
which refer to the prism shaped $L(p,p/2-1)$ at $\rho=\pi/4$
and the prism space ${\cal D}_p$.
This and the distinct behaviour between the lens shaped $L(p,p/2-1)$
at $\rho=0$ and the lens space $L(p,1)$
shown in figures \ref{fig:comp_S60_geom_part_2}(c) and
\ref{fig:comp_S60_geom_part_2}(d) can be illuminated
from the following point of view.
Consider the action of the deck group $\Gamma$
on the sphere $S_{\hbox{\scriptsize sls}}$ of last scattering having a
radius $\tau_{\hbox{\scriptsize sls}}$.
Every point $x\in S_{\hbox{\scriptsize sls}}$ on this sphere is shifted
by a certain distance $d_n^{{\cal M}}(\vec x)$ under the action of a
group element $g_n\in\Gamma$.
Similarities or differences of this transformation behaviour determine
whether the $S_{\cal M}(60^{\circ})$ differs or not for not too large
values of $\tau_{\hbox{\scriptsize sls}}$.
Since there is no direct link between the shape of the Voronoi domain
and the transformation structure on $S_{\hbox{\scriptsize sls}}$,
the following argument is not directly connected to the
well-proportioned hypothesis.

At first the transformation distances are needed.
Applying the group elements (\ref{Def:trafo_group_Lpq})
of the lens space $L(p,q)$ for $\alpha=\epsilon=0$
to a point $\vec x$, one gets the points $\vec x_n$, $n=1,\dots,p$,
which are identified with the point $\vec x$ on the covering space ${\cal S}^3$.
The distances $d_n^{L(p,q),\rho}(\vec x)$ between the point $\vec x$ and the
points $\vec x_n$ are given by
\begin{equation}
\label{Eq:distance_d_Lpq}
d_n^{L(p,q),\rho}(\vec x) \; = \; \hbox{arccos}\left(\vec x \cdot \vec x_n\right)
\hspace{10pt} , \hspace{10pt}
n=1,\dots,p
\end{equation}
\begin{eqnarray}
\nonumber
\hbox{with}\hspace{15pt} \vec x \cdot \vec x_n & = &
\frac{1}{2}(1-\cos(2\rho))\cos(q\varphi_n)\\
\nonumber
& &+\frac{1}{2}(1+\cos(2\rho))\cos(\varphi_n))\\
\nonumber
& &+\cos(2\rho)(\cos(q\varphi_n)-\cos(\varphi_n))(x_1^2+x_2^2)\\
\nonumber
& &-\sin(2\rho)(\cos(q\varphi_n)-\cos(\varphi_n))(x_0x_1+x_2x_3)
\hspace{10pt},
\end{eqnarray}
where the special case $\rho=0$ is also discussed
in \cite{Mota_Gomera_Reboucas_Tavakol_2004}.
Here the abbreviation
\begin{equation}
\label{Eq:varphi_n}
\varphi_n \; := \; 2\pi n/p
\end{equation}
is introduced.
Now we restrict the distances $d_n^{{\cal M}}(\vec x)$ to points $\vec x$ 
lying on the sphere $S_{\hbox{\scriptsize sls}}$ of last scattering.
The minimum and the maximum of the distances $d_n^{{\cal M}}(\vec x)$
taken over all points $\vec x \in S_{\hbox{\scriptsize sls}}$
are for the lens spaces $L(p,q)$  with the observer at $\rho=0$ given by
\begin{equation}
\label{Eq:distance_d_L_p_q_rho_000pi_min}
d_{n,\hbox{\scriptsize min}} ^{L(p,q),\rho=0} =
\hbox{min}\left[\varphi_n , \varphi_{\left|p-n\right|}\right]
\end{equation}
and
\begin{eqnarray}
\label{Eq:distance_d_L_p_q_rho_000pi_max}
d_{n,\hbox{\scriptsize max}}^{L(p,q),\rho=0}= \hbox{arccos}\left(w_n\right)\\
\nonumber
\hbox{with}\hspace{20pt}w_n=\cos(\varphi_n)+\sin^2(\tau_{\hbox{\scriptsize sls}})(\cos(q\varphi_n)-\cos(\varphi_n))
\hspace{10pt}.
\end{eqnarray}
Eq.\,(\ref{Eq:distance_d_L_p_q_rho_000pi_max})
simplifies for the lens spaces $L(p,p/2-1)$ to
\begin{eqnarray}
\label{Eq:distance_d_L_p_p/2-1_rho_000pi_max}
d_{n,\hbox{\scriptsize max}}^{L(p,p/2-1),\rho=0} =  \left\{\begin{array}{ll}
\hbox{arccos}\left(w_n\right)& : n \;\;\hbox{odd} \\
d_{n,\hbox{\scriptsize min}}^{L(p,p/2-1),\rho=0}& : n  \;\;\hbox{even}
\end{array}\right.\\
\nonumber
\hbox{with}\hspace{20pt}w_n=\cos(\varphi_n)-2\sin^2(\tau_{\hbox{\scriptsize sls}})\sin((p/4-1)\varphi_n)
\hspace{10pt}.
\end{eqnarray}
For the other observer position $\rho=\pi/4$ one gets for $L(p,p/2-1)$
\begin{equation}
\label{Eq:distance_d_L_p_p/2-1_rho_025pi_min}
d_{n,\hbox{\scriptsize min}}^{L(p,p/2-1),\rho=\pi/4} =
\left\{\begin{array}{ll}
\hbox{min}_{\{ i=1,2\}}\left[\hbox{arccos}\left(w^{i}_n\right)\right] &
:  n \;\;\hbox{odd},\;\;\tau_{\hbox{\scriptsize sls}}\le \pi/4 \\
\hbox{min}\left[\varphi_n ,\varphi_{\left|p/2-n\right|},
\varphi_{\left|p-n\right|}\right] &
:  n \;\;\hbox{odd},\;\;\tau_{\hbox{\scriptsize sls}}> \pi/4  \\
\hbox{min}\left[\varphi_n ,\varphi_{\left|p-n\right|}\right] &
: n \;\;\hbox{even}
\end{array}\right.
\hspace{10pt}
\end{equation}
and
\begin{equation}
\label{Eq:distance_d_L_p_p/2-1_rho_025pi_max}
d_{n,\hbox{\scriptsize max}}^{L(p,p/2-1),\rho=\pi/4} =
\left\{\begin{array}{ll}
\hbox{max}_{\{ i=1,2\}}\left[\hbox{arccos}\left(w^{i}_n\right)\right]
& : n \;\;\hbox{odd},\;\;\tau_{\hbox{\scriptsize sls}}\le \pi/4\\
\hbox{max}\left[\varphi_n ,\varphi_{\left|p/2-n\right|},
\varphi_{\left|p-n\right|}\right]& :  n \;\;\hbox{odd},\;\;\tau_{\hbox{\scriptsize sls}}> \pi/4  \\
d_{n,\hbox{\scriptsize min}}^{L(p,p/2-1),\rho=\pi/4}& : n  \;\;\hbox{even}
\end{array}\right.
\end{equation}
with
\begin{eqnarray}
\label{Eq:distance_d_L_p_p/2-1_rho_025pi_phase_w}
w^{1,2}_n & = &
\cos(n\pi/2)\cos((p/4-1)\varphi_n) \\
\nonumber
 &   &\pm 2\sin(n\pi/2)\sin((p/4-1)\varphi_n)\sin(\tau_{\hbox{\scriptsize sls}})
\cos(\tau_{\hbox{\scriptsize sls}})
\hspace{10pt}.
\end{eqnarray}

\begin{figure}
\vspace*{-15pt}
\begin{minipage}{18.0cm}
\hspace*{-15pt}
\begin{minipage}{9.0cm}
{
\includegraphics[width=9.0cm]{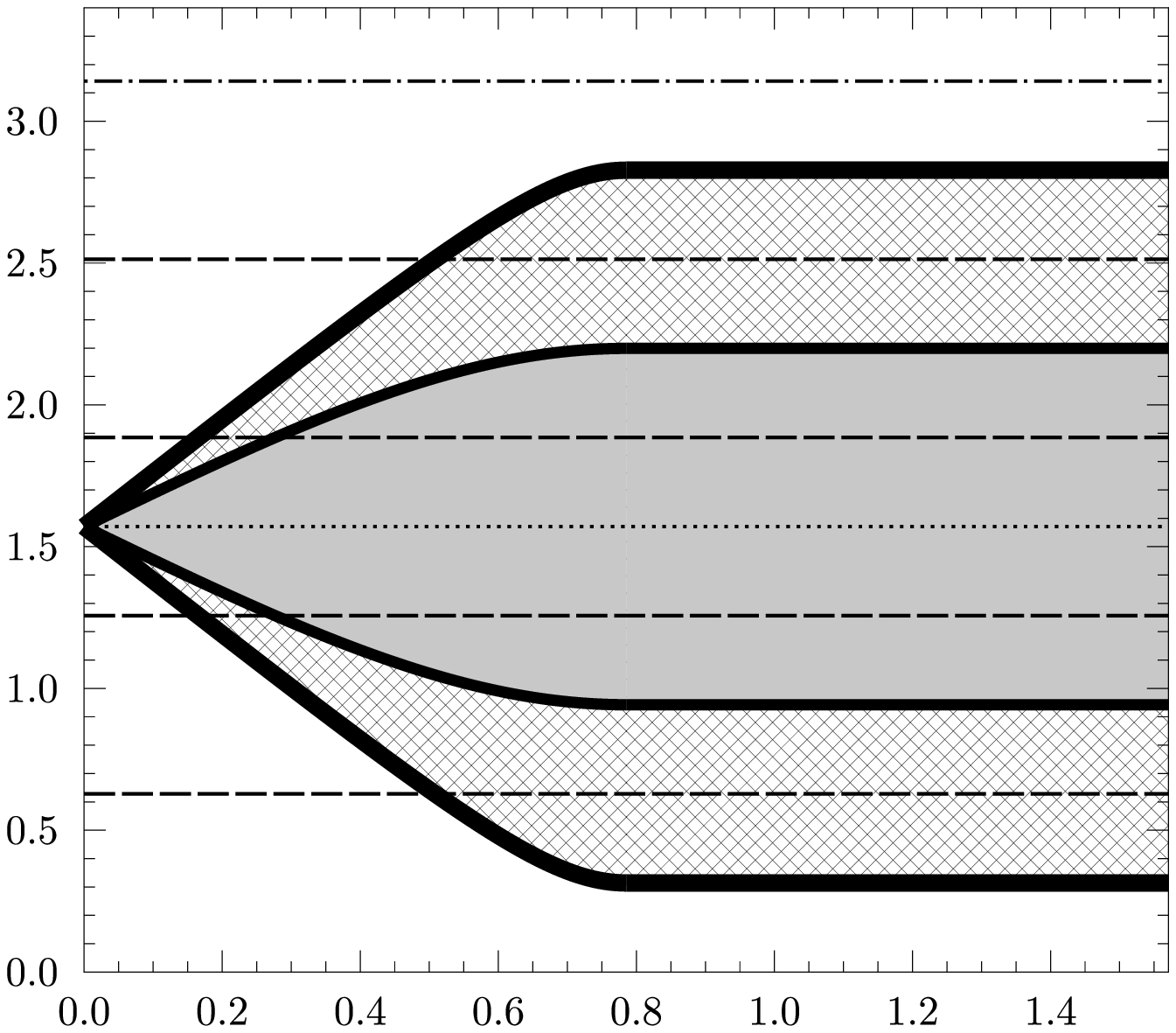}
\put(-251,185){$d_n^{{\cal M}}(\vec x)$}
\put(-45,18){$\tau_{\hbox{\scriptsize sls}}$}
\put(-165,170){(a) $L(20,9)$ at $\rho=\pi/4$}
\put(-213,183){\small{$n=10$}}
\put(-213,154){\small{$n=8,12$}}
\put(-163,125){\small{$n=6,14$}}
\put(-163,110){\small{$n=5,15$}}
\put(-163,96){\small{$n=4,16$}}
\put(-163,85){\small{$n=3,7,13,17$}}
\put(-213,66){\small{$n=2,18$}}
\put(-163,40){\small{$n=1,9,11,19$}}
\vspace*{-45pt}
}
{
\includegraphics[width=9.0cm]{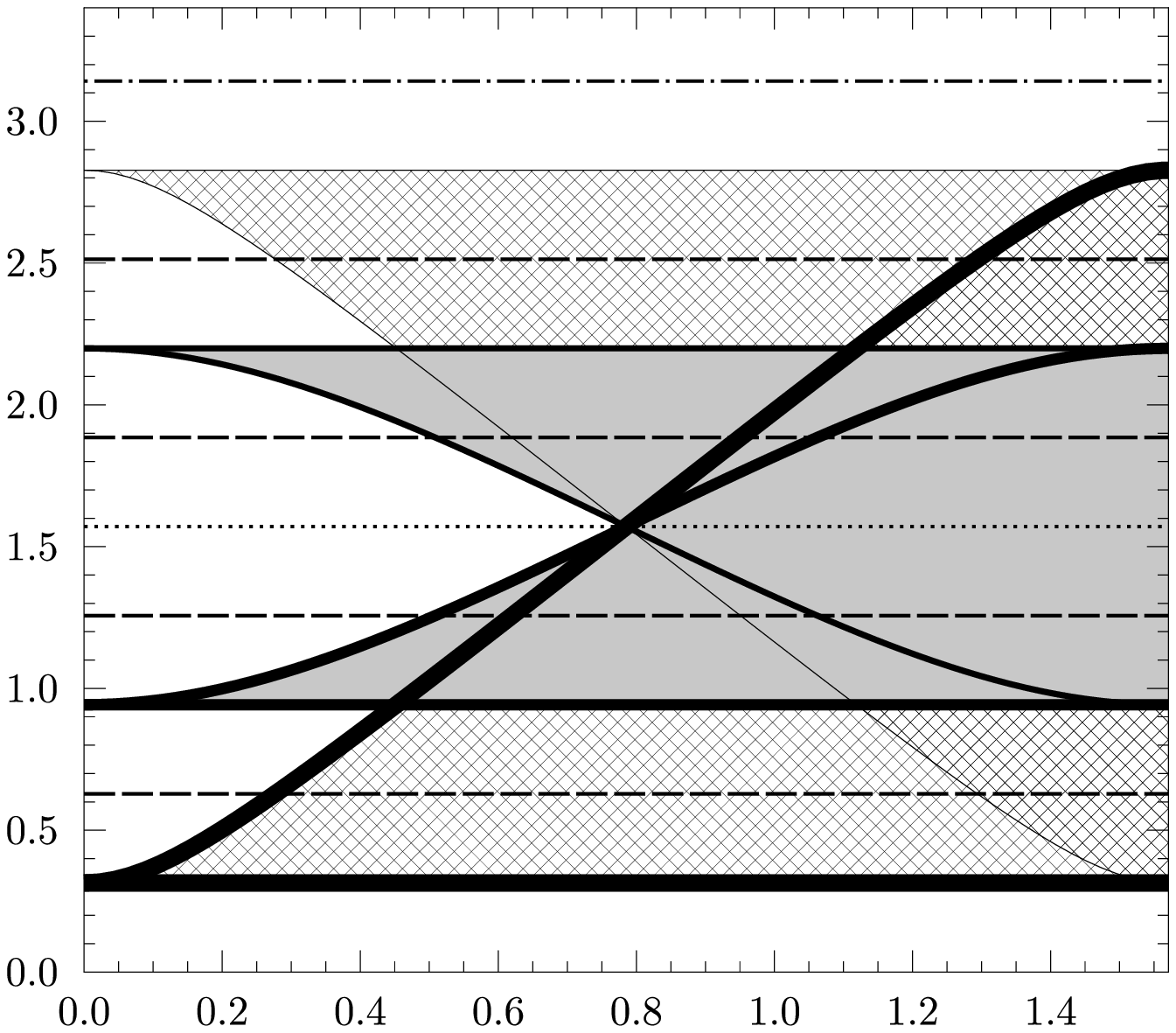}
\put(-251,185){$d_n^{{\cal M}}(\vec x)$}
\put(-45,18){$\tau_{\hbox{\scriptsize sls}}$}
\put(-165,170){(c) $L(20,9)$ at $\rho=0$}
\put(-213,183){\small{$n=10$}}
\put(-213,169){\small{$n=9,11$}}
\put(-213,154){\small{$n=8,12$}}
\put(-213,140){\small{$n=7,13$}}
\put(-213,125){\small{$n=6,14$}}
\put(-213,110){\small{$n=5,15$}}
\put(-213,96){\small{$n=4,16$}}
\put(-213,82){\small{$n=3,17$}}
\put(-213,66){\small{$n=2,18$}}
\put(-213,52){\small{$n=1,19$}}
}
\end{minipage}
\hspace*{-35pt}
\begin{minipage}{9.0cm}
{
\includegraphics[width=9.0cm]{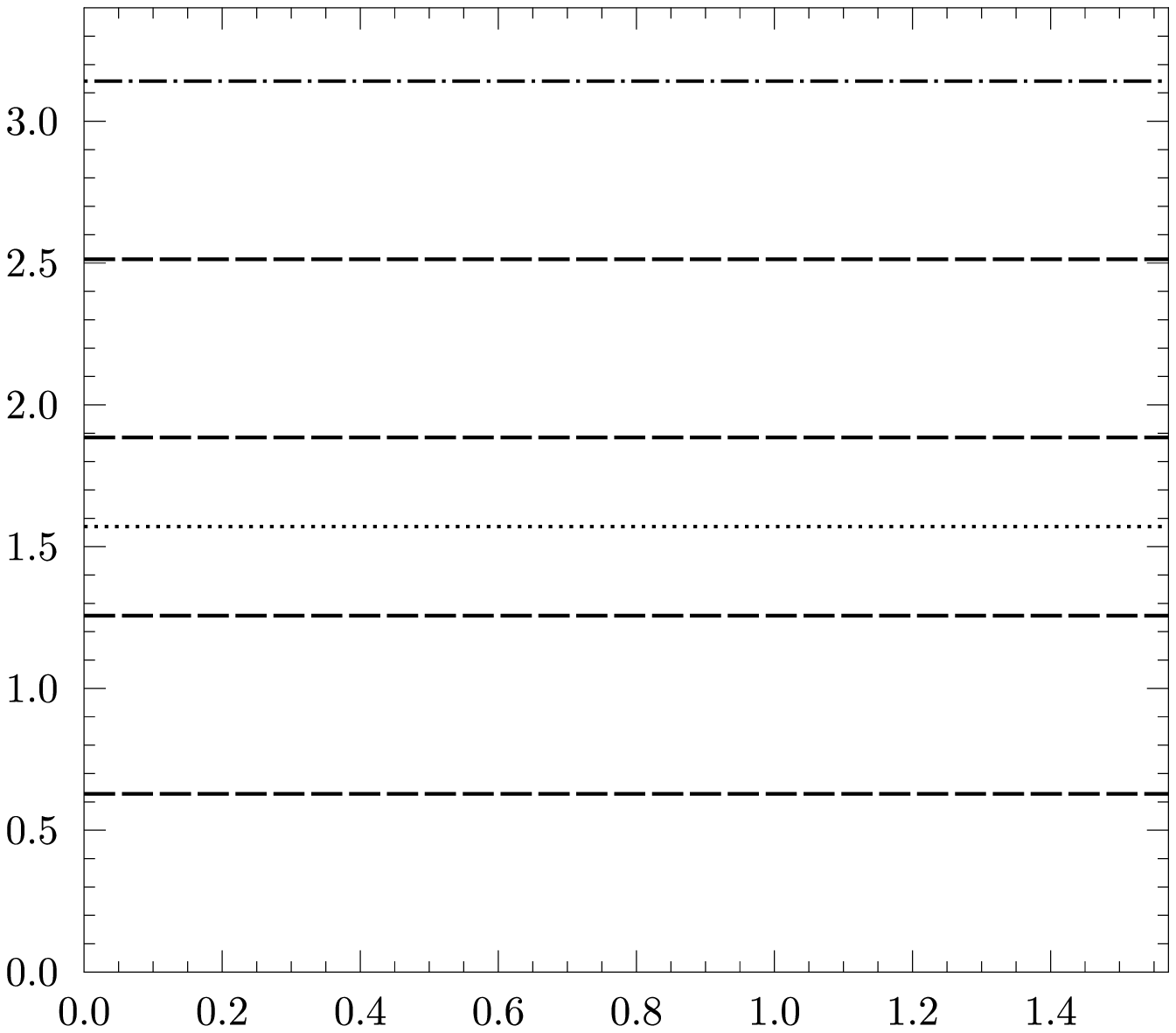}
\put(-251,185){$d_n^{{\cal M}}(\vec x)$}
\put(-45,18){$\tau_{\hbox{\scriptsize sls}}$}
\put(-165,170){(b) ${\cal D}_{20}$}
\put(-213,183){\small{$n=10$}}
\put(-213,154){\small{$n=8,12$}}
\put(-213,125){\small{$n=6,14$}}
\put(-213,110){\small{$n=1,3,5,7,9,11,13,15,17,19$}}
\put(-213,96){\small{$n=4,16$}}
\put(-213,66){\small{$n=2,18$}}
\vspace*{-45pt}
}
{
\includegraphics[width=9.0cm]{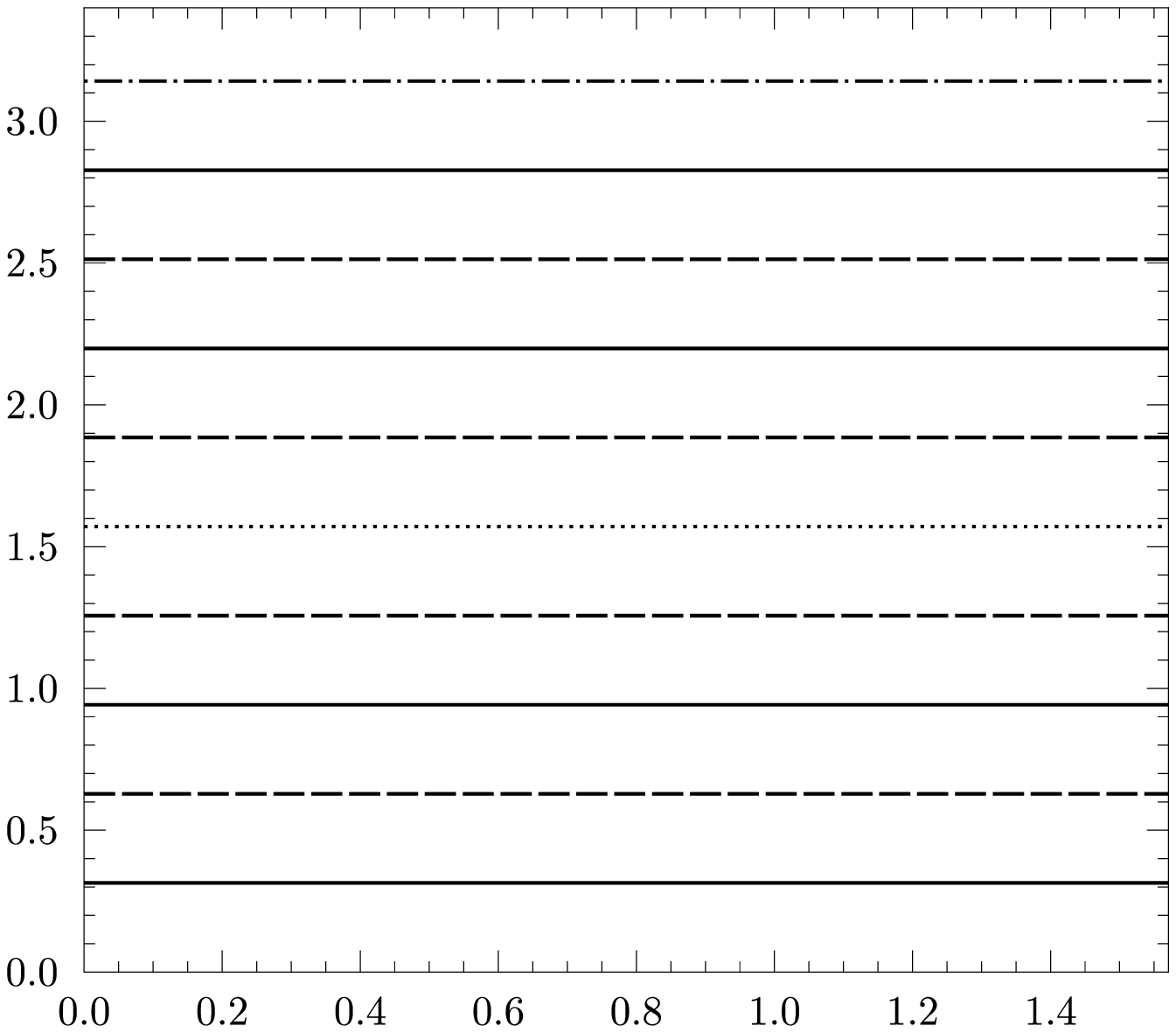}
\put(-251,185){$d_n^{{\cal M}}(\vec x)$}
\put(-45,18){$\tau_{\hbox{\scriptsize sls}}$}
\put(-165,170){(d) $L(20,1)$}
\put(-213,183){\small{$n=10$}}
\put(-213,169){\small{$n=9,11$}}
\put(-213,154){\small{$n=8,12$}}
\put(-213,140){\small{$n=7,13$}}
\put(-213,125){\small{$n=6,14$}}
\put(-213,110){\small{$n=5,15$}}
\put(-213,96){\small{$n=4,16$}}
\put(-213,82){\small{$n=3,17$}}
\put(-213,66){\small{$n=2,18$}}
\put(-213,52){\small{$n=1,19$}}
}
\end{minipage}
\end{minipage}
\vspace*{-15pt}\caption{\label{fig:comp_dist_p_20}
The transformation distances $d_n^{{\cal M}}(\vec x)$ for  points
$\vec x \in S_{\hbox{\scriptsize sls}}$ are plotted for the manifolds
$L(20,1)$, ${\cal D}_{20}$, and $L(20,9)$ with the observer position
at $\rho=0$ and at $\rho=\pi/4$.
Since $L(20,9)$ for $\rho=\pi/4$ and ${\cal D}_{20}$ have the
same prism geometry of the Voronoi domain,
one has to compare the panels (a) and (b).
The panel pair (c) and (d) is devoted to the lens shaped case.
The inhomogeneous manifold $L(20,9)$ possesses a wide distribution
of values of $d_n^{{\cal M}}(\vec x)$ for $n=1$, 3, 7, 9, 11, 13, 17, and 19 
which are depicted by the grey and shaded bands.
}
\end{figure}

The elements of the deck group of a homogeneous lens space $L(p,1)$
transform all points on the 3-sphere by the same distance
since they are all Clifford translations.
So one gets independent of $\vec x$ the distance
\begin{equation}
\label{Eq:distance_L_p_1}
d_n^{L(p,1)}(\vec x)=d_{n,\hbox{\scriptsize min}}^{L(p,1)} 
= d_{n,\hbox{\scriptsize max}}^{L(p,1)}
= \hbox{min}\left[\varphi_n , \varphi_{\left|p-n\right|}\right]
\hspace{10pt}.
\end{equation}
This general property of homogeneous manifolds occurs in
the inhomogeneous case only for special cases.
The eqs.\,(\ref{Eq:distance_d_L_p_p/2-1_rho_000pi_max}) and
(\ref{Eq:distance_d_L_p_p/2-1_rho_025pi_max}) show that
the property $d_{n,\hbox{\scriptsize max}}^{L(p,p/2-1),\rho}=
d_{n,\hbox{\scriptsize min}}^{L(p,p/2-1),\rho}=d_n^{L(p,p/2-1),\rho}(\vec x)$
is only obtained for even values of $n$ which correspond to the
cyclic subgroup $Z_{p/2}$ of the homogeneous lens space $L(p/2,1)$.
Eqs.\,(\ref{Eq:distance_d_L_p_p/2-1_rho_025pi_min}) and
(\ref{Eq:distance_d_L_p_p/2-1_rho_025pi_max}) lead
for $n=p/4$ and $n=3p/4$ to the simple result
$d_{n,\hbox{\scriptsize max}}^{L(p,p/2-1),\rho}=
d_{n,\hbox{\scriptsize min}}^{L(p,p/2-1),\rho}=\pi/2$.
This distance belongs to two group elements of the deck group $Z_{4}$
which generates the homogeneous lens space $L(4,1)$.

The deck group $D^\star_p$ of the homogeneous prism space ${\cal D}_p$ contains
the cyclic subgroup $Z_{p/2}$ and $p/4$ times the cyclic subgroup $Z_4$.
For this reason the distances $d_n^{{\cal D}_p}(\vec x)$ are given by
\begin{equation}
\label{Eq:distance_D_p_min_max_Z_p/2_Z_4}
d_n^{{\cal D}_p}(\vec x)=d_{n,\hbox{\scriptsize min}}^{{\cal D}_p} 
= d_{n,\hbox{\scriptsize max}}^{{\cal D}_p}= \left\{\begin{array}{ll}
\hbox{min}\left[\varphi_n , \varphi_{\left|p-n\right|}\right]&
: n  \;\;\hbox{even}\\
\pi/2 & : n  \;\;\hbox{odd}
\end{array}\right.
\hspace{6pt} .
\end{equation}
The distance $d_n^{{\cal D}_p}(\vec x)=\pi/2$ results from the 
$p/2$ group elements $g_n$ of the $p/4$ cyclic subgroups $Z_4$.

The distances $d_n^{{\cal M}}(\vec x)$ according to the formulae
(\ref{Eq:distance_d_L_p_q_rho_000pi_min}) to
(\ref{Eq:distance_D_p_min_max_Z_p/2_Z_4})
are shown in figure \ref{fig:comp_dist_p_20}
for the same topologies as in figure \ref{fig:comp_S60_geom_part_2}
for the special case $p=20$.
The homogeneous spaces ${\cal D}_{20}$ and $L(20,1)$ have distances 
$d_n^{{\cal M}}(\vec x)$ independent of $\tau_{\hbox{\scriptsize sls}}$
which are shown as horizontal lines in figures \ref{fig:comp_dist_p_20}(b)
and \ref{fig:comp_dist_p_20}(d).
The index $n$ of the distance $d_n^{{\cal M}}(\vec x)$ is also stated
in figure \ref{fig:comp_dist_p_20}.
The case for the inhomogeneous space $L(20,9)$ is more involved,
since only distances $d_n^{{\cal M}}(\vec x)$ with an even value of $n$ are
$\tau_{\hbox{\scriptsize sls}}$ independent.
For odd $n$ the distance $d_n^{{\cal M}}(\vec x)$ can extend over the interval
$[d_{n,\hbox{\scriptsize min}},d_{n,\hbox{\scriptsize max}}]$
which is shown as the bands in figures \ref{fig:comp_dist_p_20}(a) and
\ref{fig:comp_dist_p_20}(c).
A comparison of figure \ref{fig:comp_dist_p_20}(a) with
\ref{fig:comp_dist_p_20}(b) shows
that for $\tau_{\hbox{\scriptsize sls}}\lesssim 0.5$
the smallest distances $d_n^{{\cal M}}(\vec x)$ belonging to 
$n=2$ and $n=18$ are identical for
$L(20,9)$ at $\rho=\pi/4$ and ${\cal D}_{20}$.
Since the distance $d_n^{{\cal M}}(\vec x)$ determines the maximal wavelength
of a perturbation along the transformation direction of $g_n$,
the smallest distance is most responsible for
the suppression of the large angular CMB anisotropy.
Since it is equal for both spaces for $\tau_{\hbox{\scriptsize sls}}\lesssim 0.5$,
the CMB anisotropy behaviour is almost the same.
On the other hand,
the inhomogeneous transformations $n=1, 9, 11, 19$ belong for
$\tau_{\hbox{\scriptsize sls}}\gtrsim 0.5$ to the smallest
transformation distance, and the diverging CMB properties are also explained.
The situation is different for the spaces $L(20,9)$ at $\rho=0$ and $L(20,1)$
shown in  figures \ref{fig:comp_dist_p_20}(c) and
\ref{fig:comp_dist_p_20}(d).
Here, the smallest distance $d_n^{{\cal M}}(\vec x)$ belongs to $n=1$ and $n=19$
and is independent of $\tau_{\hbox{\scriptsize sls}}$ for $L(20,1)$,
of course.
However, for the inhomogeneous space $L(20,9)$ at $\rho=0$,
the $d_n^{{\cal M}}(\vec x)$ are distributed towards larger values, and there 
are thus directions with a less pronounced CMB anisotropy suppression.
This distinction already happens at very small values of
$\tau_{\hbox{\scriptsize sls}}$ as reflected in
figures \ref{fig:comp_S60_geom_part_2}(c) and
\ref{fig:comp_S60_geom_part_2}(d).
The transformation properties of the deck group $\Gamma$
on the sphere $S_{\hbox{\scriptsize sls}}$ of last scattering
determine the CMB anisotropy in this way.
Since they are not related to the shape of the Voronoi domain,
their consequences are independent of the well-proportioned conjecture.

\begin{figure}
\vspace*{-15pt}
\begin{minipage}{18.0cm}
\hspace*{-15pt}
\begin{minipage}{9.0cm}
{
\includegraphics[width=9.0cm]{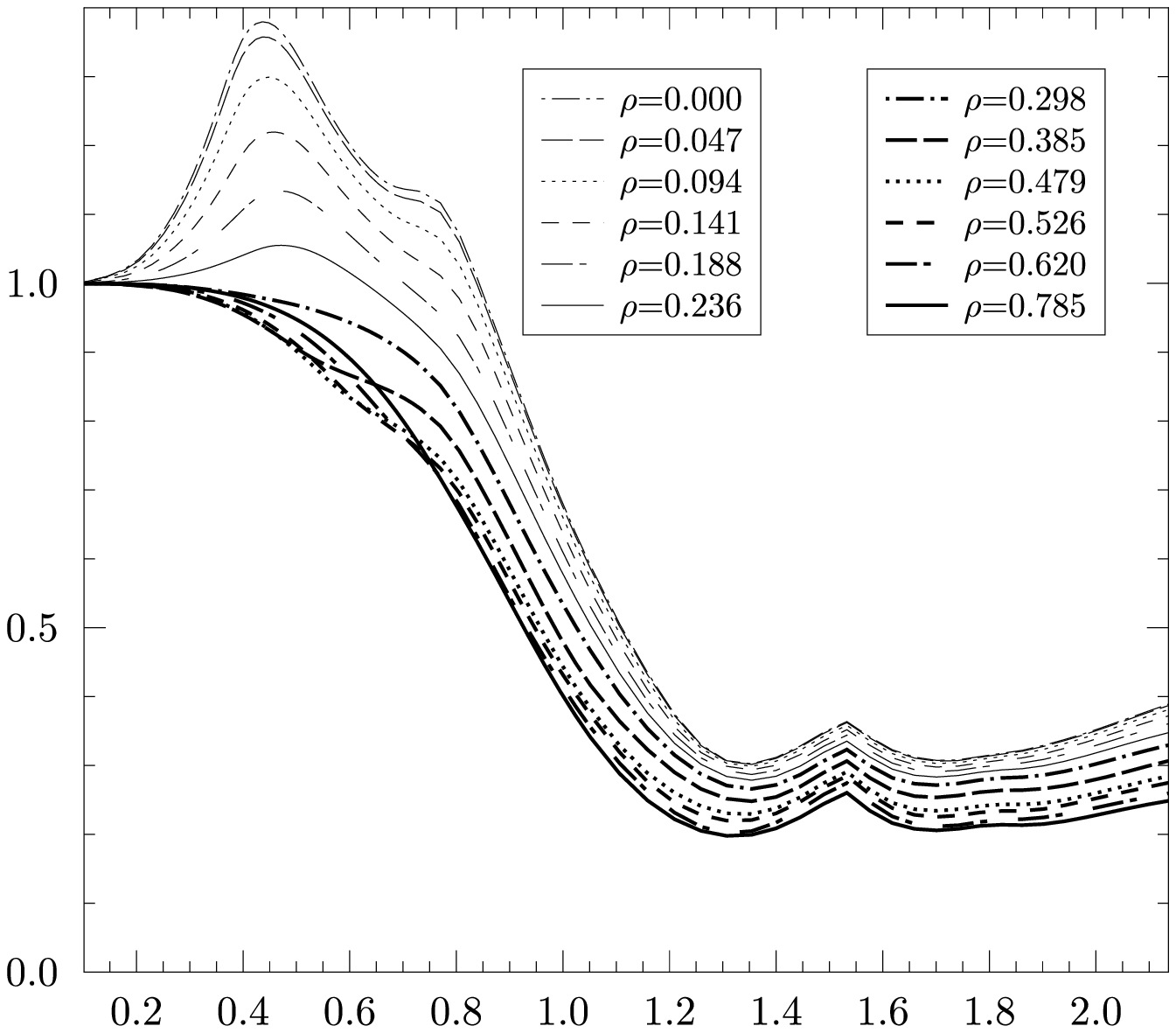}
\put(-255,162){$\frac{S_{\cal M}(60^{\circ})}{S_{{\cal P}^3}(60^{\circ})}$}
\put(-55,20){$\tau_{\hbox{\scriptsize sls}}$}
\put(-205,47){(a) $L(8,3)$}
\vspace*{-45pt}
}
{
\includegraphics[width=9.0cm]{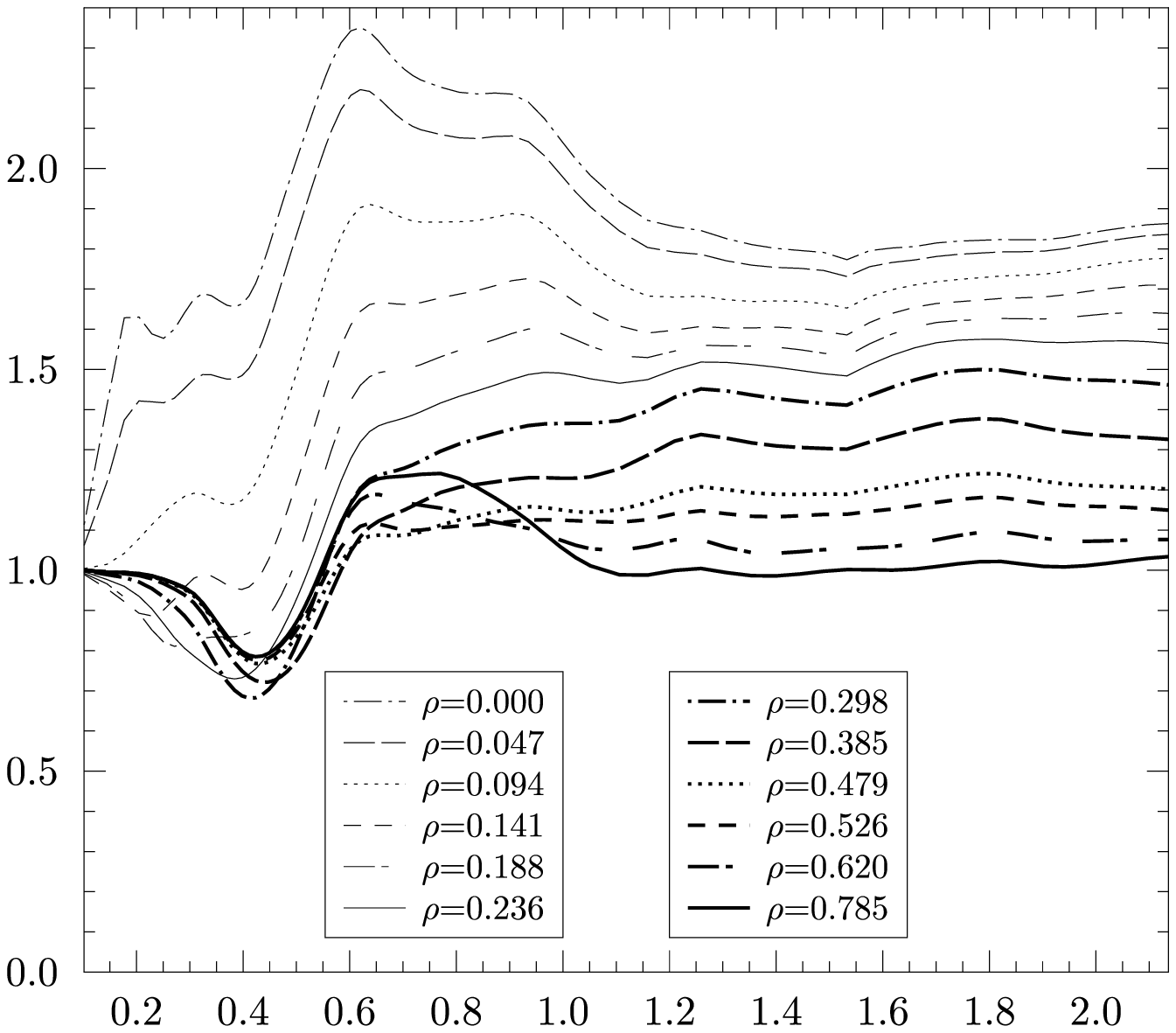}
\put(-255,176){$\frac{S_{\cal M}(60^{\circ})}{S_{{\cal P}^3}(60^{\circ})}$}
\put(-55,20){$\tau_{\hbox{\scriptsize sls}}$}
\put(-135,180){(c) $L(20,9)$}
}
\end{minipage}
\hspace*{-35pt}
\begin{minipage}{9.0cm}
{
\includegraphics[width=9.0cm]{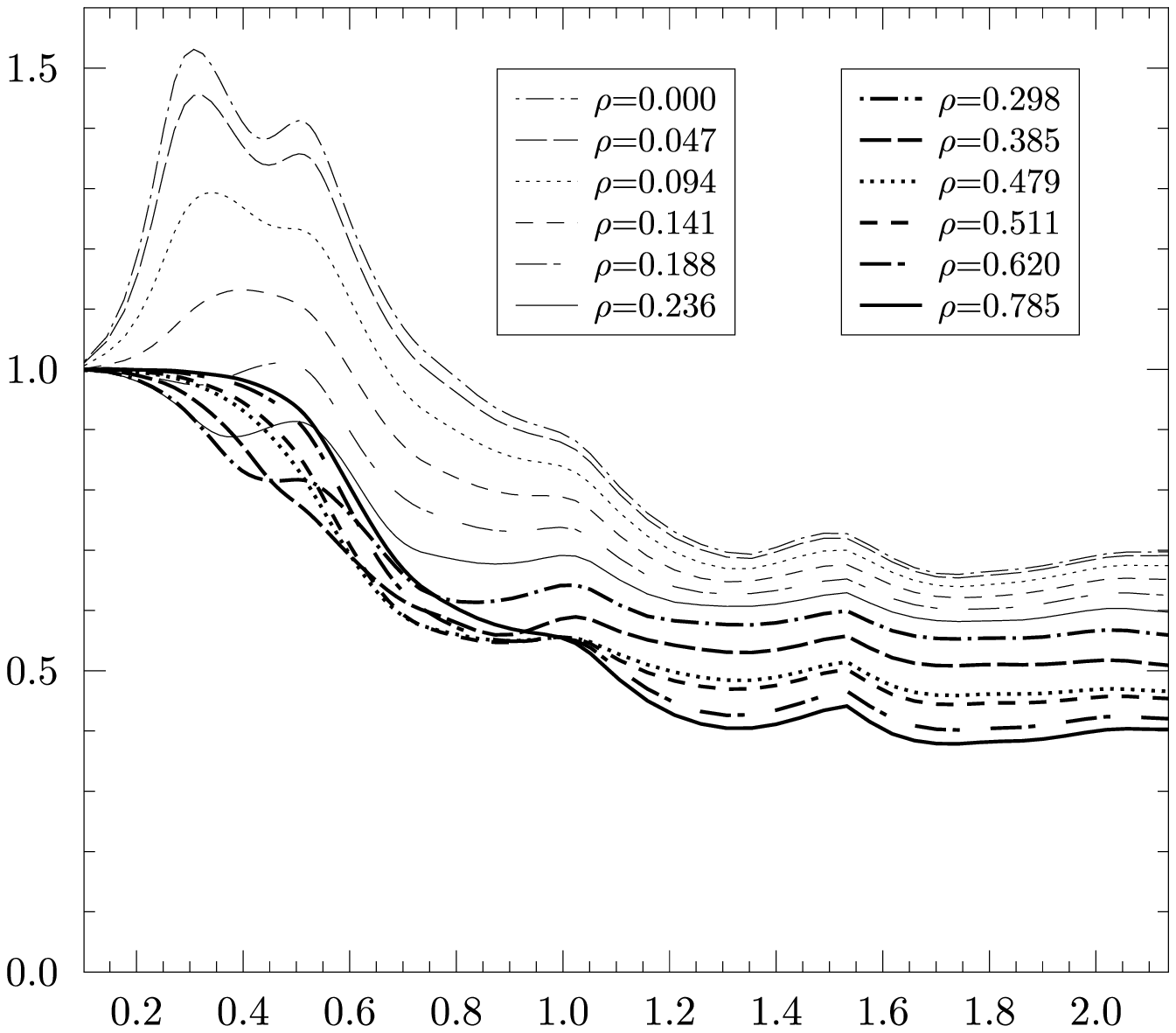}
\put(-255,162){$\frac{S_{\cal M}(60^{\circ})}{S_{{\cal P}^3}(60^{\circ})}$}
\put(-55,20){$\tau_{\hbox{\scriptsize sls}}$}
\put(-205,47){(b) $L(12,5)$}
\vspace*{-45pt}
}
{
\includegraphics[width=9.0cm]{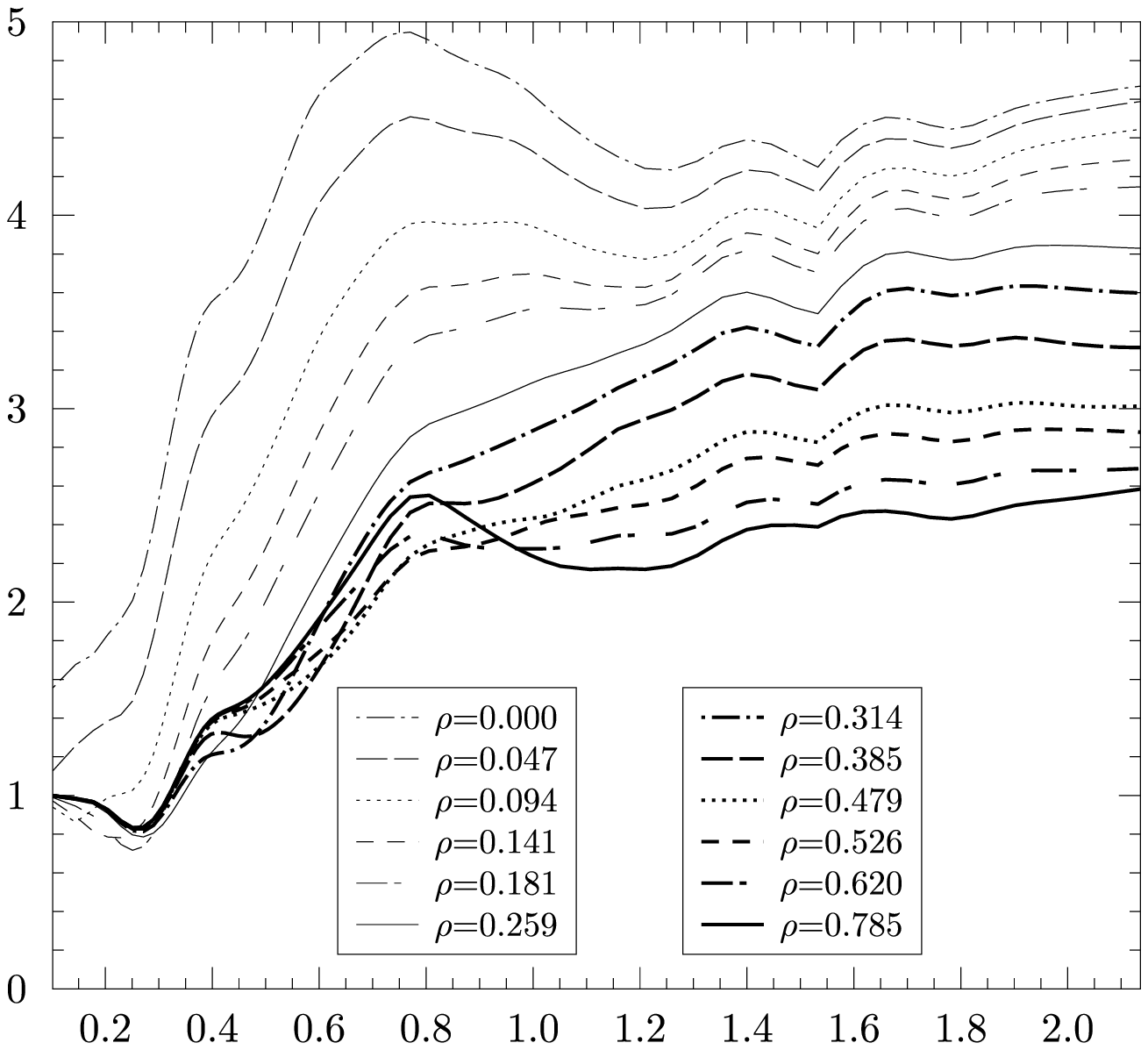}
\put(-255,176){$\frac{S_{\cal M}(60^{\circ})}{S_{{\cal P}^3}(60^{\circ})}$}
\put(-55,20){$\tau_{\hbox{\scriptsize sls}}$}
\put(-135,180){(d) $L(32,15)$}
}
\end{minipage}
\end{minipage}
\caption{\label{fig:s60_var_rho}
The variation of the $S(60^{\circ})$ statistics with respect to the 
observer position parametrised by $\rho$ is shown for the lens spaces 
$L(8,3)$, $L(12,5)$, $L(20,9)$, and $L(32,15)$.
In all four diagrams
the $S(60^{\circ})$ statistics with the largest values belongs to
the lens shaped geometry of the Voronoi domain ($\rho=0$).
}
\end{figure}


\subsection{The CMB anisotropy for general observer positions $\rho$}
\label{subsec:CMB_for_general_rho}


Up to now, only two special positions of the observer
in the inhomogeneous spaces $L(p,p/2-1)$ have been discussed
since the two positions at $\rho=0$ and $\rho=\pi/4$
lead to Voronoi domains identical to two other homogeneous spaces.
To close this gap, the variation of the CMB anisotropy with respect to
the observer position $\rho$ is shown in figure \ref{fig:s60_var_rho}
for the four spaces $L(8,3)$, $L(12,5)$, $L(20,9)$, and $L(32,15)$.
The $S(60^{\circ})$ statistics is shown for 12 values of $\rho$
for these spaces.
The band width generated by these twelve curves shows how strong
the CMB properties vary as a function of the observer position.
It turns out that the largest CMB anisotropy belongs to the
lens shaped Voronoi domain $(\rho=0)$ in all four cases.
For sufficiently large spheres of last scattering
$\tau_{\hbox{\scriptsize sls}} = 0.8\dots 1.0$,
the prism shaped Voronoi domains $(\rho=\pi/4)$ possess the smallest
CMB anisotropy.
For CMB spheres with smaller $\tau_{\hbox{\scriptsize sls}}$
other values of $\rho$ can lead to an even smaller CMB anisotropy
and thus to a better description of the CMB observations.
It is worthwhile to note that for other spaces $L(p,q)$
with $q\neq p/2-1$ and $q\neq 1$ the situation is more involved
such that this simple behaviour does not emerge in the general case.
This will be the topic of a future work.

\begin{figure}
\vspace*{-15pt}
\begin{minipage}{18.0cm}
\hspace*{-15pt}
\begin{minipage}{9.0cm}
{
\includegraphics[width=9.0cm]{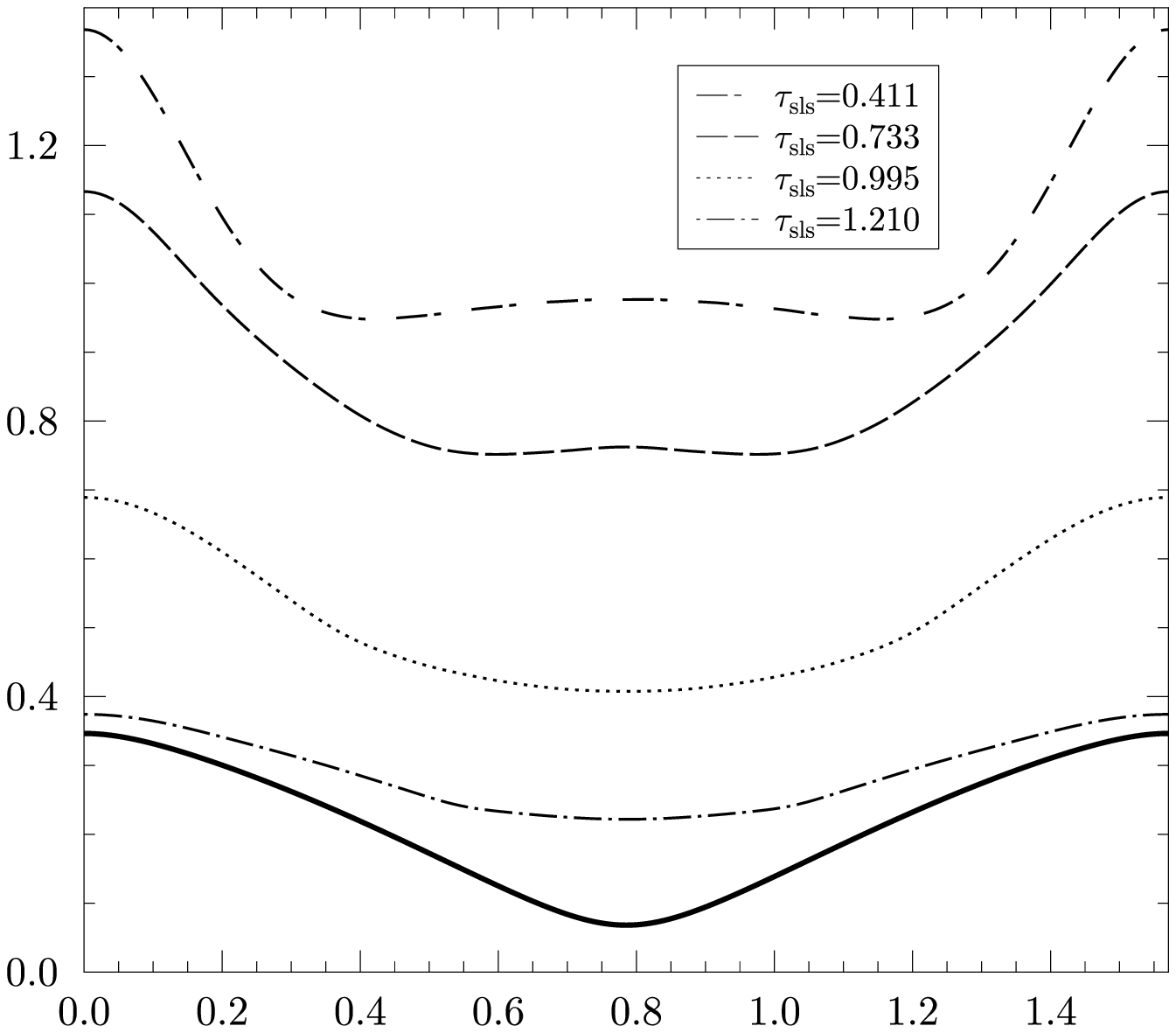}
\put(-255,182){$\frac{S_{\cal M}(60^{\circ})}{S_{{\cal P}^3}(60^{\circ})}$}
\put(-45,23){$\rho$}
\put(-185,162){(a) $L(8,3)$}
\vspace*{-25pt}
}
\end{minipage}
\hspace*{-35pt}
\begin{minipage}{9.0cm}
{
\includegraphics[width=9.0cm]{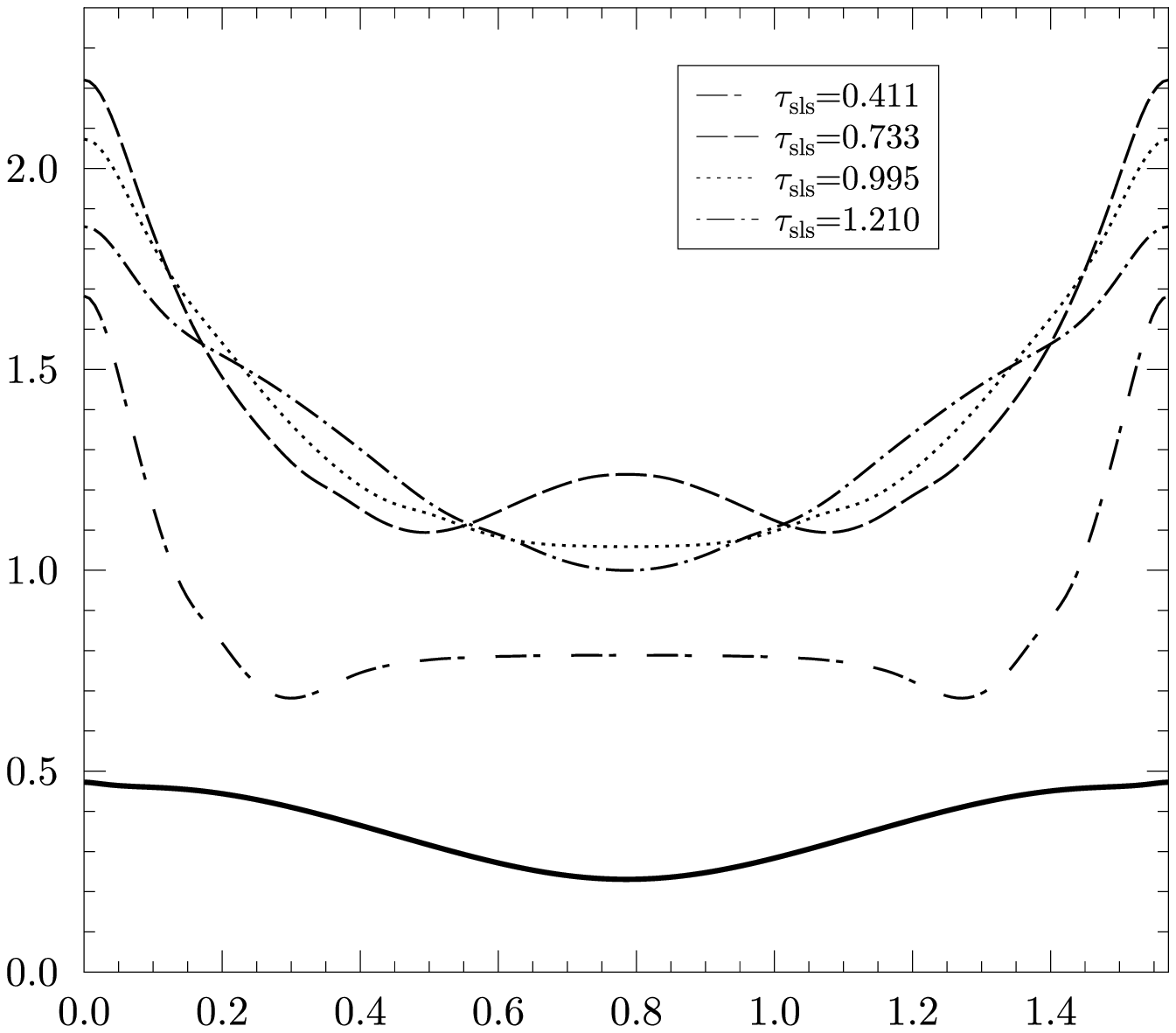}
\put(-255,182){$\frac{S_{\cal M}(60^{\circ})}{S_{{\cal P}^3}(60^{\circ})}$}
\put(-45,23){$\rho$}
\put(-185,162){(b) $L(20,9)$}
\vspace*{-25pt}
}
\end{minipage}
\end{minipage}
\caption{\label{fig:s60_ueber_rho}
The variation of the $S(60^{\circ})$ statistics with respect to the 
observer position parametrised by $\rho$ is shown 
for the lens spaces $L(8,3)$ and $L(20,9)$ for four values
of $\tau_{\hbox{\scriptsize sls}}$.
In addition, the shape measure $\sigma_{\tau}/\langle \tau \rangle$
for these two manifolds is given as a solid curve.
}
\end{figure}

Now, we return to the possible link between the geometry of the
Voronoi domain and the large scale CMB anisotropy suppression.
In sect.\ \ref{sec:shape_measure} the relative variation
$\sigma_{\tau}/\langle \tau \rangle$ is introduced as a measure
of the sphericity of the Voronoi domain.
The smaller its value is, the more well-proportioned is the Voronoi domain,
which should in turn lead to a stronger anisotropy suppression.
In figure \ref{fig:s60_var_rho} the $S(60^{\circ})$ statistics
is already shown for twelve observer positions as a function
of the cosmological parameters encoded by $\tau_{\hbox{\scriptsize sls}}$.
But in figure \ref{fig:s60_ueber_rho} the $S(60^{\circ})$ statistics
of $L(8,3)$ and $L(20,9)$ is shown as a function of $\rho$
such that it can be compared to the shape measure
$\sigma_{\tau}/\langle \tau \rangle$ (solid curve).
The shape measure is independent of the cosmological parameters and
has a minimum for the prism shaped Voronoi domain, i.\,e.\ for $\rho=\pi/4$.
As can be seen in figure \ref{fig:s60_ueber_rho},
there are values of $\tau_{\hbox{\scriptsize sls}}$ for which the minimum
in the $S(60^{\circ})$ statistics does not occur at $\rho=\pi/4$.
Thus, Voronoi domains that are more oddly shaped than the prism
produce less CMB anisotropy on large angular scales.
This again demonstrates that the geometry of the Voronoi domain
is only partially responsible for the anisotropy suppression.


\section{Summary and Discussion}


The main motivation for cosmic topology is the low power of
the CMB anisotropy that is observed at large angular scales
in the data of COBE \cite{Hinshaw_et_al_1996}
and WMAP \cite{Spergel_et_al_2003}.
The large-scale behaviour is revealed by the
temperature 2-point correlation function $C(\vartheta)$ and could be
at variance with the $\Lambda$CDM concordance model based
on a space with infinite volume as emphasised by 
\cite{Aurich_Janzer_Lustig_Steiner_2007,Copi_Huterer_Schwarz_Starkman_2008,%
Copi_Huterer_Schwarz_Starkman_2010}.
The reality of this discordance is, however, questioned in
\cite{Efstathiou_Ma_Hanson_2009,Bennett_et_al_2010}
where a reconstruction algorithm is used for the masked sky regions.
It turns out that significant power arises just in
those reconstructed regions.
Using only save data from sky regions, which are not masked,
a very low temperature correlation is obtained
for angles $\vartheta\gtrsim 60^\circ$.
The papers \cite{Aurich_Lustig_2010,Copi_Huterer_Schwarz_Starkman_2011}
conclude that it is very likely that the low power at large angles is real.
And if the low power behaviour is not a statistical fluke,
models, which naturally have few CMB correlations on large scales,
deserve a further investigation.

Because of the large number of topological spaces,
it would be desirable to have a guiding principle
that reduces the number of possibilities.
It is claimed in \cite{Weeks_Luminet_Riazuelo_Lehoucq_2005}
that well proportioned manifolds can describe the low power of the observed CMB,
but the well-proportioned conjecture is not quantified there.
To quantify this conjecture by a shape measure,
the variance of the spherical distance to the surface of the
fundamental cell defined as the Voronoi domain is introduced
in sec.\,\ref{sec:shape_measure}.
This variance is small for almost spherical Voronoi domains
and large for oddly shaped ones.
The lens spaces $L(p,p/2-1)$ are at the focus of this paper
since they provide inhomogeneous spaces so that the shape
of the Voronoi domain depends on the position of the observer.
They are thus  predestined for a test of the well-proportioned conjecture.
Furthermore, for two special positions of the observer,
they have Voronoi domains identical to those of the homogeneous
$L(p,1)$ and ${\cal D}_{p}$ spaces.

The first observer position is characterised by $\rho= 0$ and
results in a lens shaped Voronoi domain. 
Therefore, the manifold $L(p,p/2-1)$ at $\rho= 0$ has the same value 
for the shape measure as the homogeneous lens space $L(p,1)$.
The CMB anisotropies at large angular scales of both 
are compared in sec.\ \ref{subsec:cmb_correlation_and_well_prop},
and it turns out that the correlations of the CMB of $L(p,p/2-1)$
at $\rho= 0$ and of $L(p,1)$ are different contrary
to the well-proportioned conjecture.

The second observer position is at $\rho= \pi/4$,
in which case the Voronoi domain of $L(p,p/2-1)$ is prism shaped
and identical to that of the prism space ${\cal D}_p$.
For these two topologies, the $S(60^{\circ})$ statistics of the CMB
also diverges at least for large distances
$\tau_{\hbox {\scriptsize sls}}\gtrsim 0.8$ to the surface of last scattering.
Thus, a second counter-example to the well-proportioned hypothesis is found.
For this reason, the conjecture does not provide a save criterion
for the decision whether a multi-connected universe is an eligible candidate
for our Universe or not.

Interestingly, for small values of $\tau_{\hbox {\scriptsize sls}}$,
the prism shaped universe $L(p,p/2-1)$ at $\rho= \pi/4$ 
does behave almost identical to the prism space ${\cal D}_p$
with respect to the statistical behaviour of the CMB anisotropies.
This similarity can be traced back to the largest common cyclic subgroup $Z_p$
of Clifford translations that is contained in their deck groups
as outlined in sec.\ \ref{subsec:subgroups}.
If that subgroup occurs in both spaces with the same multiplicity,
the CMB anisotropies turn out to be similar.
Conversely, if it occurs with different multiplicities in the two spaces
or in only one of them, the CMB properties are different.
Furthermore, the order $p$ of the cyclic subgroup $Z_p$ is linked to
the scale on which the power of the CMB anisotropy is suppressed.
The larger the order $p$, the smaller distances $\tau_{\hbox {\scriptsize sls}}$
are necessary for a CMB anisotropy suppression and this translates
to smaller densities $\Omega_{\hbox {\scriptsize tot}}>1$.
Thus, almost flat cosmological models can be obtained
when this subgroup possesses a sufficient high group order $p$.
The distinct CMB statistics of the prism shaped universes
for large distances $\tau_{\hbox {\scriptsize sls}}$
as well as those of the lens shaped universes is explained by such
cyclic subgroups.

In addition to the lens and prism shaped fundamental cells,
sec.\ \ref{subsec:subgroups} also analyses
the minima of the $S(60^{\circ})$ statistics of the three binary polyhedral
spaces ${\cal T}$, ${\cal O}$, and ${\cal I}$ in terms of subgroups.
The superior behaviour of the Poincar\'e dodecahedron ${\cal I}$ 
is due to the sixfold occurrence of the $Z_{10}$ subgroup
of Clifford translations.

The dissimilar CMB behaviour of topologies sharing the same Voronoi domain
is evidently rooted in the transformation properties of
the deck groups on the sphere $S_{\hbox {\scriptsize sls}}$ of last scattering.
In sec.\ \ref{subsec:transfomation_on_sls}
the translation distances for points on $S_{\hbox {\scriptsize sls}}$
of the elements of the deck group
are compared between homogeneous and inhomogeneous spaces.
In those cases where the smallest translation distances are the same
between two spaces, a similar $S(60^{\circ})$ statistics is observed
and vice versa.
This provides a complementary picture to the cyclic subgroups.

Although the main topic of this paper is the study of CMB statistics
of lens and prism shaped Voronoi domains,
sec.\ \ref{subsec:CMB_for_general_rho} shows, how the CMB properties
vary by moving the observer through the $L(p,q)$ space.
This demonstrates the large range of variation for a fixed set
of cosmological parameters within a given manifold.
The comparison with the shape measure $\sigma_{\tau}/\langle\tau\rangle$
shows that there are examples where more oddly shaped Voronoi domains
exhibit less correlations in the CMB power
contrary to the well-proportioned conjecture.

Summarising, one concludes that the CMB behaviour results from
complex interwoven ingredients such that an individual analysis of
topological spaces is necessary in order to deliver a judgement
whether a topology leads to suitable CMB properties.


\section*{Acknowledgements}

We would like to thank the Deutsche Forschungsgemeinschaft
for financial support (AU 169/1-1).


\section*{References}

\bibliography{../bib_astro}

\bibliographystyle{h-physrev5}

\end{document}